\documentclass[aps,prx,twocolumn]{revtex4-1}
\usepackage{epsfig,amssymb,amsmath,mathrsfs,amsthm,graphicx,float,color}
\usepackage[matrix,frame,arrow]{xypic}
\usepackage[matrix,frame,arrow]{xy}
\usepackage{amsmath}

\newcommand{\qw}[1][-1]{\ar @{-} [0,#1]}
\newcommand{\multigate}[2]{*+<1em,.9em>{\hphantom{#2}} \qw \POS[0,0].[#1,0];p !C *{#2},p \save+LU;+RU **\dir{-}\restore\save+RU;+RD **\dir{-}\restore\save+RD;+LD **\dir{-}\restore\save+LD;+LU **\dir{-}\restore}

\newcommand{\Qcircuit}[1][0em]{\xymatrix @*[o] @*=<#1>}  
 \renewcommand{\Qcircuit}[1][0em]{\xymatrix @*=<#1>}

\newcommand{\pureghost}[1]{*+<1em,.9em>{\hphantom{#1}}}

\newcommand{\poloFantasmaCn}[1]{{{}^{#1}_{\phantom{#1}}}}

\usepackage{subfigure}
\newcommand{\N}{\mathbb{N}}

\newcommand{\R}{\mathbb{R}}
\newcommand{\C}{\mathbb{C}}

\newcommand{\set}[1]{\mathsf{#1}}

\newcommand{\spc}[1]{\mathcal{#1}}

\def\d{{\rm d}}

\newcommand{\Span}{{\mathsf{Span}}}

\def\>{\rangle}
\def\<{\langle}

\newcommand{\st}[1]{\mathbf{#1}}
\newcommand{\bs}[1]{\boldsymbol{#1}}     
\newcommand{\Sp}{\mathsf{Sp}}

\newcommand{\map}[1]{\mathcal{#1}}
\newcommand{\Tr}{\operatorname{Tr}}

\newtheorem{theo}{Theorem}

\newtheorem{lemma}{Lemma}
\newtheorem{prop}{Proposition}
\newtheorem{cor}{Corollary}
\newtheorem{defi}{Definition}

\newtheorem{eg}{Example}

\def\Proof{{\bf Proof.~}}
\def\qed{$\blacksquare$ \newline}

\begin{document}
\title{
Optimal quantum operations  at zero energy cost
} 
\author{Giulio Chiribella$^{1,2,3}$ and Yuxiang Yang$^{1,2}$}  
\affiliation{$^1$Department of Computer Science, The University of Hong Kong,
Pokfulam Road, Hong Kong  }
{\affiliation{$^2$ The University of Hong Kong Shenzhen Institute of Research and Innovation, Kejizhong 2$^{\rm nd}$ Road,  Shenzhen}
\affiliation{$^3$Canadian Institute for Advanced Research,
CIFAR Program in Quantum Information Science, Toronto, ON M5G 1Z8}
\begin{abstract}
Quantum technologies are  developing powerful tools to generate and manipulate  coherent superpositions of different energy levels. Envisaging a new generation of energy-efficient quantum devices, here we explore how coherence can be manipulated without exchanging energy with the surrounding environment. We start from  the  task of converting  a coherent  superposition of energy eigenstates into another. We identify the  optimal energy-preserving operations, both in the deterministic and in the probabilistic scenario. We then  design a recursive protocol, wherein a  branching sequence of energy-preserving filters increases the probability of success while reaching  maximum fidelity at each  iteration. Building on the recursive protocol, we  construct  efficient approximations of the optimal fidelity-probability trade-off,  by taking  coherent superpositions of  the  different branches generated by probabilistic filtering. The benefits of this construction  are illustrated in applications  to quantum metrology, quantum cloning,  coherent state amplification,  and ancilla-driven computation. Finally, we extend  our results to transitions where the input state is generally mixed and we apply our findings to the task of purifying  quantum coherence. 
\end{abstract}
\maketitle

\section{Introduction}
 Rapid experimental advances  are pushing towards  the realization of 
 new quantum technologies \cite{nobel1,nobel2,georgescu2014quantum,fuechsle2012single,barends2014superconducting,hensen2015loophole}.
     Decoherence still remains the grand challenge, but,    as   quantum technologies approach real-life applications, questions of  energy efficiency are bound to become increasingly more relevant.      Nowadays, energy efficiency is one of the major problems in    information and communication technology \cite{fettweis2008ict} and, as such, it  is   the object  of    a large amount of research, both experimentally \cite{ionescu2011tunnel,mukhanov2011energy} and theoretically \cite{berl2010energy,mittal2014survey}.  In this area, quantum technologies hold a large, relatively unexplored potential, which is likely to become critical in the long term future.       In this perspective, it is compelling to  explore the ultimate performances achieved by quantum devices with limited energy resources.    The problem is not only of fundamental interest.  Pioneering experiments in quantum optomechanics  have already  started  to develop the  tools for manipulating  quantum systems with minimal amounts of energy \cite{aspelmeyer2014cavity}. Similarly, engineered light-matter interactions in quantum dots  \cite{RMPsingle} and superconducting circuits  \cite{xiang2013hybrid}  enable the control of  dynamics  at the  level of  single quanta, offering a  promising platform for the  realization of  prototypes of energy-optimized  quantum devices. 

In order to address the question of energy efficiency,   one needs  to characterize   the quantum operations that can be performed with given energy resources.    Concretely, an energy resource is described  by the state of a  battery,~i.e.~an auxiliary system that exchanges energy with the system used as the information register.   The constraint that the battery is the {\em only} energy resource used in the processing   amounts to the requirement that  the joint evolution of register and battery  be  \emph{energy-preserving}.
   In general,  energy-preserving  evolutions need not be reversible: the register and the battery can interact non-trivially with auxiliary degrees of freedom, as long as they do not exchange  energy with it.    This scenario is  illustrated in Figure \ref{fig:battery}.     The  achievable operations  are then  modelled  as reduced evolutions of the information register, with the battery initialized in a given resource state.  In this model,  
    characterizing and optimizing  the  energy-preserving operations on the composite system of battery and information register is an essential step towards characterizing  and optimizing the achievable operations on the information register alone.    

\begin{figure}[h!]
\centering
\subfigure[]
{\label{fig:battery}
\includegraphics[width=0.9\linewidth]{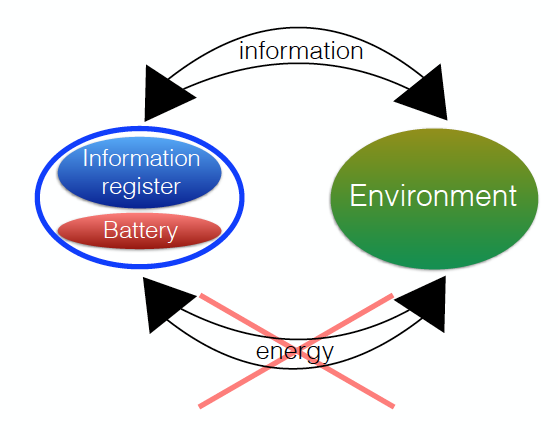}
}
\subfigure[]
{\label{fig:nobattery}
\includegraphics[width=0.7\linewidth]{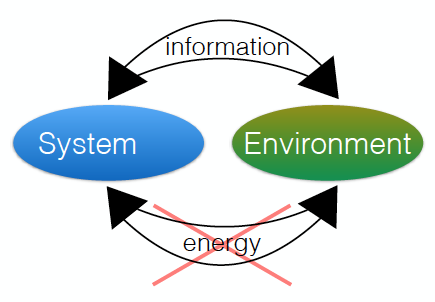}
}

\caption{{\bf  Manipulating quantum information with limited energy resources.  }     (a)  Quantum device powered by a battery.  The device contains an  information register, where data are stored, and a battery,   representing  the energy  resource(s)  used to  operate  on the data.   The information register and the battery are  allowed to interact with the surrounding environment (possibly including ancillas), as long as the interaction involves no exchange of energy.   The situation where the device uses energy from the environment to aliment its battery can be included   in the model  by  formally regarding the energy sources in the environment as part of the battery.    (b) Quantum device  in low-power mode.   In this mode, some quantum operations are performed without the aid of the battery,~i.e., relying only on the interaction between the information register (here denoted simply as ``the system")  and the environment (possibly including ancillas).    The evolution of the system is described  by an energy-preserving operation.   }
\end{figure}

The focus of this paper is the characterization and optimization of energy-preserving operations, which provide the foundation to  the  broader  programme aimed at  designing energy-optimized quantum devices.  Energy-preserving operations  are also interesting \emph{per se},  as  the operations that can be realized without the assistance of external  energy resources.  This property is  appealing in situations where a device has to switch to a ``low-power mode," as it is often the case for nowadays mobile devices and will arguably be the case also for  future devices incorporating quantum gadgets  \cite{chips}. In these situations, it may become convenient to disconnect part of the device from the battery and to let that part of the device operate in an energy-autonomous way, as illustrated in  Figure \ref{fig:nobattery}.  
Studying energy-preserving operations is also a  useful strategy to derive results about other related types of operations.  For example, the energy-preserving family includes as a special case  the operations that can be achieved with passive optical elements like beam splitters, phase shifters, and mirrors.  As a result, the optimization of a desired task over all   energy-preserving operations yields upper bounds on the performance of arbitrary quantum circuits built with passive optical elements.    Results about energy-preserving operations can  also be   used in situations involving the preservation of observables other than  the energy.  This is an important point, because   constraints on the preservation of  suitable observables occur naturally  in many applications.   An interesting example arises in quantum dots, where the implementation of logical  gates  often benefits from the existence of ``sweet spots"---special working points where the  charge noise is suppressed.   The set of gates that can be performed at the sweet spot is limited: for example,  in the three-electron exchange-only spin qubit \cite{sweet1}, only the rotations around the $z$ axis can be performed  at the sweet spot, while all other rotations  incur into undesired  noise \cite{sweet2}.  Regarding the component of the spin along the $z$ axis as the ``energy,''    it follows  that the  ``energy-preserving" channels are exactly the operations that can be performed with suppressed noise. 
  Similar physics arises in superconducting flux qubits,  where the sweet spot is with respect to magnetic noise  \cite{sweet3}.

This paper characterizes the set of energy-preserving operations and  identifies the most efficient strategies for  the manipulation of quantum states exhibiting coherence across energy levels.  We will start from the basic task of  transforming a pure superposition of energy eigenstates  into another  pure superposition.   In this context, it is interesting  to consider not only  deterministic operations, but also  probabilistic operations  arising from measurements that can be implemented at zero energy cost.          
By allowing for a non-unit probability of success, we find out that the constraint of energy preservation  can be stretched to a previously unsuspected extent.  For example, we will see  that  a beam of $N$ atoms, each of them prepared in the superposition  $ |S\>  =(|E\>  +  |G\>)/\sqrt 2$  of the ground state and the first excited state, can be probabilistically transformed at no energy cost into a stronger beam of   $N^{2-\epsilon}$ atoms in a state that is nearly identical to the state of $N^{2-\epsilon}$ identical copies of  $|S\>$, up to an exponentially small error.  
The ability to efficiently approximate forbidden transformations of coherence at zero energy cost is a new twist of the postselection approach widely applied  in quantum information \cite{weakvalues,aaronson,Bergou-Hillery1,Bergou-Hillery2,BaganDis,FiurasekEst,BaganPRA,BaganPRL,GiulioNat,chiribella2016quantum,fiurasekclon,QCMC,chiribella-xie,zhao2017quantum} and complements existing results on the resource theory of coherence  \cite{delocal,incoherence,vaccaro2008trade-off,rudolphPRX,aaberg-2014-prl,cwiklinski-2015-prl,marvian2016quantify}. 


After having characterized the structure of the optimal energy-preserving operations, we  move to  a different scenario, where the probability of success is not fixed {\em a priori}.  We consider adaptive protocols whereby the experimenter performs repeated rounds of probabilistic operations and  is free to decide on the fly whether to be content with the result obtained so far or whether to continue further.     In this scenario we design a recursive protocol, consisting of a sequence of energy-preserving binary measurements that   produce at each step  the best approximation of the target with the highest probability allowed by quantum mechanics.    Subsequent iterations of our protocol lead to an  increasing probability of success, but also to a degradation of the state of the system and, eventually, to the loss of the advantages of postselection.   
 This behavior is a consequence of the inevitable trade-off between the enhanced performance of probabilistic transformations and their reduced probability of success. The advantage of our recursive protocol is that it gives an explicit, ready-to-apply method to construct lower bounds  to the optimal trade-off curve between fidelity and probability of success, a problem that so far has been solved only in one case  \cite{BaganPRL}.   Remarkably, in this particular case we find out that our protocol reproduces the optimal trade-off curve, provided that the input state is a superposition of sufficiently many energy levels.   We conjecture that our  protocol is asymptotically optimal also in those situations where the input and the target consist of many identical  copies of pure states,  an example being the asymptotic cloning of quantum coherence.   Independently of the validity of this conjecture, the importance of the recursive protocol can be best appreciated in all those cases where the optimal trade-off  curve is not explicitly known.  To get even better lower bounds to the optimal curve, we finally introduce the   operation of  \emph{coherent coarse-graining}, which consists in joining different outcomes into a single quantum operation.  Coherent coarse-graining allows one to  keep  the same probability of success of the outcomes that are joined, while increasing the fidelity with the target.    Combined with our recursive protocol, this  operation provides a canonical  way to generate  analytical lower bounds to the optimal trade-off between fidelity and success probability, whose exact calculation is generally hard to perform without resorting to numerical optimization.   
    
 To demonstrate the broad applicability of our methods, we illustrate the  recursive protocol and its coherent coarse-graining in a number of concrete examples, including quantum  phase estimation,  energy-preserving amplification of coherent states, and the optimal design of correction operations for ancilla-driven computation.  In addition to the  applications presented explicitly in the paper, our results can be directly applied to most of  the canonical problems of quantum information processing, such as optimal state discrimination,  gate programming, entanglement conversion, universal NOT, and universal transpose---whose  implementation is  significantly affected once one imposes the requirement that  no energy should be drawn from the environment. 

Finally, we go beyond the regime of pure state transitions and extend our results to transitions where mixed states are given as inputs. For such transitions, we prove upper bounds for the performances of both deterministic and probabilistic operations, providing conditions for the saturation of the bounds.  This extension provides optimal strategies for the implementation of quantum tasks such as  purification   \cite{cirac-ekert-macchiavello,andersen,zhao2017quantum}  and  broadcasting of mixed states  \cite{dariano-macchiavello-perinotti-prl}. 



The paper is organized as follows. Section \ref{sec:general} introduces the basic framework.    In Section \ref{sec:pure} we characterize the optimal energy-preserving process.
  Using this result, we  construct the recursive protocol and study the operation of coherent coarse-graining in Section  \ref{sec:network} and apply it to several tasks in quantum information processing  (Section \ref{sec:examples}), including phase estimation (Subsections \ref{subsec:estimation} and \ref{subsec:maxcoh}), state cloning (Subsection \ref{subsec:cloning}), coherent light amplification (Subsection \ref{subsec:coherent}) and ancilla-driven computation (Subsection \ref{sec:unlearn}).The extension of our results to mixed states is discussed in Section \ref{sec:mixed}. 
  Finally, the conclusions are drawn in Section \ref{sec:conclusion}. 

\section{The energy-preserving paradigm}\label{sec:general}

In this section we introduce the framework that will be adopted in the rest of the paper. We first present the class of energy-preserving channels, which model  deterministic evolutions that can be implemented without drawing energy from the environment. 
We then move to probabilistic operations, characterizing  the stochastic evolutions that can be implemented at zero energy cost.

\subsection{Energy-preserving  channels}\label{subsec:enpreschan}


Consider a quantum system  interacting with the surrounding environment from time $t_1$ to time $t_2$ through  an interaction    Hamiltonian $H_{\rm int} (t)$, which we assume to be zero before $t_1$ and after   $t_2$.   As a result of the interaction, the system and the environment evolve jointly via the unitary operator 
\begin{align}\label{dyson} 
U   = \map T      \exp \left\{-\frac{i}{\hbar} \,     \int_{t_1}^{t_2}   \, \d t  \,   \left[  H_{\rm sys}   +   H_{\rm env}  +  H_{\rm int} (t)  \right]\right\}  \, ,
\end{align}  
where $\map T  \exp$ denotes the time-ordered exponential and $H_{\rm sys}$ and $H_{\rm env}$ are the Hamiltonians of the system and of the environment, both assumed to be time-independent.  
Regarding energy as a resource,  we require  the evolution to satisfy the condition
\begin{align}\label{totalenergy-pres}
U^\dag   \, (H_{\rm sys}  +   H_{\rm env}   )  \,   U  =  H_{\rm sys}   +  H_{\rm env}  \, ,
 \end{align}  
 meaning that the  total energy after $t_2$ is equal to the total energy before $t_1$.  A sufficient condition for the validity of Eq. (\ref{totalenergy-pres}) is  the commutation relation  
 \begin{align}\label{hamiltonian-pres}
 [  H_{\rm int}( t ),   H_{\rm sys}  +   H_{\rm env}]   =  0  \qquad \forall t\in  [t_1,t_2] \, ,
\end{align}  
which guarantees that the sum of the system and environment energies is a constant of motion during the entire evolution.  Our analysis covers this case and, possibly, more general scenarios where the sum of the system and environment energies is not a constant of motion at all times:    condition (\ref{totalenergy-pres}) is generally weaker than  condition (\ref{hamiltonian-pres}).  Note however that conditions  (\ref{totalenergy-pres})  and  (\ref{hamiltonian-pres}) define the same set of input-output evolutions from time $t_1$ to time $t_2$:  for every unitary $U$  that commutes with   $H_{\rm sys}+  H_{\rm env}$   one can always find  a suitable interaction Hamiltonian  $H_{\rm int}'(t)$ that generates $U$ as  and commutes with  $H_{\rm sys}+  H_{\rm env}$ at all times.   
 
Among the evolutions  that conserve the total energy, we are interested in those that leave  the energy of the environment untouched. Such evolutions satisfy the additional condition
\begin{align}\label{environment-pres}
U^\dag  H_{\rm env} U  =  H_{\rm env} \, .
 \end{align}  
Clearly,  the combination of Eqs. (\ref{totalenergy-pres})  and  (\ref{environment-pres}) implies that the evolution has to preserve the   energy of the system,  namely  
\begin{align}\label{energy-pres}
U^\dag  H_{\rm sys} U  =  H_{\rm sys} \, .
 \end{align}

Assuming that the environment is initially in the state $\rho_{\rm env}$,      the effective evolution of the system is described by the quantum channel  (completely positive trace-preserving map) $\map M$ defined by   
\begin{align}\label{stine}
\map M(\rho)   =  \Tr_{\rm env}  \left[  U   \,  ( \rho  \otimes  \rho_{\rm env})  \,  U^\dag\right]  
\end{align}  
where $\Tr_{\rm env}$ denotes the partial trace over the Hilbert space of the environment.
By construction, the channel $\map M$ preserves the expectation value of the system's energy. 
Even more strongly,   condition (\ref{energy-pres}) 
implies that the channel $\map M$ preserves  also the variance and all the momenta  of the Hamiltonian, namely
\begin{align}
\label{MdagH}
\map M^\dag \left(H^n_{\rm sys}\right)  =  H^n_{\rm sys}  \qquad \forall n  \in  \N  \,  ,  
\end{align}
where $\map M^\dag$  is the  completely positive identity-preserving map describing the evolution in the Heisenberg picture, defined by
\begin{align}
\map M^\dag (A)   =     \Tr_{\rm env}  [  \,     (I_{\rm sys} \otimes \rho_{\rm env})     \,  U^\dag  \,  (A\otimes I_{\rm env} ) \,  U \, ]      \, ,
\end{align} 
for arbitrary operators $A$.    When Eq. (\ref{MdagH}) is satisfied, we say that  $\map M$ is an \emph{energy-preserving channel}.  
  The energy-preserving  condition (\ref{MdagH})  is equivalent to the requirement that the  evolution $\map M$ does not affect the probability distribution of the energy,~i.e. that   one has the equality of probabilities
  \begin{align}\label{energy}
 \Tr [   P_E   \,  \map M(\rho) ]  =   \Tr[  P_E  \, \rho]   \, ,
 \end{align}
where  $\rho$ is an arbitrary state,  $E$ is an arbitrary eigenvalue of $H_{\rm sys}$, and $P_E$ is the projector on the eigenspace corresponding to $E$.   

It is easy to see that every energy-preserving channel $\map M$ is \emph{covariant} with respect to the free time evolution of the system,  that is, $\map M$ satisfies the condition  
\begin{align}
\map M(    U_t  \cdot    U_t^\dag  )  =   U_t  \,  \map M(\cdot)  \,  U_t^\dag \qquad \forall t\in  \R  \, ,
\end{align}     
with $U_t=  \exp[-itH_{\rm sys}/\hbar]$.  However, the converse is not true in general: for example, a channel that discards the input state and re-prepares an eigenstate of the energy is covariant but not energy-preserving.   
Physically, the difference between energy-preserving and covariant channels is that covariant channels preserve the sum $H_{\rm sys}  +   H_{\rm env}$, while the energy-preserving channels separately preserve the individual summands $H_{\rm sys}$ and $H_{\rm env}$.    Further discussion on the relation between energy-preserving channels and other classes of channels, such as incoherent channels and  Hadamard channels,  is presented in Appendix \ref{app:relations}.

Energy-preserving channels provide an economical  way to process information in situations where energy  becomes a scarce resource. For example, one can imagine a scenario where a microscopic machine has to perform a desired task on the system (like copying data, correcting for an error, or erasing information) without changing the energy of the surrounding environment.   
While scenarios of this kind are not a practical reality yet, prototype demonstrations of energy-preserving channels can be conceived for quantum systems with a high degree of control, such as   ion traps  \cite{ion1,ion2},   
optical cavities \cite{cavity1,cavity2},  optical lattices \cite{lattice1,lattice2}, and optomechanical systems \cite{optomechanics1,optomechanics2}.

  \subsection{Structure of the energy-preserving channels}\label{subsect:struct}
 We have seen that  every energy-preserving interaction   induces an energy-preserving quantum channel, i.e., a channel satisfying Eq. (\ref{MdagH}).
  The converse is also true:  
 given an energy-preserving channel $\map M$, one can always engineer an interaction between the system and the environment that conserves the total energy and    does not draw energy from the environment.      To establish this fact, note that the map $\map M^\dag$  satisfies the condition  
   \begin{align}\label{algpres}
 \map M^\dag  (A)  =     A  \qquad \forall  A\in\map A \, , 
 \end{align}   
where $\map A $   is the commutative algebra generated by the powers of the Hamiltonian. The algebra $\map A$ contains the identity and is closed under adjoint.  Technically,  algebras  of this kind are known as unital $*$-algebras   \cite{Bratteli_Robinson}.    
For any such  algebra,  the   maps   that satisfy Eq. (\ref{algpres}) are characterized  by a simple lemma: 

 \begin{lemma}[Lindblad \cite{Lindblad}]\label{lem:lindblad}
 Let $\map M^\dag$ be an identity-preserving completely positive map, written in the Kraus form $\map M^\dag  (A)   =  \sum_{k=1}^K  M_k ^\dag  AM_k $ and let     $\map A$ be  a unital $*$-algebra  $\map A$.  
   The map $\map M^\dag$ preserves the elements of $\map A$ if and only if each Kraus operator $M_{k}$ belongs to the commutant of  $\map A$,~i.e., to the set of operators 
 \begin{align*}
\map A'  :  =  \{    B  \in  \mathcal B  (\spc H)  \, , \,   [ A, B ]  =  0    \quad  \forall A \in  \map A  \, \} \, ,
 \end{align*}
 $\mathcal B (\spc H)$ denoting the set of bounded operators on $\spc H$. 
   \end{lemma}  
In the case of the Abelian algebra generated by the powers of $H_{\rm sys}$,    the commutation condition reduces to 
\begin{align} [  M_k,  H_{\rm sys}]  =  0 \, ,
\end{align} meaning  that each Kraus operator   $M_k$ must be of the block diagonal  form
\begin{align}\label{block}
M_k   =   \bigoplus_{E}   \,      P_E  M_k  P_E  \,  ,
\end{align} 
with the sum running over the eigenvalues of $H_{\rm sys}$.    All throughout the paper  we will assume the energy spectrum to be discrete.

As a consequence of the block diagonal form (\ref{block}),  one can realize the channel $\map M$ through an energy-preserving isometry.  
Specifically, one can express the channel as 
\begin{align}\label{Vrealization}
\map M(\rho)   =  \Tr_{\rm env}  [  V  \, \rho \, V^\dag] \, ,
\end{align} where $V$ is the isometry defined by 
\begin{align}\label{iso}
V  :  \spc H_{\rm sys}  \to \spc H_{\rm sys}\otimes \spc H_{\rm env} \, , \qquad V : =  \sum_{k=1}^K  M_k  \otimes   |    \phi_k\>   \,  , 
\end{align}
$   \{\, |\phi_k\>\,\}_{k=1}^K$ being a set of orthonormal states in the environment's Hilbert space.       
With this definition, the isometry $V$ satisfies the relation 
\begin{align}\label{Vpres}
V   H_{\rm sys}   =   (H_{\rm sys}\otimes  I_{\rm env})   \,  V   \, .
\end{align}
 In turn,  Eq. (\ref{Vpres}) implies  that the isometry $V$ can be realized via an interaction  that preserves both the energy of the system and the energy of the environment: 
 \begin{theo}\label{theo:enpres'}
Let $\map M$ be a quantum channel transforming states on $\spc H$.  Then, the following are equivalent:  
\begin{enumerate}
\item $\map M$ is energy-preserving; 
\item  $\map M$  can be realized through a joint evolution of the system together with an environment of the following form:     
\begin{align}\label{covunitary}
\map M(\rho)   =   \Tr_{\rm env}  [   U   \, ( \rho  \otimes |\phi_0\>\<\phi_0| ) \,  U^\dag] \, ,
\end{align} 
where  $|\phi_0\>$ belongs to ground eigenspace of  $H_{\rm env}$    and  $U$ is a unitary evolution that commutes with $H_{\rm sys }$ and with $H_{\rm env}$;
\item $\map M$ can be realized through a joint evolution as in Eq. (\ref{covunitary}), where the unitary $U$ is generated by an interaction $H_{\rm int}  (t)$ that commutes with $H_{\rm sys}$ and $H_{\rm env}$ at all times. 
\end{enumerate}  
\end{theo}
The proof is presented in Appendix \ref{app:proofV}.

Note the appearance of the Hamiltonian of the environment in the statement of the Theorem.   This is  natural, because in general we are dealing with the evolution of an open system.  In this situation, the Hamiltonian of the environment plays a crucial role, for it determines the minimum amount of energy one has to invest into the realization of the desired  channel.  Ideally, Theorem \ref{theo:enpres'} guarantees that such amount can be reduced to zero in the case of energy-preserving channels. Specifically, the desired evolution  can be engineered by initializing the environment in an eigenstate of its Hamiltonian and by turning on a  coupling that preserves the individual energies of system and environment, keeping the latter inside the ground space for the whole time evolution. As a result, the evolution is implemented at zero energy cost, at the price of an entropy increase in the environment, which is generally left in a mixture of states with the same energy. In other words, the environment is only used passively as  computational workspace wherein information can be stored.  
 

\subsection{Energy-preserving instruments}
While in the case of deterministic evolutions the notion of energy preservation is pretty straightforward, the situation is different for probabilistic transformations induced by quantum measurements.  
In this section we introduce a notion of \emph{probabilistic energy-preserving transformations}, which characterizes those operations that can be implemented (in principle) without paying an energy cost.

Adopting Ozawa's model of the measurement process \cite{ozawa},   we  view  probabilistic evolutions  as the result of a unitary interaction between the system and the environment, followed by the projective measurement of a ``meter observable" $O$  on the environment.  
In this model, the preservation of the energy imposes constraints on the interaction as well as constraints on the measurement. 
Like in the previous sections, we demand that the system-environment interaction preserve the total energy of the system and the environment [Eq. (\ref{totalenergy-pres})] and do not change the energy of the environment [Eq. (\ref{environment-pres})].   
As we  argued in the previous paragraph, these requirements characterize the evolutions that can be implemented at zero energy cost. 
Regarding the measurement, we demand that it should not disturb the energy of  the environment, or, equivalently, that the meter observable and the energy are compatible observables, namely  
\begin{align}\label{compatible}
[O,H_{\rm env}]  = 0  \, .
\end{align}
If this condition were not satisfied, the measurement of $O$ would collapse an eigenstate of the energy into a random eigenstate of $O$, thus altering the expectation value of the energy.  
  Observables that commute with the Hamiltonian of the environment   are the  prototype of measurements that can be performed without paying an energy cost.  The conditions (\ref{totalenergy-pres}),  (\ref{environment-pres}),   and  (\ref{compatible}) are the standard requirements put forward    in the Wigner-Araki-Yanase Theorem \cite{W,AY, ozawa2} and in all the works that followed it up   \cite{ozawaCNOT,computing3,spekkens-marvian-WAY,ahmadi}.    In this context the commutation relation (\ref{compatible})  is known as \emph{Yanase's condition} \cite{ozawa2}.  Recently, the same framework discussed here has been used as the starting point to define energy requirements for the implementation of quantum measurements  \cite{ahmadi,popescu}.  

Let us analyze the probabilistic evolutions resulting from the  requirement of zero energy cost.   According to   quantum measurement theory  \cite{ozawa,lathi_book,holevo_stat_book,heinosaar_ziman}, the measurement  of $O$ induces a stochastic evolution of the state of the  system, described by a \emph{quantum instrument}, namely a collection of quantum operations  
(completely positive trace non-increasing maps) $ \{  \map M_x\}_{x\in\set X}$   subject to the normalization condition  
\begin{align}\label{normalization}  
\sum_{x\in\set X}       \Tr   [  \map M_x  (\rho)]  =  \Tr [\rho] 
\end{align} 
for every quantum state $\rho$.    For a system prepared in the state $\rho$, the measurement  generates the outcome  $x\in \set X$ with probability 
\begin{align}\label{pxrho}  
p (x|\rho)  =   \Tr  [  \map M_x  (\rho)]   \, ,
\end{align} 
and, conditionally on outcome $x$,  returns   the  system in the state 
\begin{align}\label{rhoprime}
\rho_x'    =    \frac{\map M_x(\rho)}{\Tr[\map M_x (\rho)]}  \, .
\end{align}

In the model considered here, the set of outcomes  is the spectrum of the meter observable $O$ and the quantum operation $\map M_x$ is   defined by 
  \begin{align}\label{qostine}
  \map M_x (\rho)   =   \Tr_{\rm env}  [   (I_{\rm sys}  \otimes Q_x)  \,    U  \,(\rho \otimes \rho_{\rm env})  \,  U^\dag]  \, ,
  \end{align}
$Q_x$ being the projector on the eigenspace of $O$ with eigenvalue $x$.   Note that, by summing over all possible outcomes, one obtains 
\begin{align*}
\sum_{x\in\set X}   \map M_x  (\rho)  &  =       \Tr_{\rm env}  [     U  \,(\rho \otimes \rho_{\rm env})  \,  U^\dag] \\
 &  \equiv    \map M  (\rho) 
\end{align*} 
where $\map M$ is an energy-preserving channel.  Energy preservation for $\map M$ simply follows from just the condition $  U^\dag  \,  H_{\rm sys}  \,  U  =  H_{\rm sys}$.  Note that this conclusion is independent of the validity of Yanase's condition, because the sum over all measurement outcomes yields the identity operator, no matter what type of measurement is performed. 

The physical model discussed so far motivates the following definition:  
\begin{defi}
We say that a quantum instrument $\{\map M_x\}_{x\in\set X}$ is \emph{energy-preserving} if the average channel $\map M  :  =  \sum_{x\in\set X}  \, \map M_x$ is energy-preserving.     
\end{defi}
Energy-preserving instruments are exactly the instruments that  can be implemented (in principle) at zero energy cost. Precisely, every such instrument can be realized via an energy-preserving interaction, followed by the measurement of a meter observable that commutes with the energy:
 \begin{theo}\label{theo:enpresinst}
Let $\{\map M_x\}_{x\in\set X}$ be a quantum instrument transforming states on $\spc H$.  Then,   $\{\map M_x\}_{x\in\set X}$  is energy-preserving if and only if one has 
\[\map M_x(\rho)   =   \Tr_{\rm env}  [    (I_{\rm sys}  \otimes Q_x)    U    ( \rho  \otimes |\phi_0\>\<\phi_0| )  U^\dag]  \qquad \forall x\in\set X\]  
where   $|\phi_0\>$ is a ground state of the environment's Hamiltonian,   $U$ satisfies  the conditions $U^\dag  \,  H_{\rm sys} \,     U  =  H_{\rm sys}   $ and $U^\dag  \,  H_{\rm env} \,     U  =  H_{\rm env}   $, and $\{Q_x\}$ is a projective measurement satisfying   Yanase's condition  $[Q_x,  H_{\rm env}]=0$  $\forall x\in\set X$.
\end{theo}
The proof can be found in Appendix \ref{app:proofV}, while a simple illustration of the result is shown in the following example:  
\begin{eg}\label{eg:qubit}
Consider a system with two energy levels $E_0  =  0$ and $E_1  =  \Delta E$, corresponding to the pure states $|0\>$ and $|1\>$, respectively.   Clearly, the von Neumann instrument for the energy measurement---described by the quantum operations $\map M_x (\cdot)   =   |x\>\<  x|  \cdot  |x\>\<x|$,  $x=  0,1$---is  energy-preserving.    To implement this instrument, one can  use as environment two identical copies of the system, choose the initial state $|\phi_0\>   =     |0\>  |1\> $,    and engineer  a joint evolution $U$ satisfying
\begin{align*}
U  |0\>  |0\>  |1\>   &  =       |0\>  |0\>  |1\>   \\
U  |1\>  |0\>  |1\>   &  =       |1\>  |1\>  |0\>   \\
U  |0\>  |1\>  |0\>   &  =       |0\>  |1\>  |0\>   \\
U  |1\>  |1\>  |0\>   &  =       |1\>  |0\>  |1\>   \\
U  |0\>  |0\>  |0\>   &  =       |0\>  |0\>  |0\>   \\
U  |1\>  |0\>  |0\>   &  =       |1\>  |0\>  |0\>   \\
U  |0\>  |1\>  |1\>   &  =       |0\>  |1\>  |1\>   \\   
U  |1\>  |1\>  |1\>   &  =       |1\>  |1\>  |1\>  \, .   
\end{align*}  
By measuring  the meter observable  $M  =   |1\>\<  1|   \otimes   |0\>\<  0|$ on the environment, one then obtains the instrument $\{\map M_0 ,  \map M_1 \}$ as effective evolution of the system.  
\end{eg}

It is worth stressing that, despite the fact that the energy is preserved on average, its expectation value can fluctuate due to  postselection.     For instance, in Example  \ref{eg:qubit} one can decide to  postselect  the output state $|1\>$. 
 With probability 1/2, the postselection will transform  the state $|+\>   =  (|0\>  +  |1\>)/\sqrt 2$ into the state $|1\>$, whose energy is twice the expected energy of $|+\>$.   Still,  the transformation will take place at no energy cost, because both the interaction and the measurement of the meter observable  preserve the energy.  A further discussion on this point is provided in Appendix \ref{appendix:weak}.

Motivated by the above discussion, we put forward the following
\begin{defi}
A probabilistic transformation $\map M_0$ is  \emph{energy-preserving} iff there exists an energy-preserving instrument $\{\map M_x\}_{x\in\set X}$ and an outcome $x_0\in\set X$ such that   $\map M_0   =  \map M_{x_0}$.  
\end{defi}
Probabilistic energy-preserving transformations can be demonstrated in engineered quantum systems with high degree of control. 
For example, a proposal for an experimental amplification of weak coherent states via  probabilistic energy-preserving transformations was recently reported  by Partanen \emph{et al}  in Ref. \cite{preservingamp}, where high fidelity amplification was achieved using only passive optical elements.  

\subsection{Energy-preserving and covariant instruments: the stationary case}  

We now show that energy-preserving instruments  instruments play a central role in the optimization of probabilistic operations.  
Consider an instrument  whose outcome probabilities are not affected by time translations,  namely
\begin{align}\label{stationary}     \Tr     [ \,  \map M_x   (  U_t \rho U_t^\dag  ) ]  =   \Tr[  \,  \map M_x  (\rho)  \,]     \, , \qquad \forall x \in \set X \, , \forall t\in\R \, ,
 \end{align} 
 where $\rho$ is a generic quantum state and    $U_t\  = exp[-i t H_{\rm sys}/\hbar] $.   
 
 We call the instruments satisfying Eq. (\ref{stationary})  \emph{stationary}.  
   A common  example   is that of  stationary instruments   with outcomes   $\set X =\{\rm succ,  fail \}$.   In this case, the stationarity condition  (\ref{stationary}) simply means that the probability of implementing the desired transformation $\map M_{\rm succ}$ is the same at every time.      Stationary two-outcome instruments can be used to describe tasks like probabilistic covariant cloning and probabilistic   amplification, or more generally, any task where the goal is to probabilistically transform a set of states generated by time evolution.     An important class of stationary instruments are those that are  \emph{covariant} under time evolution, namely
\begin{align}\label{covariant}
  \map M_x   \left  (    U_t   \,\rho\,  U_t^\dag \right)    =    V_t  \,  \map M_x    ( \rho)  \,  V_t^\dag \qquad  \forall   t  \in  \R  
  \end{align}
where     $\rho$ is an arbitrary input state and   $\{V_t  ~|~  t\in\R\}$ is  the unitary representation of the translation group describing the time evolution of the output system.

Energy-preserving instruments and stationary covariant instruments are closely related.  
First of all,   every energy-preserving instrument is stationary and covariant: indeed, the block form of the Kraus operators  (\ref{block}) implies that every quantum operation  $\map M_x$ satisfies the covariance condition  with 
 \begin{align}\label{uvt}
  U_t  =  V_t     =  \exp[-i t H_{\rm sys}/\hbar]  \, .
  \end{align} 
  Moreover, energy-preserving instruments are the key probabilistic element at the basis of every stationary covariant instrument:
  \begin{prop}\label{prop:decompcov}   
Let $\{\map M_x\}_{x\in\set X}$ be a stationary covariant instrument   transforming  states on $\spc H_{\rm in}$ into states on $\spc H_{\rm out}$. Then, $\{\map M_x\}_{x\in\set X}$   can be decomposed as 
$$  \map M_x   =    \map C_x  \map P_x    \qquad\forall x\in\set X   $$ where
  $\{\map P_x\}_{x\in\set X}$ is a pure   energy-preserving instrument  transforming  
  states on $\spc H_{\rm in}$ into states on $\spc H_{\rm in}$
 and  $\map C_x$ is a covariant channel transforming states on $\spc H_{\rm in}$ into states on $\spc H_{\rm out}$. 
\end{prop}

This result, proven in Appendix \ref{app:decompcov},  provides  additional motivation to the study of energy-preserving instruments. Indeed,   there is a large class of tasks where the optimal probabilistic  strategy is  described by a stationary  covariant instrument---this is the case, e.g. of phase-covariant probabilistic cloning \cite{fiurasekclon,GiulioNat,chiribella2016quantum} and probabilistic   amplification \cite{zhao2017quantum}.  Proposition \ref{prop:decompcov} establishes that energy-preserving instruments are the canonical probabilistic element in all these tasks.  The search for the optimal quantum operation is then split into two sub-problems:  \emph{i)} the search for the optimal energy-preserving instrument and \emph{ii)} the search for the optimal deterministic operation.  
Now, the optimization of deterministic operations has been studied extensively in the literature, and the solution of problem \emph{ii)} is known in a number of relevant cases.   In all these cases, the search for the optimal probabilistic operation is reduced to the search for the optimal energy-preserving instrument.  A general method for the solution of the problem will be provided  in  Section \ref{sec:pure}.   

\subsection{Energy-preserving and  covariant instruments: the non-stationary case}
The relation between energy-preserving and covariant instruments  can  also be extended to  non-stationary cases where  the outcome probabilities  are transformed by time evolution  as  
\begin{align}\label{nonstationary}
\Tr  [   \map M_x   (   U_t    \rho  U_t^\dag)]    =   \Tr  [  \map  M_{ f_{-t} (x)}  (\rho)] \, ,\qquad \forall x\in\set X\, , \forall t\in \R \, .
\end{align}
Here   $f_t :     \set X \to \set X$ is a function  representing the action of  the time evolution on the outcomes. This means that   $f_t$ is invertible and  satisfies  the condition 
\[  f_{t_1}  \circ f_{t_2}  =  f_{t_1 +  t_2}  \, ,  \qquad \forall t_1,t_2 \in \R  \, .\]
Note that stationary instruments are included as a special case, because Eq. (\ref{stationary}) can be obtained from Eq. (\ref{nonstationary}) by setting  $f_t(x)   =  x$ for every $x$ and $t$.  

In the non-stationary case, covariant instruments are defined by the condition \cite{holevo_stat_book}
\begin{align}
\map M_x(  U_t \cdot U_t^\dag)   =     V_t  \,\map M_{f_{-t}(x)}  (\cdot) \,    V_t^\dag     \,, \qquad  \forall x\in\set  X\, , \forall t \in  \R\, .
\end{align}

An example of non-stationary covariant instrument arises in the task of phase estimation, where the time evolution is periodic and the set of outcomes is $\set X  =  [0,2\pi)$.   
A covariant measurement of phase  satisfies   Eq. (\ref{nonstationary}), with 
   \[  f_t (\theta)  =    \theta +  \omega t  \mod 2\pi    \, , \qquad \forall \theta \in  [0,2\pi) \, , t\in\R \, ,  \]
   where $\omega$ is the frequency of the oscillator.    
Another example arises in probabilistic phase estimation \cite{BaganPRL}, where the set of outcomes is
\[  \set X  = \{0,2\pi\} \cup\{\rm fail\}  \]  
 and the outcome ``$\rm fail$"  occurs  when  no  phase estimate is  produced.    In this case, it is natural to require
  \[  f_t ({\rm fail})  =   {\rm fail}    \, , \qquad  t\in\R \, ,  \]
 implying  that the probability of failure is the same at all times.  
 
For probabilistic phase estimation, the set of outcomes can be partitioned into  two orbits: one orbit containing all the outcomes in $[0,2\pi)$ and one orbit containing the  single outcome $\{\rm fail\}$. In general, for a non-stationary instrument satisfying Eq. (\ref{nonstationary}), the outcome set $\set X$ can be partitioned into disjoint orbits, as 
  \begin{align}\label{orbit}  \set X   =   \bigcup_{y  \in \set Y}    \,  \set O_y  \, ,\end{align} 
 where each $\set O_y$ is an orbit under the action of the translation group and the set $\set Y$ labels the different  orbits.    With this notation, every covariant instrument can be decomposed as follows: 
\begin{prop}\label{prop:decompcovgen}
 Let $\{\map M_x\}_{x\in\set X}$ be a covariant instrument   transforming  states on $\spc H_{\rm in}$ into states on $\spc H_{\rm out}$.  Then,  $\{\map M_x\}_{x\in\set X}$ can be decomposed as 
 \begin{align}
\map M_x    =     \map M^{(y)}_x  \circ \map P_y \, , 
\end{align}
 where $y$ is the orbit label defined in Eq. (\ref{orbit}),  $\{ \map P_y\}_{y\in\set Y}$  is    a pure, energy-preserving instrument transforming  states on $\spc H_{\rm in}$ into states on $\spc H_{\rm in}$, and  $\left\{  \map M^{(y)}_x\right\}_{x\in\set O_y}$ are covariant instruments transforming  states on $\spc H_{\rm in}$ into states on $\spc H_{\rm out}$.  
\end{prop}
The proof is provided in Appendix \ref{app:decompcovgen}. Physically, Proposition \ref{prop:decompcovgen} tells us that we can realize every  time-covariant covariant instrument through a pre-selection  implemented with energy-preserving operations, followed by  conditional measurements  that estimate the action of time translations.   

In the case of probabilistic phase estimation,   this implies that every  probabilistic phase measurement   can be broken down into a two-outcome, energy-preserving filter  $\{   \map P_{\rm succ},    \map P_{\rm fail}\}$, followed by a deterministic phase measurement  $  \{   \map M^{({\rm succ})}_x\}_{x\in [0,2\pi)}$ in the successful case.     Physically, this means that the post-selection on the measurement outcomes can be freely transformed into a  pre-selection  on the input state. Since the pre-selection can be done at zero-energy cost, our result shows that the energy cost of every probabilistic phase measurement is equal to the energy cost of a corresponding deterministic measurement.      
 
 Proposition \ref{prop:decompcovgen} has also important implications for the  search for the optimal  phase estimation strategy with a desired probability of success.    The optimization problem is  split into two sub-problems:  \emph{i)} the search for the optimal energy-preserving instrument and \emph{ii)} the search for the optimal deterministic estimation.  Since the latter is known in a number of cases  \cite{holevo_stat_book},  Proposition \ref{prop:decompcovgen}  reduces the optimization to    the search for the optimal energy-preserving instrument.   It is also important to stress that  every point in the trade-off between precision of phase estimation and probability of success  can be  explored by applying the optimal  phase measurement after a probabilistic pre-selection  done at zero energy cost. 

\section{Optimal energy-preserving operations} \label{sec:pure}

We are now ready to start the  search for the optimal  operations that  transform a coherent superposition of energy eigenstates into another.    In this section we formalize the problem and address the optimality question, providing the general form of the optimal energy-preserving operations.

\subsection{How well can we implement a desired state transition without exchanging energy?}

Regarding energy as a resource, a natural question is how well a desired task can be achieved without the assistance of external energy sources.
Consider the most basic task:   transforming a pure input state $|\varphi\>$  into a target output state $  |\psi\>$.     
For example,     the input    could be  a weak coherent state with known amplitude but unknown phase, and the target could be another coherent state with the same phase but with larger amplitude.  The problem of amplifying laser pulses using energy-preserving operations was recently studied in Ref. \cite{preservingamp}, where the authors showed that a nearly perfect amplification can be achieved probabilistically by exploiting the quantum fluctuations of the field, without drawing any energy from the outside.   Another interesting example is quantum cloning: Suppose that a spin 1/2 particle,  immersed in a magnetic field, is prepared in a superposition of spin up and spin down.  As a result, the particle will process around the direction of the magnetic field and its state will evolve in time.   How well can we copy the time information  without tapping external energy sources?   
 Note that both in the amplification and in the cloning example, the input and the target states are drawn from a \emph{set} of states: more precisely, the problem is to transform the input state $ |\varphi_t\>  =  e^{-it H_{\rm sys}/\hbar}  |\varphi\>$  into the target state  $ |\psi_t\>  =  e^{-it H_{\rm sys}/\hbar}  |\psi\>$  for an arbitrary (and possibly unknown) value of the parameter $t$.  However, since we require  our operations to be energy-preserving, we do not need to optimize them for every value of $t$:  indeed, every energy-preserving transformation that approximates the transition $ |\varphi\>  \to  |\psi\>$ will do equally well in approximating the transition $  |\varphi_t\>  \to |\psi_t\>$, due to the covariance condition of Eqs. (\ref{covariant}) and (\ref{uvt}).   This point is made clear if we measure the performance of our operations in terms of the fidelity between the target state and the actual output.  For a probabilistic transformation $\map M_x$,   the fidelity between target state and  the actual output state is 
  \begin{align*}
 F_{x,t}  :  =     \<  \psi_t  |    \rho_x'  (t)   |\psi_t\>   \, , \qquad  \rho_x'(t)   =  \frac{\map M_x ( |\varphi_t\>\<\varphi_t |}{\Tr[  \map M_x (|\varphi_t\>\<\varphi_t| ]} \, . 
 \end{align*} 
Using the covariance condition one immediately sees  that $F_{x,t}$ is independent of $t$.   Physically, this means that  energy-preserving transformations perform equally well on all possible inputs that are connected  by time evolution. 

More generally, it is interesting to consider the probabilistic transformations obtained by postselection over the outcomes of a quantum measurement.   We will call a \emph{filter}  an instrument $\{\map M_x\}_{x\in\set  X}$ along with a partition of    outcome set   $\set X$ into two disjoint subsets---$X_{\rm succ}$ and $X_{\rm fail}$---which correspond to successful  and unsuccessful instances, respectively.  Averaging the fidelity over the successful instances,  we obtain the value 
\begin{align}
F   =     \sum_{x  \in  \set{X}_{\rm succ}}   \, p(  x   |  {\rm succ})  \,    \<  \psi_0 \,|  \rho_x'  (0)\,    |\psi_0 \>  \, ,
\end{align}
where $  p(  x|{\rm succ}) $ is the conditional probability of obtaining $x$ given that a successful outcome has occurred.   
Making the filter explicit, the average fidelity can be rewritten as 
\begin{align}\label{F}
F  =   \frac{  \< \psi|  \, \map M_{\rm succ}   (  |\varphi\>\<  \varphi|)\,   |\psi\> }{p_{\rm succ}} \, ,
\end{align}
where $\map M_{\rm succ} $ is the quantum operation defined by 
\begin{align}\label{msucc}
\map M_{\rm succ}:  =  \sum_{x\in\set X_{\rm succ}}    \,  \map M_x  
\end{align} and $p_{\rm succ}$
 is the probability of success
 \begin{align}
\nonumber p_{\rm succ}   &=    \Tr  [   \map M_{\rm succ} ( |\varphi\>\<  \varphi|)]  \\
 \nonumber &=    \Tr  [   \map M^\dag_{\rm succ}  (I)  \,  |\varphi\>\<  \varphi| ]\\
& =  \<\varphi| \,   P_{\rm succ}   \,  |\varphi\>\, ,   \qquad \quad  P_{\rm succ}    :=       \map M^\dag_{\rm succ}  (I) \, .
\label{psucc}  \end{align}

In a realistic situation one will be interested not only in maximizing the fidelity, but also  in having a sufficiently high probability of success.   Requiring the success probability to be larger than a given threshold, the problem  becomes to find the energy-preserving quantum operation $\map M_{\rm succ}$ that maximizes the fidelity.  

\subsection{Deterministic transitions: optimality of  eigenstate alignment}

Let us consider first the case $p_{\rm succ} =1$.      In this case, the optimization problem has a closed-form  solution, corresponding to an operation  that we call \emph{eigenstate alignment}.   Given two superpositions of energy eigenstates, eigenstate alignment is an energy-preserving unitary operation that transforms the eigenstates in one superposition into the eigenstates in the other.  Precisely, let us decompose the pure states $|\varphi\>$  and  $|\psi\>$ as   
\begin{align}\label{statedecomp}
|\varphi\>=\sum_{E}\sqrt{p_E}|\varphi_E\>  \quad {\rm and}       \quad  |\psi\>=\sum_{E}\sqrt{q_E}|\psi_E\>
\end{align}
with  
\begin{align}\label{define-PQ} 
\begin{array}{rlrl} p_E  & =  \| P_E  |\varphi \> \|^2 \, ,  &   q_E  & =   \| P_E  |\psi \>\|^2   \, ,\\
&&&\\
   |\varphi_E\>  &:=  \frac{P_E|\varphi\>}{\|P_E|\varphi\>\|}    \, , \qquad  &    |\psi_E\> &:=  \frac{P_E|\psi\>}{\|P_E|\psi\>\|} \, ,
 \end{array}
    \end{align} 
  $P_E$ being the projector on the eigenspace of $H_{\rm sys}$ with energy $E$.      In the rest of the paper, we will extensively use the notations $p_E$ and $q_E$ for the energy spectrum of the initial and final state, respectively. With this notation, we say that a unitary operator $U$ is an  \emph{eigenstate alignment} of $|\varphi\>$ with  $|\psi\>$ if $U$ is energy-preserving and   
\begin{align}\label{eigenalign}
U |\varphi_E\>   = |\psi_E\>  \qquad \forall E  :   \,     p_E \not =  0 \,    , q_E  \not  =  0 \, .
\end{align}

It is immediate to see that an  eigenstate alignment exists for every pair of pure states.  In general, eigenstate alignments are not unique, unless the spectrum of $H_{\rm sys}$ is non-degenerate and every energy $E$ in the spectrum satisfies the conditions $p_E  \not  =  0$ and $q_E \not  =  0$.  The importance of eigenstate alignment comes from the following:   
\begin{theo}\label{theo:EAoptimal}
For $p_{\rm succ}  =1$, eigenstate alignment achieves the maximum fidelity for the transition $|\varphi\> \to  |\psi\>$.    The maximum fidelity is given by  
\begin{align}F_{\rm det}=\left(\sum_E\sqrt{p_Eq_E}\right)^2 \, .
\end{align}
\end{theo} 
The proof is provided in Appendix \ref{app:EAoptimal}. Theorem \ref{theo:EAoptimal} shows that the optimal energy-preserving channel can be chosen to be unitary without loss of generality.   In this case,  no interaction with the environment is needed.  
We stress that the optimality of unitary transformations is a specific feature of the energy-preserving framework.  Unitary transformations may not be optimal  in the broader class of phase-covariant channels---for example, they are  sometimes suboptimal for cloning qubit states on the equator of the Bloch sphere
\cite{qubit}.

\subsection{Probabilistic transitions: optimality of pure quantum operations}

Let us move to the general case $p_{\rm succ}  \le1$. 
We now show that, without loss of generality, the optimal quantum operation $\map M_{\rm succ}$ can be chosen to be \emph{pure},~i.~e.~of the form
 $\map M_{\rm succ} (  \cdot)  =  M_{\rm succ}  \cdot M_{\rm succ}^\dag$
for some suitable operator $M_{\rm succ}$.   To prove this result, we provide a construction that transforms any given quantum operation  $\map M_{\rm succ}$ into a pure quantum operation  $\map M_{\rm succ}'$ with the same probability of success and possibly a higher fidelity.  
The construction is based on a new ingredient that we name the \emph{L\"uders reduction}.  

\subsubsection{L\"uders reduction}
The L\"uders reduction transforms a given quantum operation into a pure quantum operation with the same probability of success.  
Specifically, the L\"uders reduction of a quantum operation $\map M$ is the pure quantum operation  $\map P$ defined by  
\begin{align}\label{luders}
\map P  (\cdot)   =   \sqrt  P  \cdot \sqrt P  \, ,       \quad  P    =  \map M^\dag (I) \, .   
\end{align}
When $P$ is a projector, the quantum operation  $\map P$ coincides with the state transformation defined by L\"uders in his treatment of the measurement process \cite{luders}.  When $P$ is not a projector,   $\map P$ is often called the ``generalized L\"uders transformation" associated with $P$  \cite{busch}.

By construction, a quantum operation and its L\"uders reduction have  the same  probability: 
For every quantum state $\rho$ one has  
\begin{align}
\nonumber \Tr [  \map P  (\rho)]  &  =  \Tr   [     \sqrt {P}   \rho  \sqrt{P}]  \\
 \nonumber   &  =  \Tr   [     P   \rho]  \\
\nonumber   &  =  \Tr  \left[  \map M^\dag  (I)    \rho  \right]  \\
\label{ludersprob}     &  =   \Tr [  \map M  (\rho) ]   \, .
\end{align}
Among the quantum operations that happen with the same probability as $\map M$,   the L\"uders reduction can be characterized as the ``least noisy," meaning that every other quantum operations can be reproduced by applying a noisy channel to the output of $\map P$:  
\begin{prop}\label{prop:ludersdecomp}
Every quantum operation $\map M$ can be decomposed as  $ \map M   =   \map C  \circ \map P $ where $\map P$ is the L\"uders reduction of $\map M$ and $\map C$ is a suitable quantum channel.  
Moreover, if $\map M$ is energy-preserving, then also $\map P$ and $\map C$ are energy-preserving. 
\end{prop} 
The proof is provided in Appendix \ref{app:proofProp}.      Using this result, the search for the optimal quantum operation is split into two different problems:  the search for an optimal \emph{pure} operation $\map P$ and the search for the optimal \emph{deterministic} operation $\map C$.   Note that this conclusion  applies not apply only to the problem of transforming pure states,  but also to the optimization of transitions  involving mixed states.

\subsubsection{Increasing the fidelity without changing the success probability}

Combining the L\"uders reduction and eigenstate alignment we can turn every quantum operation  $\map M_{\rm succ}$ into a pure quantum operation $\map M_{\rm succ}'$ with the same success probability  and a possibly higher fidelity.   
The idea is simple:  by Proposition  \ref{prop:ludersdecomp}, every energy-preserving  quantum operation    $\map M_{\rm succ}$   
can be decomposed as  
\begin{align*}
\map M_{\rm succ}   = \map C \circ  \map P_{\rm succ} \, ,
\end{align*}
where $\map C$ is an energy-preserving channel and $\map P_{\rm succ}$ is the L\"uders reduction given by  
\begin{align*}
\map P_{\rm succ}    (\cdot)   =    \sqrt{ P_{\rm succ}}   \cdot     \sqrt{ P_{\rm succ}} \, ,  \qquad       P_{\rm succ}  =  \map M_{\rm succ}^\dag (I) \, .
\end{align*}
When the  L\"uders reduction takes place, the input state $|\varphi\>$ is transformed into the pure state 
 \begin{align}\label{phiprime}
|\varphi'\>    =    \frac{  \sqrt{ P_{\rm succ} } |\varphi\>}{  \|   \sqrt {P_{\rm succ}}  |\varphi\>  \|}  
\, .
\end{align}   
Now,  the probability of success depends only on the operator $P_{\rm succ}$.  
Fixing $P_{\rm succ}$, we know that the optimal energy-preserving channel for the transition $|\varphi'\> \to |\psi\>$ is given by eigenstate alignment  (Theorem \ref{theo:EAoptimal}).   Hence, the fidelity for the quantum operation $\map M_{\rm succ}$ cannot be larger than the fidelity of the quantum operation\begin{align}\label{Msuccprime}
\map M_{\rm succ}'     =\,      \map U  \circ \map P_{\rm succ} ,       
\end{align}
 where $\map U $ is the unitary channel corresponding to the eigenstate alignment of $|\varphi'\>$ with  $|\psi\>$.  Note that $\map M_{\rm succ}'$ is energy-preserving, because it is the composition of two energy-preserving operations.    Summarizing, we have proven the following
  \begin{prop}\label{prop:better}
 For every energy-preserving operation $\map M_{\rm succ}$, the energy-preserving operation $\map M_{\rm succ}'$ defined in Eq. (\ref{Msuccprime}) has the same success probability and at least the same fidelity in the implementation of the  state transition $|\varphi\>\to|\psi\>$.   
 \end{prop}
 Explicitly,  the success probability and the fidelity of $\map M_{\rm succ}'$ are given by 
 \begin{align}
 \label{psuccaa}   p_{\rm succ}      
&=   \sum_E    p_E  \,  \< \varphi_E  |   P_{\rm succ}  |  \varphi_E\>
 \end{align} 
 and
 \begin{align}
  \label{Faa}
F  =  \left(  \sum_E  \sqrt{ p_E'  q_E}  \right)^2\, , \qquad  p_E'    =   \frac{p_E   \, \<\varphi_E | \, P_{\rm succ}\, |\varphi_E\>}{p_{\rm succ}} \, ,
 \end{align} 
 where $p_E$ and $q_E$ are the probabilities in the input and output states, as defined in Eq. (\ref{define-PQ}).  The above expression of the fidelity follows directly from the application of Theorem \ref{theo:EAoptimal} to the transition  $|\varphi'\>\to |\psi\>$.

 Now, since turning a quantum operation into a pure quantum operation can only increase the fidelity, we proved the following
\begin{cor}[Optimality of pure quantum operations]
For every fixed value of the success probability, the energy-preserving   operation that maximizes the fidelity can be chosen to be pure without loss of generality. 
\end{cor}

\subsubsection{Optimal quantum operations from Lagrangian optimization}  \label{subsec:optimal}

The optimization of the fidelity for given success probability can be completed by  Lagrangian optimization.
Let us define the coefficients  $$x_E:=  \<  \varphi_E  |  \,  P_{\rm succ} \, |\varphi_E  \>   \, .$$ 
With this definition,    the probability of success (\ref{psuccaa})  and the fidelity   (\ref{Faa})  can be expressed as 
\begin{eqnarray}
p_{\rm succ}&=&\sum_E  \, p_E     x_E  
\end{eqnarray}
and    
\begin{eqnarray}\label{lagrangefid}
F=p_{\rm succ}^{-1}\left(\sum_E \sqrt{x_E p_Eq_E}\right)^2  \, 
\end{eqnarray}
respectively.  
By Lagrangian optimization, we  obtain that  the optimal filter has a simple structure:  the energy spectrum is  partitioned into two disjoint subsets, $\set S_0$  and $\set S_1$, and  the coefficients of the optimal transformation are given by  
\begin{eqnarray}\label{general}
x_E   =   
\left \{   
\begin{array}{ll}  
 1    \qquad  &    E  \in\set S_0   \\
        \frac{p_{\rm succ}   -    p(\set S_0)}{  1-  q(\set S_0) }         \,    \frac{q_E}{p_E}     \qquad   &E  \in\set S_1 
\end{array}
\right.
\end{eqnarray}
where      $p(\set S_0)  :  =   \sum_{E\in\set S_0}   p_E$ and $q  (\set S_0) :  =   \sum_{E\in\set S_0}   q_E$.  
In other words, the optimal filter is completely determined by the choice of the set $\set S_0$.   Inserting Eq. (\ref{general}) into  Eq. (\ref{lagrangefid}),  the maximization of the fidelity is reduced  to      the maximization of the quantity 
  \begin{align}
  \Omega  [  S_0]   =  \sum_{E\in\set S_0}   \sqrt{p_E  q_E}  +  \sqrt{   [  p_{\rm succ}   -   p (\set S_0)]  \, [1-q(\set S_0)]} \, . 
  \end{align}    
  
  Examples of quantum operations of the form  (\ref{general}) can be  found in Ref. \cite{BaganPRL}, which focused on the specific problem of  phase-estimation.  
More examples   will be provided in section \ref{sec:examples}.

\medskip 

\subsection{The ultimate limits of probabilistic energy-preserving processes}

So far we considered the optimization of the fidelity for  fixed   success probability.  We now remove the constraint and focus only on the maximization of the fidelity.  The problem is interesting because it highlights  the quantum limits  to what is  logically possible, no matter how small the probability \cite{calsamiglia-2014-natphys}.  

The ultimate limits for energy-preserving operations is characterized by the following
\begin{prop}\label{cor:maxfid}
Let $|\varphi\>$ and $|\psi\>$ be two generic pure states of a finite-dimensional quantum system. For the transition $|\varphi\>\to |\psi\>$,  the maximum of the fidelity over all energy-preserving quantum operations       is 
\begin{align}\label{fmax} F_{\max}     =      \sum_{E    \in  \Sp (\varphi) \cap  \Sp(\psi) }     \,     q_E   \, , 
\end{align}
where $q_E$ is the probability defined in Eq. (\ref{define-PQ}) and $\Sp(\chi)$ denotes the energy spectrum of a generic state $|\chi\>$, defined as 
$$   \Sp (\chi)   :  =  \{   E ~|~  \<   \chi  |  P_E   |\chi\>  \not  =  0\} \, .$$
For a quantum operation achieving fidelity $F_{\max}$ the maximum probability of success is given by  
\begin{align}\label{psuccmax} p^{\max}_{\rm succ}   =   \left(  \min_{E  \in  \Sp(\varphi) \cap  \Sp(\psi)   }  ~    \frac {p_E}{q_E}     \right)  \,    F_{\max}  \, ,  
\end{align}
where $p_E$ is the probability defined in Eq. (\ref{define-PQ}). 
The quantum operation achieving maximum fidelity with maximum probability is pure and its Kraus operator satisfies the condition
\begin{align}\label{Moptimized} M  |\varphi_E\>   =                  \left[     \min_{E'    \in  \Sp(\varphi) \cap  \Sp(\psi)    }   \sqrt{     \,   \frac {     p_{E'}   }{q_{E'}}     }\right]   \,              \sqrt{\frac{q_E}{     p_E   } }   ~  |\psi_E\>    
\end{align}
for every energy $E$  in  $\Sp(\varphi)$. 
\end{prop}

The proof is provided in Appendix \ref{app:ultimate}.

\section{Multiround recursive protocols}\label{sec:network}
In the previous section we provided a recipe to find the  protocol  that achieves maximum fidelity for  a fixed value of the success probability.  By definition,  the resulting protocol is  taylor-made to that specific value of the probability.    However, in many situations it is useful to have  a more flexible protocol, where  the experimenter can  make successive attempts at realizing the desired transformation and is free to decide on the fly when to stop.  
In this section we analyze protocols of this form, which we refer to as  \emph{recursive protocols}.     
Under the energy-preserving constraint, we identify  the protocol that produces   the best possible approximation of the target state at each step.  
   It is important to stress that  the protocol does not require an actual experimenter to read the outcomes and to make decisions: in principle, all the measurements and conditional operations can be implemented by a fully quantum machine operating in an energy-preserving fashion. 

\subsection{The optimal  recursive protocol}\label{subsec:RUS}
  
  Given a sequence  of   $K$  binary  filters
 with outcomes $\{\rm succ, \rm fail\}$, consider the  protocol defined by the following instructions: 
\begin{enumerate}
\item Set $k=1$.
\item   If $k\le K$, then apply the $k$-th filter; else terminate.
\item If the outcome  is $x= \rm succ$, then terminate.
\item If the outcome is $x=\rm fail$, then replace $k$ with $k+1$ and  go back to instruction 1. 
\end{enumerate} 
Recursive protocols of this form are an example of  ``quantum loop programs",  studied in Ref.  \cite{loop}.  All these protocols  can be can be visualized  as  decision trees  of the following form
\begin{equation*}
\begin{aligned}\Qcircuit @C=1em @R=0.2em @!R
{
   & \multigate{1}{   {\rm Filter~ 1}} & \qw\poloFantasmaCn{\rm succ}&\qw&    &\\
  & \pureghost{  {\rm Filter~ 1}  } &\qw\poloFantasmaCn{\rm fail}&\qw &\multigate{1}{  {\rm Filter~ 2}  } &\qw\poloFantasmaCn{\rm succ} &\qw& \\
  & & & &\pureghost{  {\rm Filter ~2}} & \qw\poloFantasmaCn{\rm fail}&\qw&\\
  & & & & & &\vdots&\\
  & & & & & & \qw\poloFantasmaCn{\rm fail} &\multigate{1}{  {\rm Filter~ K}  }&\qw\poloFantasmaCn{\rm succ}& \qw  \\
  & & & & & & &\pureghost{ {\rm Filter ~K}} & \qw\poloFantasmaCn{\rm fail} &\qw
}
\end{aligned}
\end{equation*}
Protocols of this form have been employed for different purposes, including  entanglement concentration \cite{divincenzo}, implementation of quantum gates \cite{RUS1,RUS2,RUS3} and ancilla-driven computation \cite{ADQC}.     One such protocol that is particularly similar to ours is  \emph{quantum rejection sampling} \cite{rejectionsampling}. There,  the goal is to generate a  target superposition  $|\psi\>$   using a black box $U_\varphi$ that prepares another   superposition  $|\varphi\>$  from a fixed state $|0\>$.  The difference between rejection sampling and our problem is that in our case we do not have the black box $U_\varphi$ at disposal. Instead, we have  the coherent superposition  $|\varphi\>$, which is a strictly weaker resource than the gate $U_{\varphi}$, due to Nielsen and Chuang's no-programming Theorem \cite{nielsen1997programmable}.

 In our case, the goal of the protocol is to transform a coherent  superposition of energy eigenstates   into another. 
Of course, at each step there will be a trade-off between the fidelity with the target and the probability of success.      In the simplest scenario, the protocol can be designed to attain the absolute maximum of the fidelity at each round, and to do so with  maximum probability of success.    
An experimenter following such a protocol  will  have the guarantee that the best possible performance is achieved in each individual round.

We consider  the transition  $|\varphi\>  \to |\psi\>$    in the case of states $|\varphi\>$ and $|\psi\>$ in a finite dimensional Hilbert space $\spc H  \simeq \C^d$,  $d<  \infty$, or, more generally, states whose energy spectra intersect on a finite set of points, with $|   \Sp (\varphi)  \cap \Sp(\psi)|  \le d$.   For the optimal protocol we make a list of desiderata in decreasing order of priority: 
for every $k\in\{1,\dots, K-1\}$
\begin{enumerate}
\item at the $k$-th round, the successful quantum operation should transform the input state $\rho^{(k)}$  into the target $  |\psi\>$ with  maximum fidelity;
\item the optimal transition $\rho^{(k)}  \to  |\psi\>\<\psi|$ must be achieved with maximum probability of success;
\item the unsuccessful quantum operation   at the $k$-th round should produce the state  $\rho^{(k+1)}$ that leads to maximum fidelity for the transformation  $\rho^{(k+1)} \to |\psi\>\<\psi|$ at the $(k+1)$-th round; 
\item at the final round ($k=K$) the successful quantum operation should achieve maximum fidelity with maximum probability and, conditional on the fulfillment of this requirement, the unsuccessful quantum operation should achieve maximum fidelity.
\end{enumerate}

The derivation of the optimal protocol is rather technical and is provided in Appendix \ref{app:derivation}.   In the following we present the final result of the optimization and discuss the implications of our findings.  

  At the $k$-th round, the optimal binary filter consists of two pure quantum operations, $\map B^{(k)}_{\rm succ}  (\cdot)   =  B^{(k)}_{\rm succ}   \cdot  B^{(k)  \dag}_{\rm succ}  $ and $\map B^{(k)}_{\rm fail}  (\cdot)  =   B^{(k)}_{\rm fail } \cdot  B^{(k)\dag}_{\rm fail}$.   Since  all quantum operations are pure, the state of the system is pure at every round.  The input state at 
  the $k$-th round, denoted   by $|\varphi^{(k)}\>$, can be expanded as
$$\left|\varphi^{(k)}\right\>   =   \sum_{E}   \sqrt{  p^{(k)}_{E}}   \,  \left |\varphi_{E} \right \> \, , $$
where the energy eigenstates are the same as in the decomposition of $|\varphi\>$, cf. Eq. (\ref{define-PQ}).  
With this notation, the successful quantum operation is determined in an essentially unique way by  the condition   
\begin{align}
\nonumber   B^{(k)}_{\rm succ}   \left| \varphi_E  \right  \>       =         &  \left[     \min_{E'  \in  \Sp(\varphi^{(k)})  \cap \Sp(\psi) }   \sqrt{     \,   \frac {     p^{(k)}_{E'}   }{q_{E'}}     }\right]          \sqrt{\frac{q_E}{     p^{(k)}_E   } }  ~  |\psi_E\>     \\
 \label{bksucc}     & \qquad  \qquad \qquad \qquad\quad \forall E  \in \Sp(\varphi^{(k)})\, .
\end{align}  
Here the only freedom is in the definition of   the operator in the subspace spanned by   energy eigenstates outside the spectrum of $|\varphi^{(k)}\>$.  The form of Eq. (\ref{bksucc}) follows directly from the requirements 1 and 2 in our list of desiderata (cf.    Proposition \ref{cor:maxfid}).     Similarly, the unsuccessful quantum operation is determined in an essentially unique way by the condition
\begin{align}\label{bkfail}
B^{(k)}_{\rm fail}   =   \sqrt{  I-    B^{(k)\dag}_{\rm succ}   B^{(k)}_{\rm succ}} \, .
\end{align}
The form of Eq. (\ref{bkfail}) follows from the requirement 3 in our list.    Remarkably, the quantum operation  $\map B^{(k)}_{\rm fail}$ does not  maximize only the fidelity achievable from the input state $\rho^{(k+1)}$, but also the \emph{probability}    that maximum fidelity is achieved.

\subsection{Fidelity and success probability}

The optimal protocol is specified  recursively by  equations \eqref{bksucc}  and \eqref{bkfail}.  Making the dependence on the input and target states  explicit,  it is possible to derive closed formulas for the fidelity and the success probability.  
 To this purpose,    we need to introduce some notation.   Given a pair of pure states $|\varphi\>$ and $|\psi\>$ and given the corresponding probabilities $p_E$ and $q_E$ defined in Eq. (\ref{define-PQ}), we define the ratio    
$ p_E/q_E  $ for all  values of the  energy in $  \Sp (\varphi) \cap \Sp(\psi)$.  Then, we arrange the  values  of the ration $r_E$  in increasing order as 
\[ 0<r_1  < r_2  < \dots <  r_L  \, ,  \] where  $r_L$ is the maximum ratio.  Clearly,  by the assumption of finite dimensionality,  $L $ satisfies the relation 
$$L  \le  |  \Sp (\varphi) \cap \Sp(\psi)  |  \le d  \, .$$ 
For every possible value  $r_i$,  we consider the set of energy eigenvalues  $\set R_i$ defined as
\begin{align}\label{R}
\set R_i     :=   \left \{   E  \in  \Sp(\varphi)  \cap \Sp (\psi)  ~  \left |~  \frac {p_E}{q_E}   =   r_i \right. \right\}  
\end{align}  
and we denote the union of the first $k$ sets by 
\begin{align}\label{U}
\set U_k  : =     \bigcup_{i=1}^k   \,  \set R_i \, . 
\end{align}
With this definition,  the fidelity and the success probability at the $k$-th step can be expressed as
\begin{align}\label{fidelity1}
F_{\max}^{(k)}=
\sum_{E   \in   \Sp(\varphi)  \setminus    \set U_{k-1} } \, q_{E}
 \,  .
 \end{align}
and
\begin{eqnarray}\label{prob1}
p_{\rm succ}^{(k)}= (r_k-r_{k-1})  \,  \cdot \, F^{(k)}_{\max}  \, , 
\end{eqnarray}
respectively.   The proof is presented in Appendix \ref{app:Mk}.   Note that the fidelity is strictly decreasing  with $k$, reaching zero for $k   =   L$.  In other words, it is useless to consider protocols with more than $L$ rounds. 

The explicit expressions given by Eqs.  (\ref{fidelity1}) and (\ref{prob1})  turn out to be  very useful 
 for  studying the trade-off between fidelity and success probability.   
 Indeed, they allow us to evaluate the probability  that the protocol succeeds in one of the first $T$ rounds, given by 
 \begin{eqnarray}
\nonumber 
p_{{\rm succ}}  (T)  &:=&\sum_{k=1}^{T}p_{\rm succ}^{(k)}   \\
  &   =&    \sum\limits_{E\in        \set U_{T-1}  }  p_E    +    r_T    \,  F_{\rm max}^{(T)}
    \label{Ptot}  
   \, ,
\end{eqnarray}
and to observe its scaling with  the average fidelity achieved in the first  $T$ steps, given by 
 \begin{eqnarray}
F(T)&:=  \frac { \sum_{k=1}^{T}    \,  p_{\rm succ}^{(k)}   ~F_{\max}^{(k)}  } { p_{\rm succ}(T)  } \, . 
\end{eqnarray}
 The trade-off curve between $F(T)$ and $p_{\rm succ} (T)$  will be illustrated in  section \ref{sec:examples} for a number of concrete examples.

\subsection{Output states and termination time of the protocol}

In addition to  the fidelity and success probability, it is useful  to know what states are produced  at every step of the protocol.   Assuming that the total number of rounds is upper bounded as  $K\le L$,  the explicit expression of the output state produced at the $k$-th round can be obtained as follows.  
We   regard the recursive  protocol as a quantum instrument, with outcomes in the set $\{1,\dots ,  K+1\} $.  The outcome $k$ corresponds to the pure quantum operation    with Kraus operator 
\begin{eqnarray}\label{Mk}
M_k:=   \left  \{  
\begin{array}{ll} 
B^{(k)}_{\rm succ}B_{\rm fail}^{(k-1)}\cdots B_{\rm fail}^{(1)}   \quad   & k=1,\dots, K, \\   
& \\
B^{(K)}_{\rm fail}B_{\rm fail}^{(K-1)}\cdots B_{\rm fail}^{(1)}&  k =   K+1
\end{array}  \, .
\right.
\end{eqnarray}
For $k  \le K$,  the  Kraus operators   are characterized in Appendix \ref{app:Mk}. Using this characterization, we show that 
 the output state in the case of success at  the $k$-th round is  given by  
\begin{align}   
 \nonumber |  \psi^{(k)}  \>       & :  =    \frac{  M_k |\varphi\>}{  \|  M_k  |\varphi\>  \|} \\
 \label{outputk} & =   \frac 1 {\sqrt{  F_{\max}^{(k)}}} \sum_{E    \in   \Sp(\varphi)  \setminus   \set U_{k-1} }   \sqrt{  q_E}  \,   |\psi_E\>     \, .
  \end{align}

Note that   $|\psi^{(k)}\>$ is a truncated version of the target state, with the energy spectrum  deprived of all the values in   $\set U_{k-1}$ and of all the values that are not in in the spectrum of $|\varphi\>$.   The energy spectrum of the output state is eroded from one step to the next: each iteration of the protocol produces a state with a strictly lesser amount of coherence in the energy eigenbasis.    Due to the assumption of finite dimensionality,   the  process  of erosion  terminates in a finite number of steps, equal to $T_{\max}$. 
Protocols with more than $T_{\max}$ rounds terminate after $T_{\max}$ steps, meaning that the probability of success satisfies  
$$   p_{\rm succ}  (T)   =  1  \qquad \forall\ T  >  T_{\max} \, .$$
 
The fact that the  protocol is guaranteed to terminate in a finite time is  an appealing feature. 
 It is worth stressing that the termination time $T_{\max}$ is upper bounded by the number of distinct energy levels of the system, which   can be much smaller than the dimension of the Hilbert space, as in the following  
 
 \begin{eg}
 Consider  the case of $N$ identical non-interacting systems of dimension $d$.   In this case the  total Hamiltonian is the sum of the single-system Hamiltonians, and its number of energy levels   is upper bounded by the number of partitions of $N$ into $d$ non-negative numbers (see e.g. \cite{GiulioNat}).   We then have that the number of rounds needed   to terminate is upper bounded as 
 $$  T_{\max} \le  \begin{pmatrix}  d+N-1  \\  N    \end{pmatrix}      <   (N+1)^{d-1}  \, ,  $$ 
i.~e.~ by a  polynomial  in $N$.     Even if the probability of success in the first round is exponentially small in $N$, as in the case of quantum super-replication \cite{GiulioNat,chiribella2016quantum}, the recursive protocol is guaranteed to reach unit probability  in  a polynomial  number of iterations.  
\end{eg}

\subsection{Increasing the fidelity of the recursive protocol}
At every iteration of the recursive protocol, the total probability of success increases, while the average fidelity decreases.    In general, the relation between fidelity and probability of success is not optimal, because the histories leading to successful outcomes are mixed incoherently: at the $T$-th step,  the successful quantum operation has the form 
\begin{align}
\map M^{(T)}  (\rho)    =   \sum_{k=1}^T   \, M_k  \rho M_k^\dag  \, ,
\end{align}
where $M_k$ are the Kraus operators defined in Eq. (\ref{Mk}).   Now,  we have  a systematic method to increase the fidelity while keeping the same  success probability: the method is to take the L\"uders reduction of    $\map M^{(T)}$ and  to perform eigenstate alignment on the output.     The following paragraphs highlight the main features of this method.

\subsubsection{Coherent coarse-graining}  
The L\"uders reduction transforms the quantum operation $\map M^{(T)}$ into the pure quantum operation   
$$\map P^{(T)}    (\cdot)   = \sqrt{ P^{(T)}}  \,(\cdot )\,  \sqrt{ P^{(T)}} $$ with 
$$  P^{(T)}   =   \map M^{(T)  \dag}  (I)   =  \sum_{k=1}^T  \,  M_k^\dag M_k\,.  $$  
The technique of joining different quantum operations into a single one will be an important tool in the following. For this reason, it is convenient to have a name for it: 
\begin{defi}
We call $\map P^{(T)}$ the \emph{coherent coarse-graining} of the quantum operations $\{\map M_k~|~  k   =1 ,  \dots,  T\}$.     
\end{defi}

An intuitive way to visualize coherent coarse-graining  is through a generalization of the double slit experiment. Consider an  interference experiment   involving multiple slits \cite{sorkin1994quantum}.  When detectors are placed on the slits, the passage of a particle through the $k$-th slit will trigger the occurrence of the  quantum operation  $\map M_k$.  When the detectors at the first $T$ slits are removed, the passage through these slits will result into the coherently coarse-grained operation $\map P^{(T)}$.    

Note that the flexibility of the recursive protocol is lost after coherent-coarse graining: when multiple histories are coherently combined, it is not possible anymore to choose on the fly when to stop the protocol. Still, the advantage of coherent coarse-graining is that it provides a heuristic way  to construct lower bounds on the probability/fidelity trade-off.  

\subsubsection{Eigenstate alignment}
By construction, coherent coarse-graining does not change the probability of success.   The fidelity is then increased by eigenstate alignment, achieved by any energy-preserving unitary $U$ such that 
$$  U  |\varphi_E\>  =  |\psi_E\>  \qquad \forall  E \in  \Sp(\varphi)  \cap \Sp(\psi) \, .$$
Note that the operation of eigenstate alignment does not depend on how many rounds of the protocols are coarse-grained.  The operation could be performed  even before the filter is applied, provided that one suitably adapts the definition of the filter.   

When combined, coherent coarse-graining and eigenstate alignment yield the pure quantum operation 
\begin{align}\label{coarsesolution}
\map M^{(T)\prime}  (\cdot)  =     M^{(T)\prime}     \cdot M^{(T)\prime\dag}   \, , \qquad   M^{(T)\prime}   :   =    U  \sqrt{   \sum_{k=1}^T     M_k^\dag M_k} \, ,
\end{align} 
whose action on the energy eigenstates is  given by 
\begin{align}\label{MT}
M^{(T)\prime}|\varphi_E\>    =   
\left\{  
\begin{array}{ll}
|\psi_E\>   \qquad  &  E   \in  \set U_T    \\  \\
\sqrt{  r_T    \,   \frac {q_E}{p_E}}   |\psi_E\>    &     E   \not \in \Sp (\varphi)  \setminus  \set U_T   \,  ,
\end{array}  
\right.
\end{align}
(see Eq. (\ref{solution1}) of Appendix \ref{app:Mk} for the explicit derivation). 
   For $T$ larger than  the termination time $T_{\max}$,  our construction eventually yields the \emph{optimal} energy-preserving channel for the transition $|\varphi\>  \to  |\psi\>$ (cf. Theorem \ref{theo:EAoptimal}).

\subsubsection{The performance of the coherently coarse-grained protocol}

Since taking eigenstate alignment as an obliged step in the optimal operation,  we refer to the quantum operation  $\map M^{(T)\prime}$   simply as a \emph{coherent coarse-graining}  (of the first $T$ steps of the protocol).  By construction, the probability of success of the quantum operation $\map M^{(T)\prime}$  is equal to the probability that the original (non coarse-grained) protocol, succeeds within $T$ steps [cf. Eq. (\ref{prob1})]. 
On the other hand, the fidelity  can be evaluated explicitly by using Eq. (\ref{MT}), which yields 
\begin{align}\label{F'}
F' (T)=\frac{\left[   \sum\limits_{E\in\set  U_T }\sqrt{p_E q_E}     +    \sqrt{  r_T }   \,   \sum\limits_{ E   \in \Sp(\varphi)  \setminus \set U_T }    q_E  \right]^2}{   \sum\limits_{E\in\set{U}_T}  p_E    +    r_T          \sum\limits_{E\in  \Sp(\varphi)  \setminus \set U_T}q_E}.
\end{align}

Performing  the operation of coherent coarse-graining for different values of $T$ one can obtain a sequence of filters that approximate the optimal curve of the fidelity-probability trade-off.    The improvement due to  coherent coarse-graining will be illustrated in  the next section with a number of  concrete examples.




\section{Applications}\label{sec:examples}
In this  section we apply the recursive protocol and the method of coherent coarse-graining to  the tasks of  phase estimation, cloning of quantum clocks, phase-insensitive amplification of coherent states,  and approximate correction in ancilla-driven quantum computation.   

\subsection{Quantum metrology with probabilistic energy-preserving operations}\label{subsec:estimation}

Here we apply the recursive protocol to the  task of phase estimation \cite{holevo_stat_book,helstrom}.  
 The main idea is the following: When the phase is encoded in a quantum state in a suboptimal way, one can try to improve the precision of phase estimation by first transforming the state into the optimal state.   Of course, such transition cannot take place deterministically---for otherwise  the original state would have been already optimal.  
However, a probabilistic protocol can produce good approximations of the optimal state and, conditionally on the success of the probabilistic transformations, it can enable an improved phase estimation.   In the following we will use our recursive protocol  to scan the trade-off curve between fidelity and probability of success. 

To illustrate the idea, we consider the simple case where the phase is encoded into the state of a two-level quantum system, as 
\[|\varphi_\theta\>  =    e^{-i\theta   Z}     |\varphi\>       \qquad \theta \in [0,2\pi) \, , \]
with $  Z  =  |0\>\<0| -  |1\>\<1|$  and $|\varphi\>  =(|0\>+e^{i\theta}|1\>)/\sqrt 2$.   
We assume that $N$ identical copies of the state  are available and search for the optimal strategy to estimate $\theta$.   
To quantify the precision, we use the gain function  $G(\theta, \hat \theta)$ defined by  \cite{holevo_stat_book}
\begin{align*}
G(\theta,\hat{\theta}):=\frac{1+\cos(\theta-\hat{\theta})}{2}  \, .
\end{align*}
Note that the gain function assigns a larger gain when the estimate $\hat \theta$ is closer to the true value $\theta$, attaining the maximum value $1$ if and only if $\hat \theta  = \theta$. 
Then, the goal is to find the estimation strategy that maximizes the average gain
\begin{equation}\label{GDEF}
\begin{split}
\<G\>&:=\int\frac{d\theta}{2\pi}\int\frac{d\hat{\theta}}{2\pi}  \, G(\theta,\hat{\theta})  ~  \< \varphi_\theta|^{\otimes N}   \,  E_{\hat \theta}  \,  |\varphi_\theta\>^{\otimes N} 
\end{split}  \, .
\end{equation}
where $\left\{E_{\hat{\theta}}  \,  \right\}$ is the Positive Operator-Valued Measure (POVM) describing the estimation strategy. 

For  phase  estimation with pure states, the optimal POVM has been derived by Holevo \cite{holevo_stat_book}. Specifically, for  pure states of the form 
\begin{align}\label{cnform}  |\Phi_\theta \>    =    \sum_{n=0}^{N}   c_n \,  e^{-i\theta  n}  \,  |n\>  \, ,  \qquad  c_n \ge 0  \, ,\forall  n\in  [0,N] \, , 
\end{align}
Holevo's POVM yields the gain 
\begin{equation}\label{G}
\begin{split}
\<G_{\rm det}\>&=\frac{1}{2}+\frac{1}{2}\<\Phi_0| \Delta|\Phi_0\>\\
\Delta_{mn}&=\frac{1}{2}\left[\delta_{m(n-1)}+\delta_{m(n+1)}\right] \, . 
\end{split}
\end{equation}
 In our case, the above expression yields the value  
\begin{align}
\label{detG1} 
\<G_{\rm det}  \>  &=\frac12+\frac{1}{2^{N+1}}\sum_{n=0}^{N-1}\sqrt{{N\choose n}{N\choose n+1}}  \\
\nonumber 
  &  = 1-  O\left(\frac 1 N\right) \, .   
\end{align}

Now, when the unknown  phase shift $e^{-i\theta  Z}$ is probed $N$ times,  one can obtain a much better estimate  by preparing the optimal input state, which in this case is the ``sine-state"  \cite{sin,berry-wiseman-2000-prl}
\begin{align}\label{EstTarget}
|\varphi_{{\rm opt},\theta}\>=\sqrt{\frac{2}{N+1}}  \, \sum_{n=0}^{N}\sin\left(\frac{n\pi}{N+1}\right) e^{i\theta n} \, |n\> \, .
\end{align}
This state achieves the Heisenberg scaling $\<G  \>   = 1-  O(1/N^2)$.  
 In the following, we will use our recursive protocol to  transform the state $|\varphi_\theta\>^{\otimes N}$ into approximations of the optimal state $|\varphi_{\rm opt,\theta}\>   $, which will then be used for state estimation. 

  Note that  the output states of our protocol are of the form (\ref{EstTarget}) at every step. Thanks to this fact, we can apply Holevo's recipe (\ref{G}) to compute the optimal gain.  Precisely, the gain  at the $k$-th step is given by
\begin{align}\label{Gn}
\left\<G^{(k)}  \right\>=\frac{1}{2}+\frac{1}{2} \,  \left\<\psi^{(k)} \right|  \, \Delta  \,\left|\psi^{(k)}  \right\>  \,  ,
\end{align}
where   $\left |\psi^{(k)}  \right\>$ is the output state at the $k$-th step, given by  Eq. (\ref{outputk}).   Averaging over the first  $T$ steps  we obtain the gain
\begin{align}
\<  G_T \>&:=\frac{\sum_{k=1}^{T}p_{\rm succ}^{(k)}\<G^{(k)}\>}{p_{\rm succ}(T)}  \label{Gav1}
\end{align}
where $p_{\rm succ}^{(k)}$ is the probability of achieving success at the $k$-th step and  $p_{\rm succ}  (T)   =  \sum_{k=1}^T  \,  p_{\rm succ}^{(k)}$.       The value of the gain can be explicitly calculated using Eqs.  (\ref{Ptot}),    (\ref{outputk}), and (\ref{solution1}). In Figure \ref{fig:EstTradeoff} we show the estimation gain for $N=30$ copies of the input state and for $K=  27 $ iterations of the recursive protocol.

\begin{figure}[t!]
\centering
\includegraphics[width=0.9\linewidth]{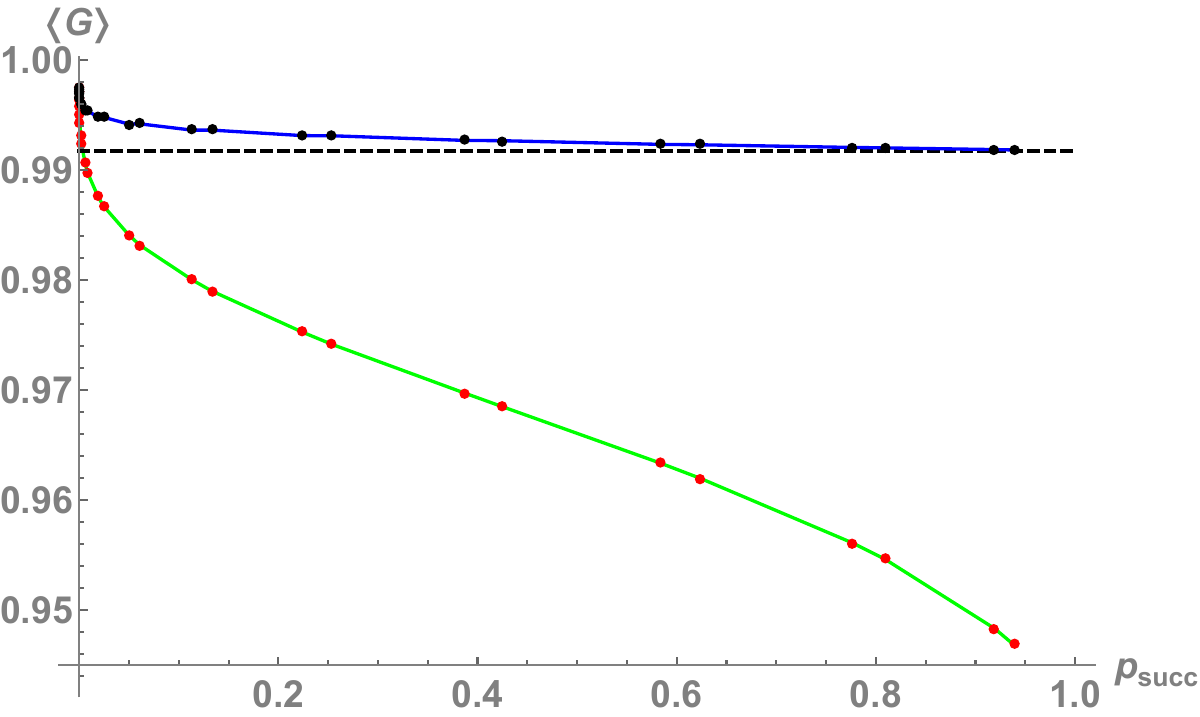}\\
\caption{{\bf Probabilistic phase estimation via the  recursive protocol and its coherent coarse-graining.} The figure shows the trade-off between success probability and  average gain for  phase estimation with  the  qubit state $|\varphi_\theta\>^{\otimes N}$ for $N=30$. The green solid line (with numerics represented by red dots) shows the trade-off between estimation gain and success probability for a recursive protocol with $K=  27$ rounds, with the $T$-th point corresponding to the  first $T$  steps.   The blue solid line (with numerics represented by the black dots) shows the trade-off for filters generated by coherent coarse-graining, with the $T$-th point corresponding to the coherent coarse-graining of the first $T$ steps.   Note that the gain for the coherent coarse-graining remains higher than the optimal deterministic estimation's gain (the black dashed line) even when the protocol becomes ``almost deterministic" (i.e. the probability of success tends to one), while the gain for the recursive protocol drops down quickly with the growth of the success probability.  }
\label{fig:EstTradeoff}
\end{figure}

The performance of the recursive protocol can be compared with the performances of its coherent coarse-graining.   By coherently coarse-graining over the first $T$ rounds, we obtain the average gain given by   
\begin{align}\label{Gavcoarse1}
\<G'_{T}\>=\frac{1}{2}+\frac{1}{2}  \,   \left\< \psi'  (T )  \right|   \Delta  \left|\psi'(T)  \right\>  \, ,      
\end{align}
with
\[\qquad \left|\psi'(T)  \right\>  
=   \frac{  M^{\prime (T) }  |\varphi\>^{\otimes N}  }{  \|  M^{\prime (T)}  |\varphi\>^{\otimes N}  \| }\]
and $M^{(T)}$ as in Eq. (\ref{coarsesolution}).    The estimation gain of the coherent coarse-graining is  plotted in Figure \ref{fig:EstTradeoff}.   
In addition,  Figure \ref{fig:EstGandP} shows the scaling of the  gain and  the success probability with the number of copies $N$.

\begin{figure}[t!]
\centering
\subfigure[]{\label{fig:EstGain1}
\includegraphics[width=0.9\linewidth]{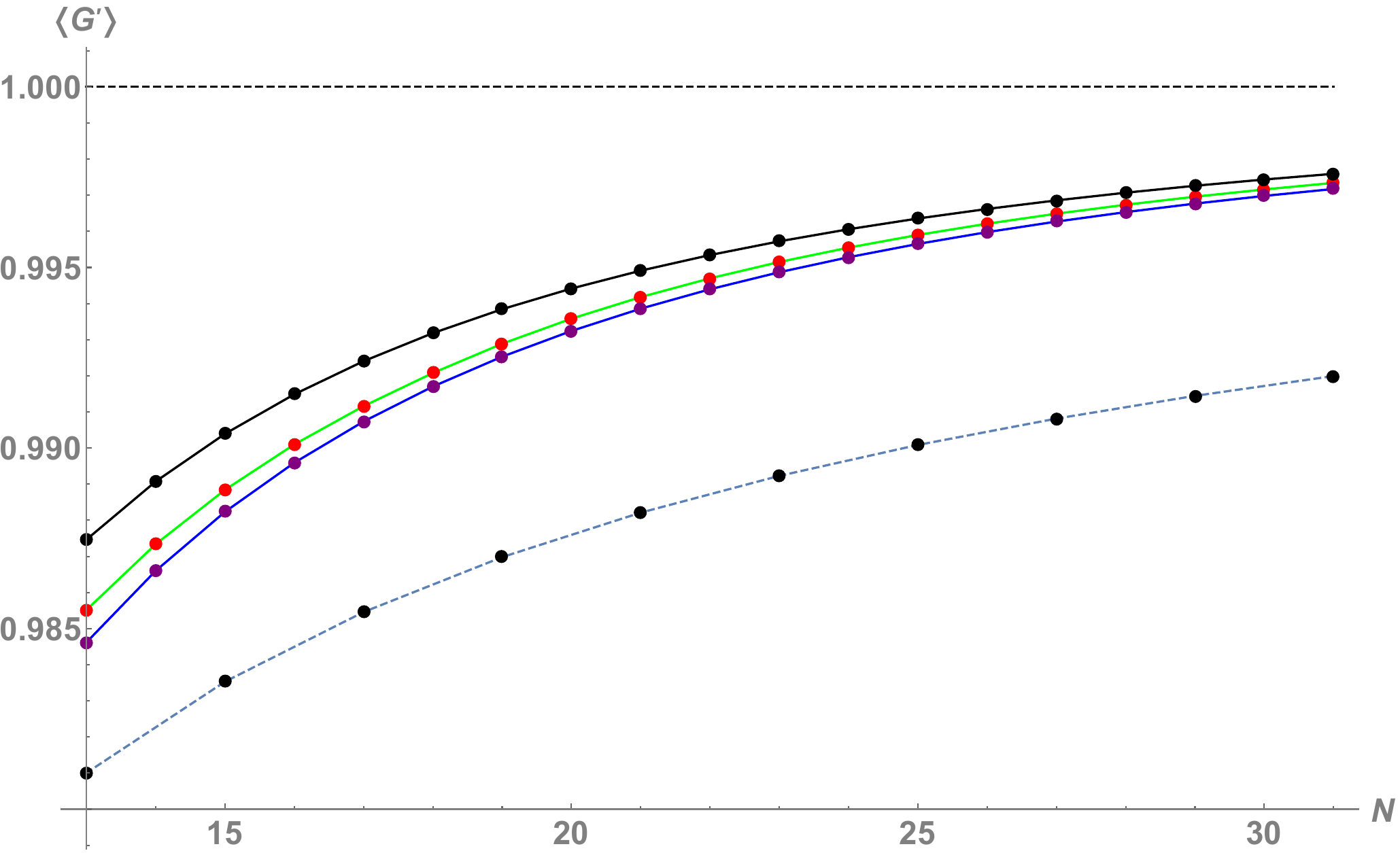}
}
\subfigure[]{\label{fig:EstProb1}
\includegraphics[width=0.9\linewidth]{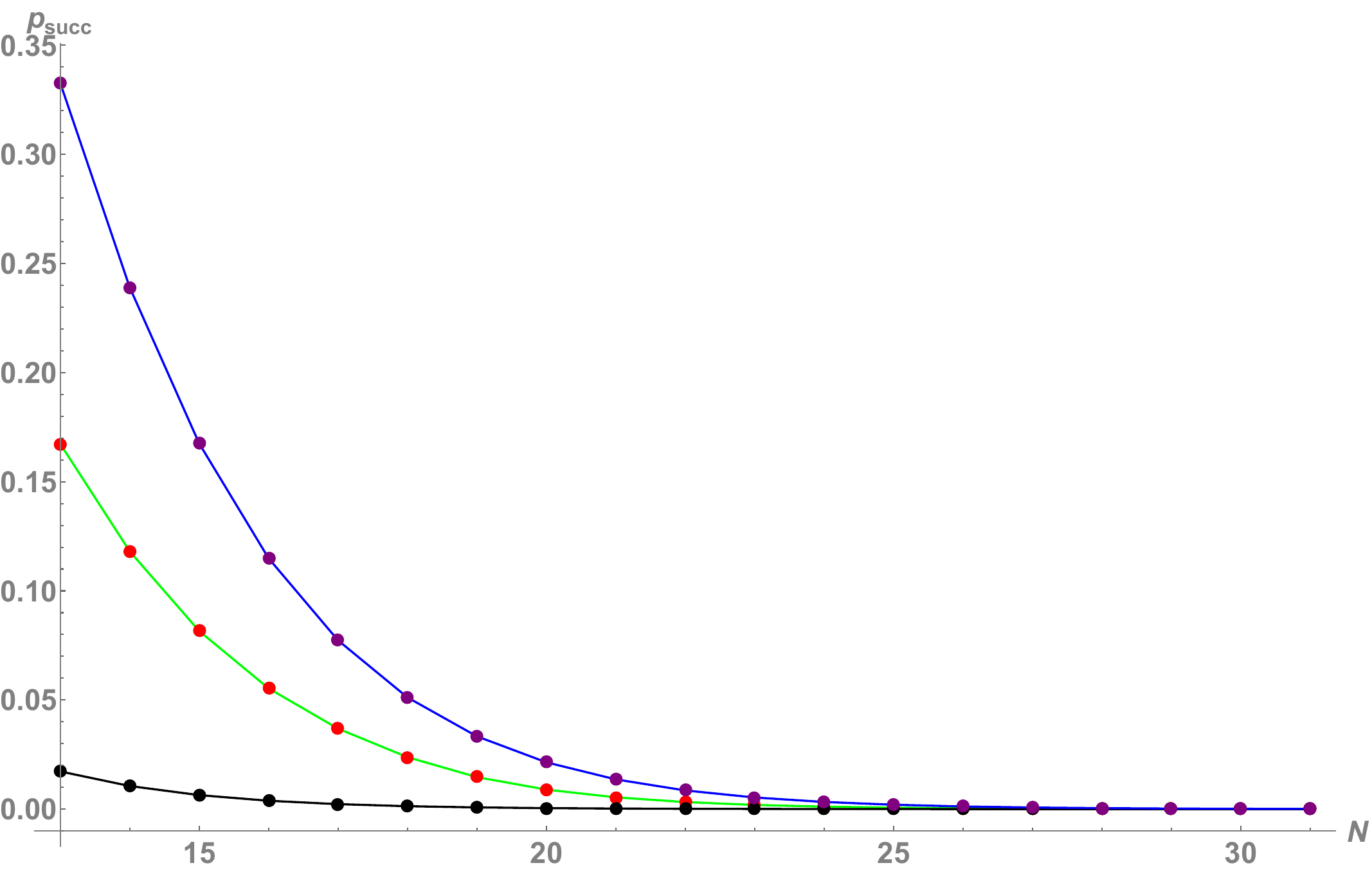}
}
\caption{{\bf Scaling of the gain and success probability for coherently coarse-grained protocols.  }  Figure  \ref{fig:EstGain1}  shows the estimation gain  as a function of the number of copies $N$, for three coherently coarse-grained protocols corresponding to different values of $T$, including  $T=1$ (black line with black dots), $T=3$ (green line with red dots) and $T=5$ (blue line with purple dots).    The dashed line with black dots represents the optimal deterministic gain $\<G_{\rm det}\>$.    Figure  \ref{fig:EstProb1} shows the decrease of the total success probability as a function of $N$ for different values of $T$, including $T=1$ (black line with black dots), $T=3$ (green line with red dots) and $T=5$ (blue line with purple dots).  }
\label{fig:EstGandP}
\end{figure}

\subsection{Converting coherence into metrological precision}\label{subsec:maxcoh}

In the previous Subsection we analyzed  the problem of phase estimation with $N$ identical qubits.   
Here we will consider a one-shot scenario, where the phase has to be estimated from a single copy of the state  
\begin{align}\label{EstResource}
|\varphi_\theta\>:=\frac{1}{\sqrt{N}}\sum_{n=0}^{N-1}  \,  e^{-in\theta} \,   |n\> \, ,
\end{align}
consisting of a uniform superposition of  the eigenstates of the Hamiltonian $H  =    \sum_{n=1}^{N-1}  \,   n\,  |n\>\<n|$.  The above state is the \emph{maximally coherent state} \cite{incoherence} in the energy eigenbasis, that is, it is the most valuable state in the resource theory of coherence.  Interestingly, it is \emph{not} the optimal state for phase estimation.  
Indeed, the estimation gain for the  maximally coherent state can be evaluated with Eq. (\ref{G}), which in this case yields 
\begin{align}\label{detG}
\<G_{\rm det}  \>=1-\frac1 { 2N} \, .
\end{align}
When the number of copies is asymptotically large, the gain approaches its maximum value with the standard quantum limit scaling $1/N$, rather than the Heisenberg scaling $1/N^2$. 

We now  explore how the maximally coherent state can be transformed into approximations of the optimal state for phase estimation.   The performances of the recursive protocols and of its coherent coarse-graining can be evaluated using Eqs. (\ref{Gav1}) and (\ref{Gavcoarse1}). 
  When the number of iterations $T$ is small compared to the number of energy levels $N$, the average gain has the simple analytical expression  
\begin{align}\label{asyG}
\<G_T\>=1-\frac{\pi^2}{2N^2}\left[T(T-1)+\frac12\right]+O\left[\left(\frac{T}{N}\right)^{3}\right]  \, .
\end{align}
Note that the gain exhibits   Heisenberg scaling with the number of energy levels $N$, with a constant that grows quadratically with the number of rounds $T$.    The success probability can also be evaluated analytically in the regime $  N \gg  T$ and its value is given by
\begin{align}\label{asyP}
p_{\rm succ}(T)=\frac{1}{2}+\frac{\pi^2}{N^2}\left[T(T-1)+\frac18\right]+O\left[\left(\frac{T}{N}\right)^{3}\right]   \, .
\end{align}  
From the above expressions, one can clearly see the trade-off between gain and success probability,  which can be made explicit in the trade-off curve    
\begin{align}\label{curve}
\<G_T\>=1-\frac{3\pi^2}{16N^2}-\frac{p_{\rm succ}(T)-1/2}{2}  \, ,  \qquad N \gg T \,  . 
\end{align}
In Figure \ref{fig:EstTradeoff} we illustrate the trade-off between the probability of success and the average gain for $N  =61$.    The recursive protocol manages to increase the probability of success by approximately $30\%$ from the first round to the 14-th, while keeping the average gain above the deterministic gain.   In Figure \ref{fig:EstGandP} we show the scaling of the  gain and  the success probability with the dimension $N$   for different values of $T$. 

\begin{figure}[h!]
\centering
\includegraphics[width=0.9\linewidth]{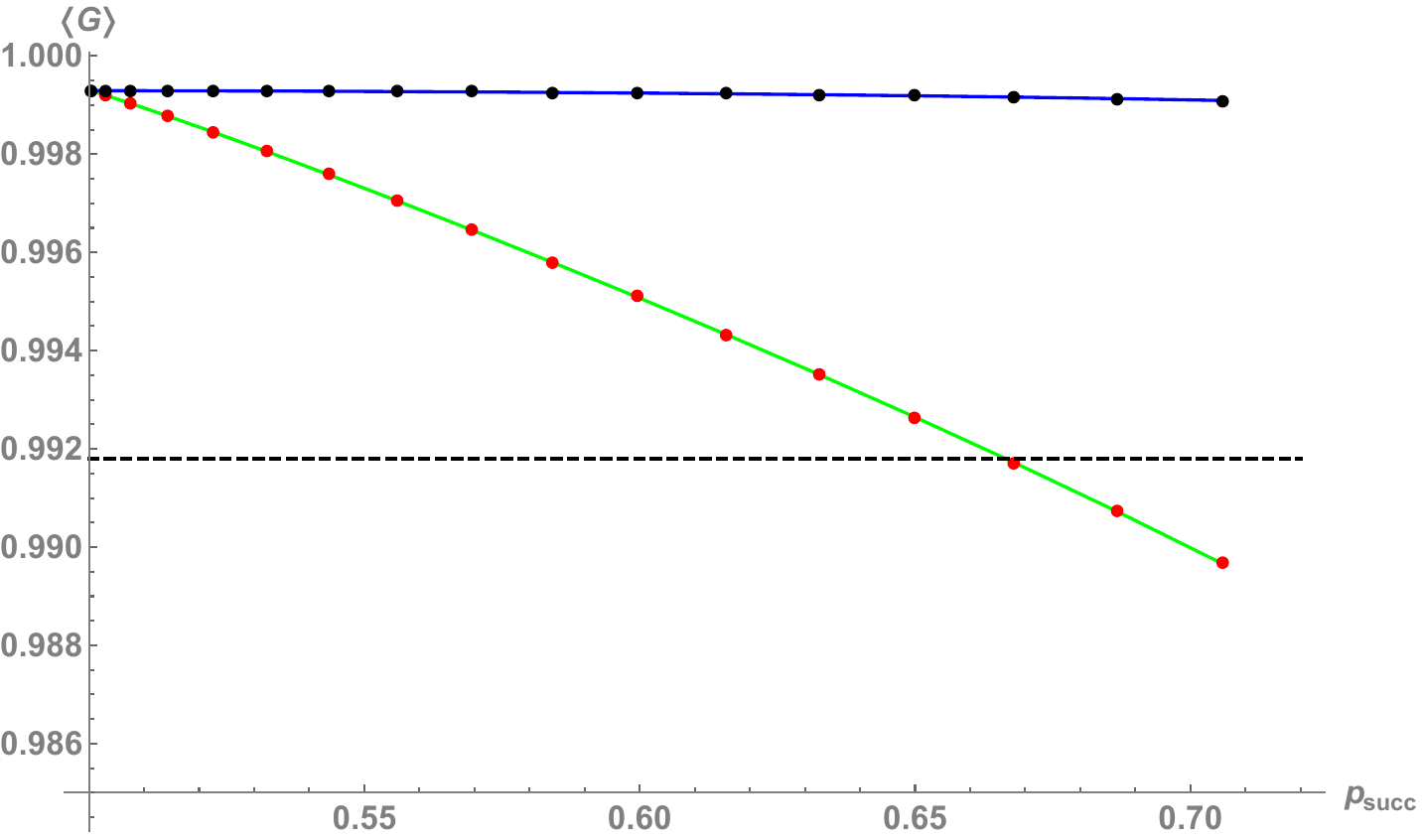}\\
\caption{{\bf Probabilistic phase estimation via the  recursive protocol and its coherent coarse-graining.} The figure shows the trade-off between success probability and  average gain for  phase estimation with   maximally coherent states with $N=61$. The green solid line (with numerics represented by red dots) shows the probability-gain trade-off for  $K=  17$ rounds of the recursive protocol. At the first round the protocol reaches the maximum possible gain, equal to $G_{\max}  = 99.9\%$, in agreement with the analytical expression of Eq.  (\ref{asyG}).   The blue solid line (with numerics represented by the black dots) shows the trade-off for filters generated by coherent coarse-graining, with the $T$-th point corresponding to the coherent coarse-graining of the first $T$ steps of the recursive protocol.   For the first $K=17$ rounds the estimation gain of coherent coarse-graining remains approximately equal to  $G_{\max}  = 99.9\%$, although eventually it is bound to decrease to the optimal deterministic value  $\< G_{\rm det}\>   =  99.2\%$  (black dashed line).  }
\label{fig:EstTradeoff}
\end{figure}

\begin{figure}[h!]
\centering
\subfigure[]{\label{fig:EstGain}
\includegraphics[width=0.9\linewidth]{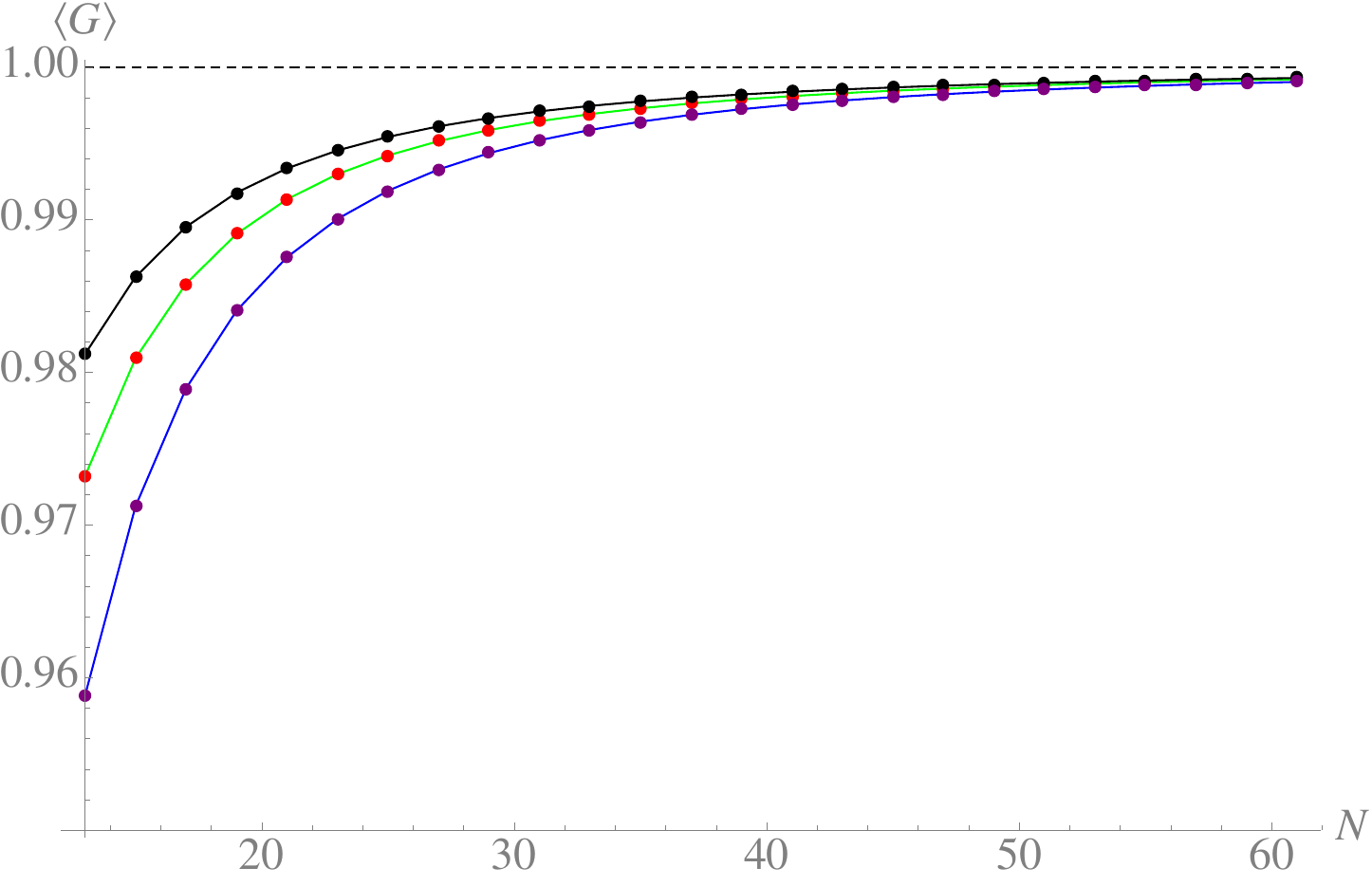}
}
\subfigure[]{\label{fig:EstProb}
\includegraphics[width=0.9\linewidth]{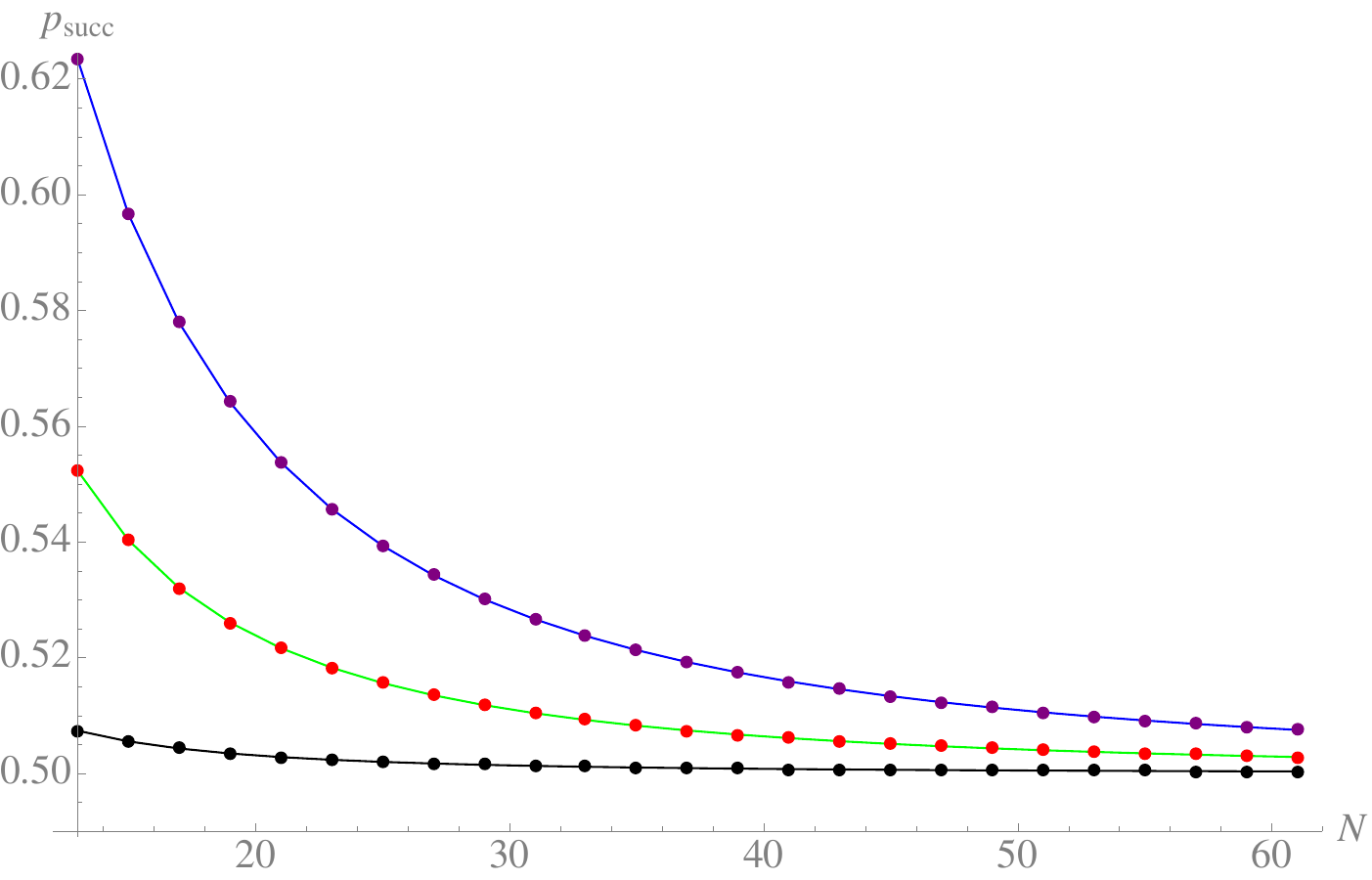}
}
\caption{{\bf Scaling of the estimation gain and success probability for the recursive protocol.} Figures  \ref{fig:EstGain} and  \ref{fig:EstProb} show the average gain $G$  and   the total success probability (\ref{fig:EstProb}) as a function of the energy scale $N$ for different values of $T$, including $T=1$ (black line with black dots), $T=2$ (green line with red dots) and $T=3$ (blue line with purple dots).  }
\label{fig:EstGandP}
\end{figure}

Let us evaluate  now the performance of   coherent coarse-graining.     In the $N\gg  T$ regime, the gain has the analytical expression  
\begin{align*}
\< \, G_T' \, \> =1-\frac{\pi^2}{4N^2}\,  \left \{1+4 \, \left[\,  p_{\rm succ}(T)-\frac 12\,  \right]^2  \right\} \, . 
\end{align*}
The trade-off between estimation gain and probability of success is illustrated in Figure \ref{fig:EstTradeoff} for $N=61$. 
Also in this case,  one can observe that coherent coarse-graining offers a better trade-off curve than the recursive protocol.  

Finally, it is interesting to compare the performance of the coherent coarse-graining with the optimal trade-off curve between gain and probability of success, which is known explicitly in the case under consideration    \cite{BaganPRL}.    Remarkably, the comparison shows that  for large $N$ the coherent coarse-graining yields exactly the optimal estimation  strategy of Ref.  \cite{BaganPRL}.   In other words, in this case the coherent coarse-graining of our recursive protocol is asymptotically optimal.    
At this point, a natural question is whether coherent coarse-graining   \emph{always} gives the optimal fidelity/probability trade-off.    The answer turns out to be negative: by evaluating Eq. (\ref{Gavcoarse1})  for small values of $N$  (e.g., $N=10$)   we find out that the average gain of the coherent coarse-graining  sometimes falls below the threshold of the optimal deterministic gain in Eq. (\ref{detG}), clearly indicating sub-optimality in the non-asymptotic regime. 

\subsection{Energy-preserving  cloning of quantum coherence}\label{subsec:cloning}
 
Here we consider the problem of quantum cloning  \cite{rmp,cerfreview}, where the task is to transform  $N$ identical  copies of an unknown quantum state into  a  larger number  $M  \ge N$ of approximate copies.  
In most cases, the problem  has been addressed without imposing any constraint on the cloning process, except for its compatibility with the laws of quantum mechanics.    Instead, here we consider copy machines that have to work without any supply of energy for the outside.  Consider for example  a scenario where one wants to clone the state of a quantum clock \cite{GiulioNat}, given by   
  $$   |\psi_t\>    =   e^{    -  it H/\hbar}   |\psi\> \, , $$  
  where $H  =  H^\dag$ is a suitable Hamiltonian.     Here the time parameter $t$ is assumed to be unknown  and the copy machine is required to work equally well for every value of $t$.     In order to produce copies without requiring energy from the outside, the machine has to process the $N$ input clocks  jointly with a state of $M-N$   ``blank clocks," which provide no information about time,    but possess sufficient energy to enable the desired transition.  Indeed, in order to approximate $M$ perfect copies of the state $|\psi_t\>$ the machine should at least  be able to produce  output  states that have energy close to $M  \<  \psi|  H|\psi\>$, meaning that the blank clocks  should have energy close to $(M-N)   \<  \psi|  H|\psi\>$.   The problem of energy-preserving cloning of clock states is equivalent to the  problem of cloning coherence:  denoting by $  |\beta\> $ the blank state,  the cloning machine attempts at   converting the state $   |\psi\>^{\otimes N}  \otimes |\beta\>$ into the state $ |\psi\>^{\otimes M}$.     Choosing the blank state to be an eigenstate of the energy, we have that  maximizing  the fidelity for the  transition   $   |\psi\>^{\otimes N}  \otimes |\beta\>   \xrightarrow  {} |\psi\>^{\otimes M}$ under the energy-preserving restriction is equivalent to maximizing  the fidelity of cloning for every instant of time.  
   
In the following we analyze in detail the simplest example of energy-preserving cloning of quantum coherence:    we consider $N$ two-level systems, each of them with Hamiltonian $H=\frac{\hbar\omega}{2} \,  Z  $ and  initially prepared in the coherent superposition $|+\>  =  (|0\>+|1\>)/\sqrt{2}$.
 The question is how well one can produce $M  >  N$ approximate copies without paying an energy cost.  For simplicity, we assume that the difference $M-N$ is even:   under this assumption, we can choose  the blank state  to be an energy eigenstate   with  energy  \emph{exactly} equal to zero.      Specifically, we choose the    state $|\beta\>=  |M-N, 0\>$, belonging to  the symmetric  eigenbasis
\begin{align*}
|L, m\>:=\frac{ \sum_{\pi\in\set{S}_{L}}   U_\pi|0\>^{\otimes (L+m)/2} \,  \otimes \, |1\>^{\otimes (L-m)/2}}{  \sqrt{  L!   \,   [( L+m)/2]! \,  [  (L-m)/2]!}}   \, ,
\end{align*}
where $\set{S}_{L}$ denotes the group  of  permutations of an $L$-element set and $U_{\pi}$ is the unitary that permutes $L$ Hilbert spaces according to the permutation $\pi$. 

We now apply  our  recursive protocol,  producing at each step an approximation of the desired $M$-copy state.      Let us  expand the states $ |\psi\>^{\otimes N} $ and $|\psi\>^{\otimes M}$  as 
$$|\psi\>^{\otimes L}=2^{-L/2}\sum_{m=-L}^{L}\sqrt{{L\choose \frac{L-m}2}}    \, |L,m\>  \qquad   L  =  M,N  \, , $$
 then use the formulas for the fidelity and success probability derived in Section~\ref{sec:network}. 
  At the first step of the protocol,  the successful quantum operation  produces an output state with the maximum possible fidelity, given by  
 $$F^{(1)}_{\max}=\frac{1}{2^M}\sum_{n=-N}^{N}{M\choose \frac{M-n}{2}}.$$
The above  fidelity turns out to be equal to  the absolute maximum of the fidelity achievable over all covariant quantum operations, derived by Fiura\`sek in Ref.  \cite{fiurasekclon}. For large $N$,  the fidelity is  close to 1 whenever $M$ is small compared to $N^2$, thus allowing one   to achieve quantum super-replication \cite{GiulioNat}. It is well known   that the price of super-replication is a probability of success vanishing exponentially fast with $N$  \cite{GiulioNat}.   The main interest of our recursive protocol lies in the fact that it allows us to increase the probability of success.    In a protocol with $K>1$ steps, the average fidelity decreases at each step, while the probability of success increases. The trade-off between the fidelity and the probability of success  is illustrated  in Figure~\ref{fig:qubit} for the case of $N=  80$, $M = 400$ and $K=  32$, using Eqs. (\ref{fidelity1}) and (\ref{prob1}).   In addition,  we compare the fidelity of the recursive protocol with that of its coherent coarse-graining, given by Eq. (\ref{F'}).    As already observed,  the coherent coarse-graining achieves a  higher fidelity, while keeping  the same success probability. In the figure we also plot the optimal fidelity in  the deterministic case  (black dashed line in Figure~\ref{fig:qubit}).       The deterministic fidelity (derived in Theorem \ref{theo:EAoptimal}) coincides with the fidelity for phase-covariant cloning \cite{qubit}, meaning that the optimal cloner can be realized in an energy-preserving fashion. 

Figure \ref{fig:qubit}  well illustrates  the advantages of the recursive protocol.  At the  first round the fidelity is very high, but the success probability has the  extremely tiny value $p_{\rm succ}^{(1)}  =  6\times10^{-20}$.   The   subsequent rounds of the  protocol   increase the  success probability dramatically, reaching a probability of approximately   $23\%$ at the 31-st step.  The fidelity for the recursive protocol remains higher than the optimal deterministic fidelity up to almost the very last step. An even better performance is attained through coherent coarse-graining.

\begin{figure}
\centering
\includegraphics[width=0.9\linewidth]{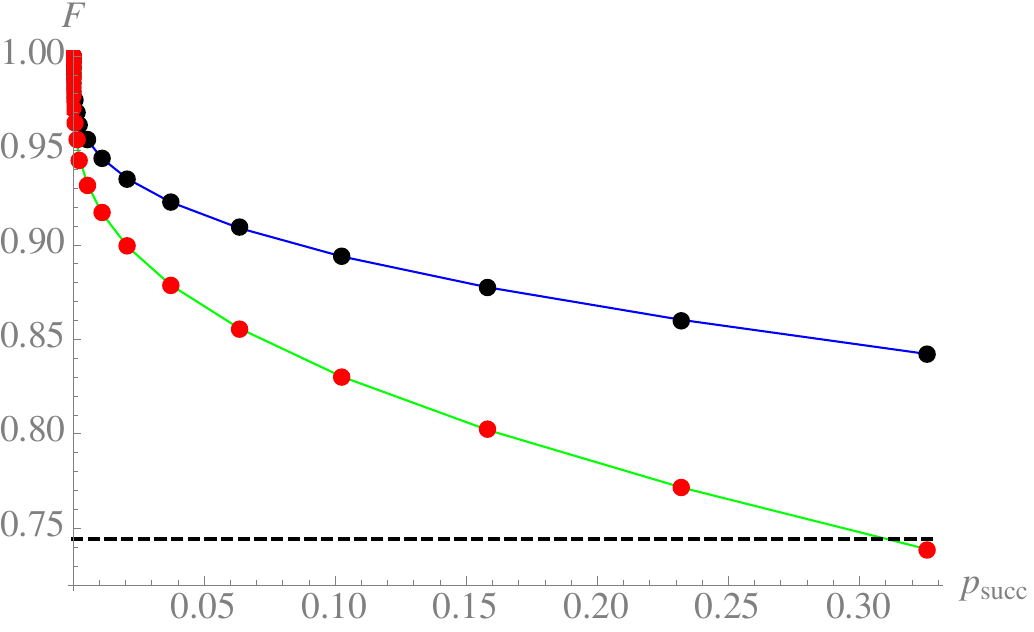}\\
\caption{{\bf Energy-preserving cloning of quantum clocks via  the  recursive protocol and its coherent coarse-graining.} The figure shows the trade-off between  success probability and fidelity  for the $N$-to-$M$ cloning of the clock state $|\psi_t\>  = (e^{i\omega t/2}|0\>+e^{-i\omega t/2}|1\>)/\sqrt{2}$ with $M=400$ and $N=80$. The green solid line (with numerics represented by red dots) shows the probability-fidelity trade-off for a recursive protocol with $K=  32$ rounds. The blue solid line (with numerics represented by the black dots) shows the trade-off for filters generated by coherent coarse-graining, with the $T$-th point corresponding to the coherent coarse-graining of the first $T$ steps  of the recursive protocol.   Finally, the black dashed line represents the fidelity for the optimal deterministic cloning protocol.  Notice that the recursive protocol maintains fidelity larger than the optimal deterministic fidelity for all steps up to the last.  }
\label{fig:qubit}
\end{figure}

\subsection{Probabilistic energy-preserving amplification of coherent light}\label{subsec:coherent}

In quantum optics the energy-preserving instruments are those that preserve the average photon number.   In the single-mode scenario, the number observable is non-degenerate and  the energy-preserving quantum operations have diagonal  Kraus operators in the Fock basis $\{|n\>\}$. In the following we consider the application of the recursive protocol to the amplification of the coherent state of light 
\begin{align*}
&|r_1    \>\xrightarrow{  \quad }|r_2  \>  \qquad 0\le r_1\le r_2      \,  .
\end{align*}
Note that, since we require the amplification map to be part of a number-preserving quantum instrument, our protocol defines a phase-insensitive amplifier \cite{insensitive}, which works equally well for the transition 
\begin{align*}
&|r_1  e^{i\theta}  \>\xrightarrow{ \quad }|r_2  e^{i\theta}  \>  \qquad 0\le r_1 \le r_2   \, ,          
\end{align*}   
where $\theta$ is an arbitrary angle.  

Amplifying a coherent state without increasing its photon number seems to be a daunting task. However, the fact that the number is preserved only on average grants us the opportunity  to reach high fidelity in a probabilistic fashion. In the case of amplifiers, the trade-off between  success probability and   fidelity is essentially a trade-off between success probability and  photon number modulation. 

Since the Hilbert space is infinite dimensional,    our recursive protocol cannot be applied directly.   To overcome the obstacle, we define a threshold $N$ and assume that the successful operations project the input state inside the subspace spanned by Fock states with number smaller than $N$.  Practically,  for $N  \gg    r_2^2$, the projection can be done without affecting the fidelity.    The fidelity  at the $k$-th round, given by  Eq. (\ref{fidelity1}), can be lower  bounded as
\begin{align}\label{coherentfid}
F^{(k)}_{\max}\ge1-e^{-r_2^2}\left(\frac{r_2^2e}{N-k+1}\right)^{N-k+1} 
\end{align}
when $N-k+1>r_2^2$.
On the other hand,   the probability of success in Eq. (\ref{prob1}) can be expressed as
\begin{align}\label{coherentprob}
p_{\rm succ}^{(k)}=\left\{\begin{matrix} e^{r_2^2-r_1^2}\left(\frac{r_1}{r_2}\right)^{2N} \,   F^{(1)}_{\max}& \quad k=1\\
\\
e^{r_2^2-r_1^2}\left(\frac{r_1}{r_2}\right)^{2N-2k+2}\left[1-\left(\frac{r_1}{r_2}\right)^2\right]   \,  F^{(k)}_{\max}&\quad k>1 \end{matrix}\right.
\end{align}
\begin{figure}
\centering
\includegraphics[width=0.9\linewidth]{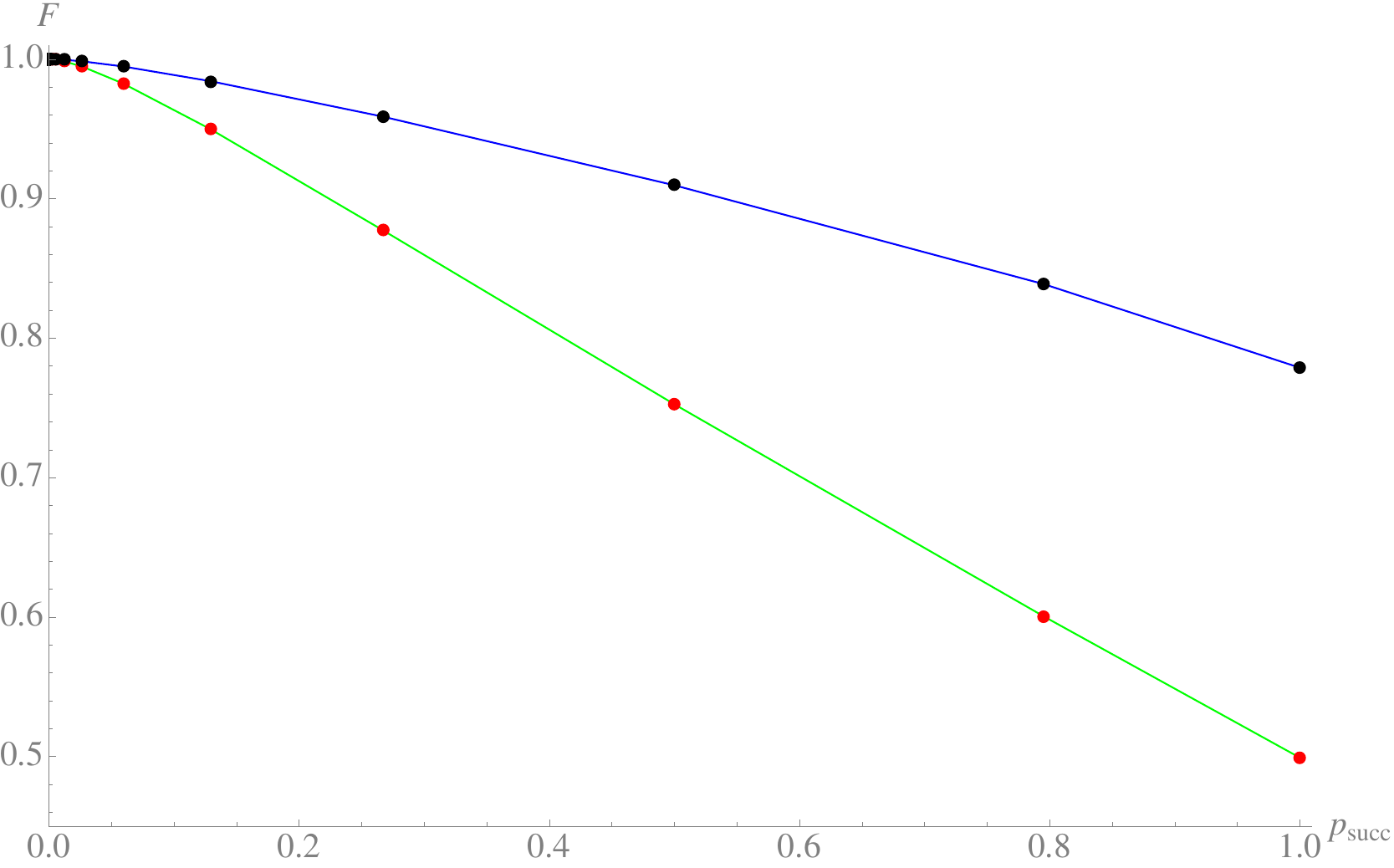}\\
\caption{{\bf Energy-preserving amplification of coherent light via the recursive protocol and its coherent coarse graining.} The figure shows the trade-off between  success probability and the average fidelity for the amplification $|r_1 \>\to|r_2 \> $ with $r_1=1$, $r_2=1.5$ and $N=80$. The green solid line (with numerics represented by red dots) shows the probability-fidelity trade-off for a recursive protocol with $K=81$ rounds. The blue solid line (with numerics represented by the black dots) shows the trade-off for filters generated by coherent coarse-graining.  Note that the difference between the two curves becomes large as the success probability tends to 1.    For unit probability the recursive protocol and its coherent coarse-graining give fidelities   $F_{\rm det}  =  49.9\%$ and   $F_{ {\rm det}}'  =   |\<r_1|r_2\>|^2  =77.9\%$, respectively.  }
\label{fig:amplify}
\end{figure}

Interestingly, the successful quantum operation at the first round of our protocol ($k=1$) coincides with the optimal probabilistic amplifier for coherent states with known amplitude   \cite{fiurasekclon,caves}, which indeed can be implemented with energy-preserving operations.    
Specifically, evaluating  Eqs. (\ref{coherentfid}) and (\ref{coherentprob}) for $k=1$ one retrieves the expressions for the optimal fidelity and success probability appearing in Eqs. (6.24) and (6.36) of Ref. \cite{caves} 
For the subsequent rounds of the recursive protocol  ($k>1$), the input state is not coherent anymore and the successful quantum operation differs from the optimal coherent-state amplifier.

In Figure \ref{fig:amplify} we show the performance of the recursive protocol and its coherent coarse-graining for the amplification of coherent states  from $r_1=1$ to $r_2=1.5$. The threshold in the Fock space  is chosen to be $N=80$ and the protocol is applied recursively for $K=81$ rounds. From the plot it can be seen that  the filters generated by coherent coarse-graining reach a relatively high fidelity, compatibly with the strong constraint set by the number-preserving condition.  For instance, the coherent coarse-graining of the recursive protocol with $K=80$ succeeds with probability $p_{\rm succ} =79.6\%$ and  reaches  fidelity $F=  83.9\%$.    When the probability reaches 1, the fidelities of the recursive protocol and its coherent coarse graining become    $F_{\rm det}  =  49.9\%$ and   $F_{ {\rm det}}'  =   |\<r_1|r_2\>|^2  =77.9\%$, respectively.     The latter is well above the fidelity of the optimal amplifier for \emph{arbitrary} coherent states, which is given by $F_{\rm universal}=  4/9 $  \cite{namikirapid,chiribella-xie}.

\subsection{Energy-preserving correction in ancilla-driven quantum computation}\label{sec:unlearn}

In ancilla-driven quantum computation \cite{ADQC}  the evolution of the system is determined by the outcomes of   measurements on the ancilla.  Ideally, the goal is to implement  measurements that induce unitary gates on the system.  To achieve this goal, the measurements should not not extract any information about the state of the system: the probability of each outcome should be the probability that a particular unitary gate is applied to the system  \cite{LostFound}.   However, in many non-ideal situations the measurement extracts some information, thus inducing a non-unitary evolution on the system.   When this is the case, one can attempt to correct the unwanted non-unitarity by performing additional measurements.  
This type of correction has been studied in Refs. \cite{trade-off,unlearn}, 
where a number of different strategies have been proposed.

Here we  consider  the problem in the energy-preserving setting:  suppose that a quantum system  with $d$  non-degenerate energy levels interacts with an ancilla via an energy-preserving unitary evolution.  Then, the ancilla undergoes the measurement  of an observable that is compatible with the energy.  As a result, the system evolves randomly according to an energy-preserving instrument $\{\map M_x\}_{x\in\set X}$.   We assume that the measurement on the ancilla is a rank-1 projective measurement and, therefore, the quantum operations $\{\map M_x\}$ are pure.   For every given $x  \in\set X$, the problem is to correct the quantum operation $\map M_x$, making it as close as possible to a desired energy-preserving unitary gate  $U_x$.      As a correction we allow ourselves to use an energy-preserving filter, with quantum operations $\left\{\map N^{(x)}_{\rm succ},  \map N^{(x)}_{\rm fail}\right\}$.   Due to the presence of the filter, an initial pure state $|\eta\>$ is transformed probabilistically into the pure state 
$$|\eta_x\>   =      \frac{  N^{(x)}_{\rm succ}   \,  M_x   \,  |\eta\> }{\|   N^{(x)}_{\rm succ}   \,  M_x   \,  |\eta\> \|}   \, .$$
To evaluate the quality of the correction, we consider the fidelity between  $|\eta_x\>$ and the target state   $  U_x  |\eta\>$,   averaging  over all possible pure input states.  
    Assuming that initially the state $|\eta\>$   is drawn at random according to the Haar measure, the conditional probability distribution over the pure states is given by  
$$    p(\eta|  x,  {\rm succ} )   \,  \d \eta   =   \lambda_x  \,        \left\|      N^{(x)}_{\rm succ}   \,  M_x   \,  |\eta\>  \right\|^2       \,  \d \eta   \,  ,  
$$      
 where $M_x$ and $  N^{(x)}_{\rm succ}$ are the Kraus operators of $\map M_x$ and $\map N^{(x)}_{\rm succ}$, respectively, and $\lambda_x $  is the normalization constant $\lambda_x: = \left( \int        \|     N^{(x)}_{\rm succ}   \,  M_x   \,  |\eta'\>  \|^2    \,  \d  \eta' \right)^{-1}$.  
 Hence, the average fidelity over all pure states is given by  
 \begin{align}\label{Fx}
\nonumber F_x  & : =  \int  \d \eta  \,   p(\eta|  x, {\rm succ})    \,      \left  |    \<   \eta_x|  U_x  |\eta\>  \right|^2 \\ 
 \nonumber &  =  \frac{ \int  \d \eta  \,      | \<  \eta|   U_x^\dag    N^{(x)}_{\rm succ}   M_x  |\eta\>  |^2   }{  \int\d  \eta'     \|     N^{(x)}_{\rm succ}   \,  M_x   \,  |\eta'\>  \|^2   }   \\
 &   =  \frac{  F_{0}^{(x)} \cdot d+1 }{  d +1 }
 \end{align}
where $F_{0}^{(x)}$ is the fidelity given by 
$$  F_{0}^{(x)}  =   \frac{   \left  | \<  e_0 |    \,      U_x^\dag  N^{(x)}_{\rm succ}    M_x      \,|e_0  \>\right|^2}{\|   N^{(x)}_{\rm succ}  M_x   |e_0\> \|^2}    \qquad  |e_0\>     =    \frac{  \sum_{n=1}^d    |n\>}{\sqrt d}\,. $$
Maximizing the average fidelity   is then equivalent to finding  the optimal quantum operation for the transformation
$$     |\varphi_x\>  :=    \frac{  M_x    |e_0  \>}{  \|   M_x    |e_0  \>  \|} ~ \xrightarrow{\quad} ~ |\psi_x\>   :  =    U_x  |e_0\> \, .   $$
 The maximization under the energy-preserving constraint is exactly the problem solved in this paper.  In particular, for every outcome $x$ we can use our recursive protocol to obtain a high-fidelity approximation of the desired transformation. In this context, it is immediate to realize that our protocol provides an approximate correction strategy, with the property that    the overall quantum operation acts exactly like the target gate $U_x$ in a suitable subspace, whose dimension shrinks at every step.    
 
For concreteness, let us see explicitly how the protocol works in a concrete example. We choose the quantum operation $\map M_x$ with Kraus operator $  M_x   =   \sum_{n=1}^d    \,     \mu^{n/2} \,   |n\>\<n|$. 
The  fidelity at the $k$-th  step is given by  
$$    F^{(k)}_x    =  \frac{d+2  -  k}{d+1}  \,  ,$$    
while the probability of success, averaged over all pure states, is given by  
$$ p_{\rm succ}^{(k)}   =\left\{\begin{array}{ll}       \,  \mu^{d-k}(1-\mu)^2(d+1-k)/(1-\mu^d)&\quad k>1   \, \\ \\
\mu^{d-1}(1-\mu)d/(1-\mu^d)&\quad k=1\,.\end{array}\right.
$$ 

\begin{figure}
\centering
\includegraphics[width=0.9\linewidth]{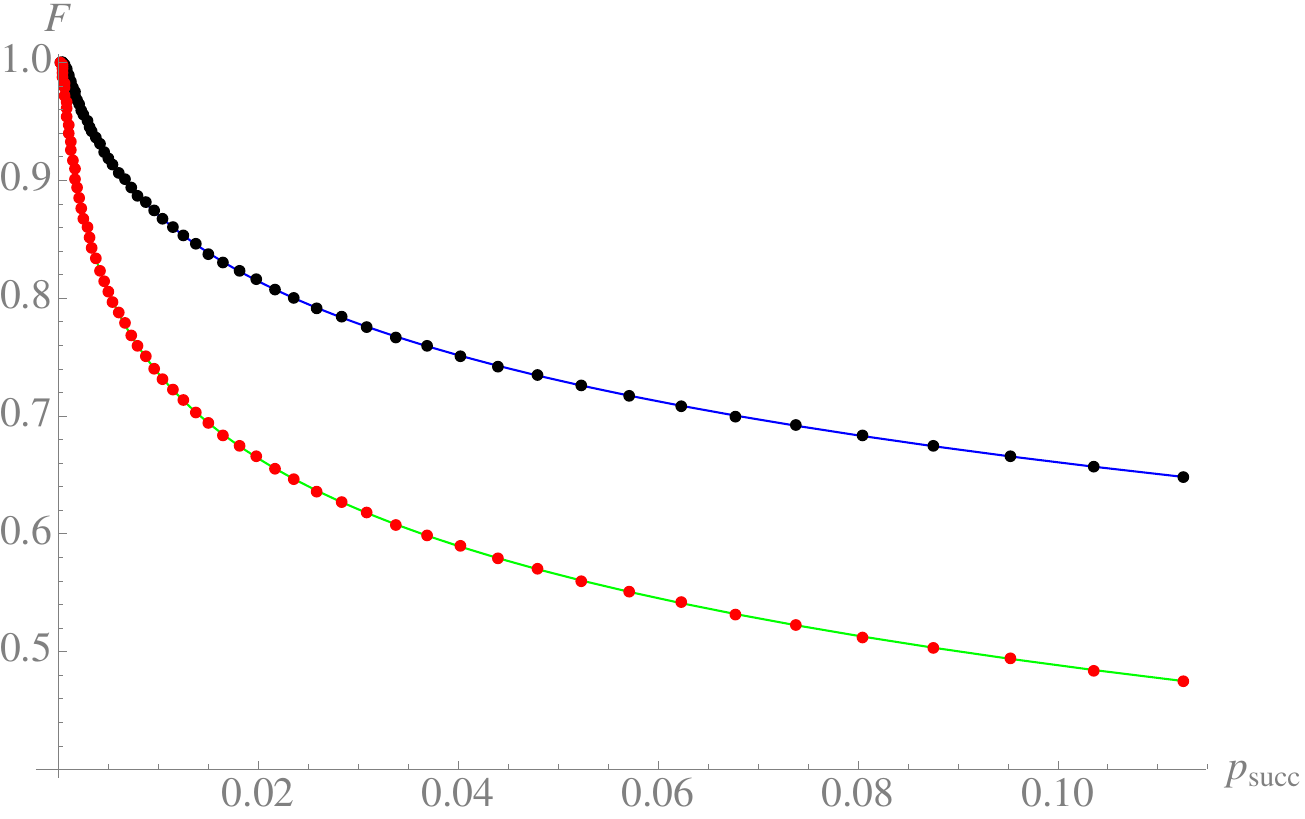}\\
\caption{{\bf Energy-preserving correction of a quantum operation via the recursive protocol and its coherent coarse graining.} The figure shows the trade-off between  success probability and  average fidelity  for unlearning the quantum operation with Kraus operator $  M_x   =   \sum_{n=1}^d    \,     \mu^{n/2} \,   |n\>\<n|$ with $d=100$ and $\mu=0.9$. The green solid line (with numerics represented by red dots) shows the probability-fidelity trade-off for a recursive protocol with $K=  70$ rounds. The blue solid line (with numerics represented by the black dots) shows the trade-off for filters generated by coherent coarse-graining.}
\label{fig:unlearn}
\end{figure}

The features of the recursive protocol and of its coherent coarse-graning are  illustrated in  figure  \ref{fig:unlearn} for  $d=100$, $\mu=0.9$ and $K=70$.  The probability of success increases from a very small value ($p_{\rm succ}^{(1)}  =  3\times10^{-4}$) to approximately $14\%$ at the 68-th step, at the cost of a reduced fidelity.

\section{Extension to mixed states} \label{sec:mixed}
So far we considered transitions between pure states.   
However, for many practical applications it is important to consider transformations where the input states are mixed,~e.~g.~due to the presence of decoherence.   Normally the target  state is still pure, since  ideally one would like to remove the noise from the output.    This is the case for tasks like mixed state purification \cite{cirac-ekert-macchiavello,andersen,zhao2017quantum}, super-broadcasting \cite{dariano-macchiavello-perinotti-prl}, and for the evaluation of the  corresponding quantum benchmarks \cite{polzik-NJP}.   In all these cases, our techniques provide general upper bounds on the fidelity of the optimal energy-preserving operations.     The bounds are achievable for a quite large class of states, which includes all the thermal states associated with stoquastic Hamiltonians \cite{bravyi2008thecomplexity,bravyi2009complexity}, such as the Hamiltonians of flux qubits in Josephson junctions \cite{burkard2004multilevel} and the Hamiltonians of  (bosonic) Bose-Einstein condensates \cite{ceperley1995path}.     

\subsection{Deterministic transitions and eigenstate alignment}

Given a generic mixed state  $\rho$, we search for the best energy-preserving approximation of the transition $\rho  \to  |\psi\>\<\psi|$, where $|\psi\>$ is a target pure state.   To this purpose, it is convenient to decompose the initial state into blocks corresponding to the different energy eigenvalues, namely 
\begin{align}\label{blockrho}
\rho  =  \sum_{E,E'}    \,  \rho_{E,E'}  \qquad \rho_{E,E'}  =    P_E   \rho  P_{E'} \, . 
\end{align}
With this notation, we have the following: 
\begin{theo}\label{theo:EAoptimal_mixed}
For $p_{\rm succ}  =1$, every energy-preserving approximation of the transition $\rho  \to |\psi\>\<\psi|$ satisfies  the bound  
\begin{align}\label{mixedbound}
F_{\rm det}\le   \sum_{E,E'  \in  \Sp(\rho)}\sqrt{q_Eq_{E'}}  \,   \|     \rho_{E,E'}  \|_1  \, ,
\end{align}
where $   \|  A   \|_1   =    \Tr [  \sqrt{  A^\dag A} ]$ is the trace norm, $q_E = \| P_E  |\psi \>\|^2 $,   and $\Sp(\rho)$ is the \emph{energy spectrum of $\rho$},  defined as 
\begin{align}
\Sp (\rho)   =   \{  E~|~      P_E   \rho  P_E  \not  =  0  \, \}.
\end{align}       
The bound is achievable if the input state $\rho$ is \emph{block positive},~that is, if there exist orthonormal bases for the energy eigenspaces  such that, when the matrix elements are taken in those bases,  each matrix $[\rho_{E,E'}]$  is positive semidefinite.  
 In this case, every quantum channel $\map A$ satisfying the condition
\begin{align}\label{EA_mixed}
\map A  (\rho)   =    \sum_{E,E'}    \, \|\rho_{E,E'}\|_1  \,  |\psi_E\>\<\psi_{E'}|  \qquad \forall E,E'  \in \Sp(\rho)
\end{align} 
 is optimal where $|\psi_E\>$ and $|\psi_{E'}\>$ are defined by Eq. (\ref{define-PQ}).  
\end{theo} 

Note that, when the input state is pure, the bound (\ref{mixedbound}) coincides with the optimal fidelity.  
We refer to every channel  $\map A$ satisfying Eq. (\ref{EA_mixed}) as an \emph{eigenstate alignment of the mixed state $\rho$ to the pure state $|\psi\>$}.  Note that  eigenstate alignment may not be a unitary operation anymore, because it may have to  send different eigenstates with energy $E$ to the fixed eigenstate $|\psi_E\>$.    

The proof of Theorem \ref{theo:EAoptimal_mixed} is provided in Appendix \ref{app:optprob_mixed}. Three applications of the Theorem are as follows:  

\begin{eg}[Non-degenerate Hamiltonians]     When the Hamiltonian $H$ is non-degenerate, the bound (\ref{mixedbound}) becomes  
\begin{align}
F\le    \sum_{E,E'}   \,  \sqrt{ q_E  q_{E'}}  \,  |   \<  \varphi_E  |   \rho | \varphi_{E'}  \> | \, ,
\end{align}
where  $\{|\varphi_E\>\}$ is the  energy eigenbasis.  Note that, since the Hamiltonian is non-degenerate, the choice of eigenbasis is unique  up to phase transformations $|\varphi_E\>   \mapsto   |\varphi_E'\>    =   e^{i\theta_E} \,  |\varphi_E\>$.  
The bound is achievable if, for a suitable choice of phases,  one has 
\begin{align}\label{pureinphase}  
\<  \varphi_E  |\rho_E  |\varphi_{E'} \> \ge 0  \qquad \forall  E, E'  
\end{align}  
Mixed states of this form were called \emph{pure in phase}  by D'Ariano \emph{et al}  \cite{d2000isotropic}, who considered them   in the context of phase estimation.  Such states  play an important role in the area of quantum Hamiltonian complexity, where they arise as thermal states of \emph{stoquastic Hamiltonians} \cite{bravyi2008thecomplexity,bravyi2009complexity},~i.~e. Hamiltonians  with non-positive matrix elements in a given basis.   
  Physically, we can consider a scenario  where   the system starts  in the thermal state of a stoquastic Hamilonian and subsequently undergoes a rapid change of Hamiltonian to a diagonal one, making the initial thermal state a non-trivial resource.   
 \end{eg}

\begin{eg}[Pure states subject to random time evolution]
Suppose that the system, initially prepared in a pure state $|\varphi\>$,  has evolved under its  free Hamiltonian for a time $t$ which is  not perfectly known,~e.~g.~due to the finite time resolution of the clocks available in the laboratory.  
Then, the system is effectively described by the mixed state 
 \[    \rho  =  \int  \d t  \,       \pi(t)   \,    U_t   |\varphi\>\<\varphi|  U_{t}^\dag \, , \qquad U_\tau   =  e^{-i  t H_{\rm sys}/\hbar }\, ,  \]
 $\pi(t)$ being the probability distribution for $t$.   
In this case, the bound (\ref{mixedbound})  becomes  
\[  F  \le   \sum_{E,E'}   \,   \sqrt{q_E q_{E'}  p_E  p_{E'}}  \, \left|  \widetilde{\pi} (E-E')\right| \, , \]
where $p_E$ and $q_E$ are defined as in Eq. (\ref{statedecomp}) and $\widetilde {\pi}$ is the Fourier transform of $\pi$.
  If the Fourier transform is positive (i.~e.~if the noise has positive spectrum), then the bound is attainable by every unitary operation  $U$ satisfying the  eigenstate alignment condition  [Eq. (\ref{eigenalign})].
  \end{eg} 
  
  \begin{eg}[Multiple copies of qubit mixed states]\label{eg:mixed}
  Consider a system of $N$ non-interacting qubits, each having  the same Hamiltonian  $H  =   E_0  \, Z/2$.   Then, the total Hamiltonian of the system is degenerate and has the block diagonal form  
  \[  H_{\rm sys}    =  E_0  \,  \bigoplus_{l}  \,   \left( J_z^{(l)}  \otimes    I_{d^{(N)}_l}\right)  \, , \] 
  where $l$ is the quantum number of the total angular momentum,  $J^{(l)}_z$ is the $z$ component of the angular momentum operator in the subspace with quantum number $l$, and $I_{d^{(N)}_l}$ is the identity on a multiplicity space  $\spc M_l^{(N)}$, of dimension  
  $$d_l^{(N)}=\frac{4l+2}{N+2l+2}{N\choose N/2+l} \, .$$ 
   From the above decomposition it is clear that the eigenvalues of the energy are given by $E_m  = E_0  m$, where  $m$ are the eigenvalues of the $z$ component of the total angular momentum operator.  A basis for the corresponding eigenspace is given by  the vectors   
  \[|\varphi_{m,l,n}\>   =    |l,m\>  \otimes | \mu_{l,n}  \> \, ,\] 
  where  $l$ goes from $|m|$ to $N/2$,   $|l,m\>$ is the eigenstate of $J^{(l)}_z$ with eigenvalue $m$, and $\{|\mu_{l,n}\>~|~  n=  1, \dots  ,  d_l^{(N)}\}$ is a basis for  the multiplicity space.   
  Now, suppose that each qubit is initially prepared in the state 
  \[\omega  =   \frac{e^{\beta \, X}}{\Tr[e^{\beta\,  X}]}  \, ,  \qquad \beta \ge 0 \, , \quad  X  = \begin{pmatrix}   0&1\\
  1&0  \end{pmatrix} \, .    \] 
  With this choice, the state $\omega^{\otimes  N}$ satisfies the condition for the achievability of the bound (\ref{mixedbound}): indeed, one has  
  \begin{align}
  \nonumber 
   P_{E_m}    \omega^{\otimes N}   P_{E_{m'}}   & =  \bigoplus_l   \,   \frac{\<  l,m  |    e^{2\beta  J_x^{(l)}} |l,m'\> }{{\Tr[e^{\beta  X}]^{N}}}  \\   
   &  \qquad \qquad  \, \times     \left(   |l,m\>\<l,m'|  \otimes I_{d_l^{(N)}}\right)  \, ,
  \label{blockmix}
 \end{align}
and   $ \<  l,m  |    e^{2\beta  J_x^{(l)}} |l,m'\>\ge 0$ since the matrix $J_x^{(l)}$ has positive matrix elements. 
Hence, the matrix elements of the operator $P_{E_m}    \rho   P_{E_{m'}} $   in the basis $\{  |l,m\>  \otimes |\mu_{l,n}\>  ~|~   l \ge \max\{ |m|, |m'|\}  \, ,   n  = 1, \dots  d_l^{(N)} \}$ form a non-negative matrix. 
%
For the transition $\omega^{\otimes N}  \to |\psi\>\<\psi|$, eigenstate alignment is not a unitary operation, because all basis vectors with  energy $E_m$ are transformed into $|\psi_{E_m\>}$.   
\end{eg}

\subsection{The ultimate probabilistic fidelity}

We now provide the exact value of the maximum  fidelity for the transition $\rho  \to  |\psi\>\<\psi|$ when no restriction is imposed on the probability of success. 
\begin{theo}\label{prop:optprob_mixed}
For a finite-dimensional Hilbert space, the maximum fidelity over all energy-preserving operations is given by  
\begin{align}
\nonumber 
&F_{\max}      =    \left  \|   A \right\|_{\infty} \, ,  \\  
&A   =    \sum_{E,E'}  \,   \sqrt{q_E  q_{E'}}  \,   |\psi_E\>\<\psi_{E'}|  \otimes    \left( \rho_{E,E}^{T}\right)^{-\frac 12}  \rho_{E,E'}^T \left( \rho_{E',E'}^{T}\right)^{-\frac 12}   \, ,
\label{mixedFmax}
\end{align} 
where $  \|   A  \|_\infty   =  \max_{  \|  |\psi\>   \|=1}  \,  \|   A  |\psi\>  \|$ denotes the operator norm  and $\rho^T$ denotes the transpose of $\rho$.     
For a quantum operation achieving fidelity $F_{\max}$, the maximum probability of success is equal to  
\begin{align}\label{maxprobmixed}
p^{\max}_{\rm succ}  =   \max_{\sigma}  \min_E  \frac 1 {   \left \|  \left(  \rho_{E,E}^T\right)^{-1/2}    \sigma_E   \left(\rho_{E,E}^T\right)^{-1/2}  \right\|_\infty  \ }     \, ,
\end{align} 
where the maximum $\max_\sigma$  runs over all density matrices  $\sigma$ with support contained in the eigenspace of $A$ with maximum eigenvalue and $\sigma_E  :  =  \Tr_{\rm 1}  [    (P_E \otimes I)     \sigma ]$, $\Tr_1$ denoting the partial trace over the first Hilbert space.  

\end{theo}  

The proof, provided in appendix \ref{app:optprob_mixed}, includes the explicit construction of the optimal strategy and an expression for the maximum probability of success. 


\subsection{Purification of coherence at zero energy cost}
  
We now illustrate the application of our techniques to the concrete  problem of purifying a mixed state   \cite{cirac-ekert-macchiavello,andersen}. Suppose that we are given $N$ identical quantum systems, each prepared in the same  mixed state, which happens to  possess a non-zero amount of coherence in  across different energy levels. Can we collect  the coherence present in the $N$ systems and concentrate it in a single system?   And can we do it without drawing energy from the outside?  Mathematically, the  task is to implement the transition 
\[  \omega^{\otimes N}  \to    |\phi\>\<\phi|  \otimes |\beta\>\<\beta| \, ,\] 
where  $\omega$ is the initial mixed state, $|\phi\>  =   \sum_{n=1}^d  \, |n\> /\sqrt d$ is the maximally coherent state,  and $|\beta\> $ is an eigenstate of the energy, in which $N-1$ systems are meant to be left.     In the following we discuss the qubit case ($d=2$) and we choose the mixed state to be  
  \[\omega=\frac{e^{\beta  X}}{\Tr\left[e^{\beta   X}\right]} \, . \]    
  This state can be thought as the thermal state of the initial  Hamiltonian $  H_{\rm in}=  -   X$ and represents a non-trivial resource if the Hamiltonian is suddenly changed into $H  =   Z$.    
   For simplicity, we choose $N$ to be odd, so that $|\beta\>$ can be chosen to be an eigenstate with zero energy.

Let us  consider first the deterministic transitions.   The performance of the optimal energy-preserving channel is   determined  by  Theorem  \ref{theo:EAoptimal_mixed}, which leads to the 
expression 
\begin{align*}
F_{\rm det}    =    \frac 1 2  \sum_{m,m'  =\pm  \frac 12 }\,   \sum_{l  =\frac 12}^{\frac N2}     \,  \frac{  d_l^{(N)}  \, \<  l,m  |    e^{2\beta  J_x^{(l)}} |l,m'\> }{{\Tr[e^{\beta  X}]^{N}}}   \, .
\end{align*} 
Here the right-hand side follows from Eqs. (\ref{mixedbound}) and  (\ref{blockmix}),  using the fact that $\omega^{\otimes N}$ is block positive, as observed in Example \ref{eg:mixed}.    The optimal deterministic  fidelity  is plotted in Figure \ref{fig:Fmixed} for various values of $N$ and $\beta$.  
Note that, quite counterintuitively,  the deterministic fidelity \emph{decreases} with the growth of $N$. The origin of this behavior can be found in the constraint of energy preservation. Essentially, a deterministic energy preserving operation cannot do anything better than realigning  the blocks corresponding the values $m  =  \pm 1/2$  to the corresponding eigenstates.  However, when  $N$ grows, the blocks  are spread over an increasing number of values of $m$, so that the weight of the $m=  \pm  1/2$ component becomes less and less significant. As a result, the deterministic fidelity vanishes in the limit $N\to \infty$.   While the state $\omega^{\otimes N}$ contains an increasing amount of coherence, collecting this coherence from the high-energy blocks requires an exchange of energy with the surrounding environment.

For probabilistic strategies, the situation is different: the limitation due to energy-preservation can be partially  lifted and the  fidelity approaches unit as  $N$ increases.                  To evaluate the maximum fidelity, we have to compute the norm of the operator $A$ defined in Eq. (\ref{mixedFmax}).   In the case at hand, we have 
\begin{align*} A    & =  \frac 12  \,  \sum_{m,m'  =  \pm \frac 12}  \sum_{l=\frac 12}^{\frac N2}\frac{\<l,m|  e^{2\beta  J_x^{(l)}}  |l,m'\>}{\sqrt  {\<l,m|  e^{2\beta  J_x^{(l)}}  |l,m\>   \<l,m'|  e^{2\beta  J_x^{(l)}}  |l,m'\> } }   \\
  &  \qquad \qquad    \qquad     \,   \left|\frac 12, m\right\>\left\<\frac 12, m'  \right|   \otimes  |l,m\>\<l,m'|  \otimes I_{d_l^{(N)}} \, . 
  \end{align*} 
Taking the operator norm, we then obtain  \[  F_{\rm prob}    =    \max_{l  \in  \left\{ \frac 12\, , \dots \, ,  \frac N 2  \right\}  }    \frac{ 1+  a_l}2    \, ,   \]
with 
 \[    \qquad    a_l  =     \frac{\<l,\frac 12|  e^{2\beta  J_x^{(l)}}  |l,-\frac 12\>}{\sqrt  {\<l,\frac 12|  e^{2\beta  J_x^{(l)}}  |l,\frac 12\>   \<l,-\frac 12|  e^{2\beta  J_x^{(l)}}  |l,-\frac 12\> } }    \] 
 The optimal probabilistic fidelity is plotted in Figure \ref{fig:Fmixed} for different values of $N$ and $\beta$.  

\begin{figure}[h!]
\centering
\subfigure[]
{\label{fig:Fmixed}
\includegraphics[width=0.9\linewidth]{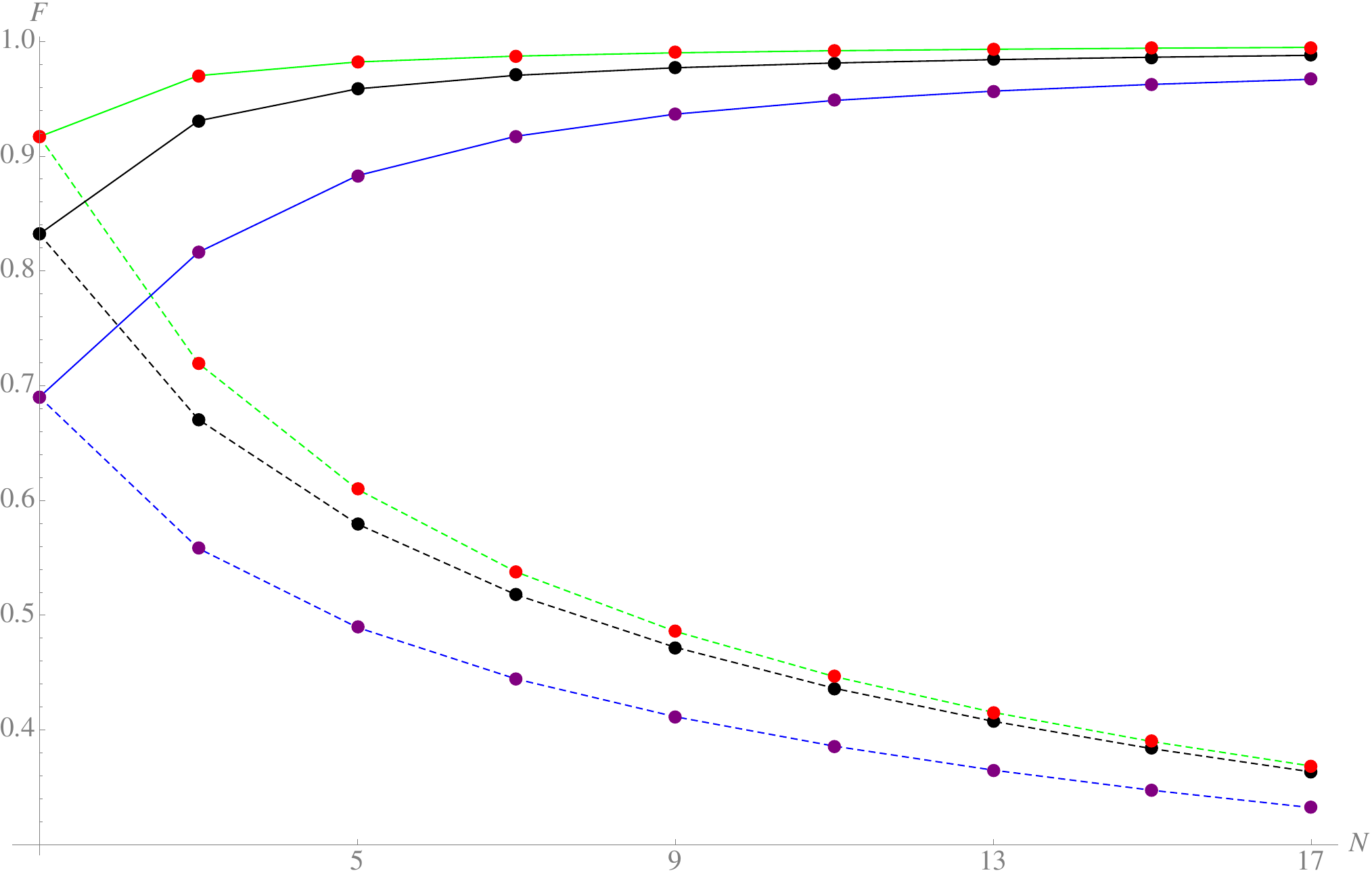}
}
\subfigure[]{\label{fig:Pmixed}
\includegraphics[width=0.9\linewidth]{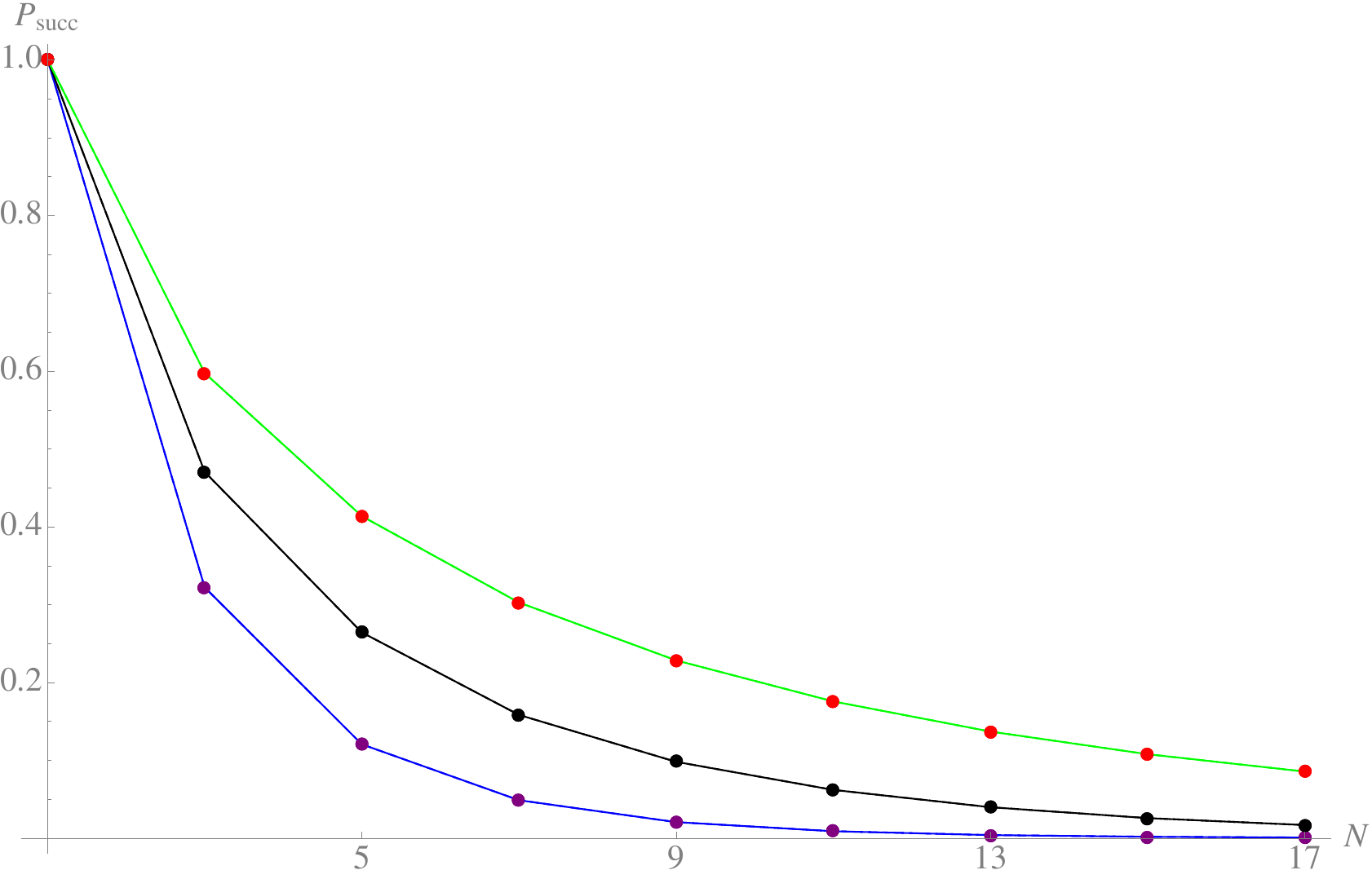}
}
\caption{{\bf Optimal    purification of coherence via energy-preserving operations.} Panel \ref{fig:Fmixed} shows the optimal  fidelities as function of the number of input copies $N$. Here different colors represent different values of $\beta$---specifically,  $\beta=0.4$ (blue line with purple dots), $\beta=0.8$ (black line with black dots) and $\beta=1.2$ (green line with red dots). The dashed lines represent the deterministic fidelities, while the solid lines represent the probabilistic fidelities. While the probabilistic fidelities increase with $N$,  the deterministic fidelities decrease, due to the restriction imposed by energy preservation. 
Panel \ref{fig:Pmixed} shows the maximum success probability for the quantum operations with maximum fidelity.   The color code is the same as in  panel  \ref{fig:Fmixed}. 
}
\label{fig:mixed}
\end{figure}
Finally, Theorem \ref{prop:optprob_mixed} allows us to evaluate the maximum probability of success for the quantum operations that achieve maximum fidelity.    According to the Theorem,  it is enough to characterize the density matrices that have support inside the eigenspace of $A$ with maximum eigenvalue. Such density matrices  have the form  $\sigma  =  |\Phi_{N/2}\>\<  \Phi_{N/2}|  \otimes \tau$, where $|\Phi_{N/2}\>$ is the maximally entangled state
\[|\Phi_{N/2}\>     =    \left(  \left|\frac12,  \frac12\right\> \left |  \frac N 2,\frac12\right\>  +   \left|  \frac 12,-\frac 12   \right\> \left |  \frac N2, -\frac 12\right\>\right)/\sqrt 2 \, .\]     
 Hence, we have   the relation 
 \begin{align*}   \sigma_m   &  =   \Tr_1\left [   \left( \left|  \frac 12,  m  \right  \>  \left\< \frac 12 , m  \right|  \otimes I   \right)  \sigma\right]   \\
 &    =       \frac 12     \,  \left|\frac N 2,m\right\>\left\<\frac N2,m\right|  \otimes \tau  
 \end{align*}
which can be inserted into Eq. (\ref{maxprobmixed}), yielding
\begin{align*}
p_{\rm succ}^{\max}    
&  =    \max_\tau   \min_{m=  \pm \frac 12} \frac {2     \left\< l,  m \right |      e^{2  \beta    J_x^{(l)}  } \,  \left| l,m   \right\> }{  \Tr [  e^{\beta   X}]^N  \,    \| \tau  \|_\infty  }  \\
&  =     \frac {2   d_l^{(N)}    \left\< l,  -\frac 12 \right |      e^{2  \beta    J_x^{(l)}  } \,  \left| l,-\frac 12  \right\> }{  \Tr [  e^{\beta   X}]^N }  \, .
\end{align*}
A plot of the probability of success as a function of $N$ and $\beta$ is shown in Figure \ref{fig:Pmixed}.

\section{Conclusions}\label{sec:conclusion}

We  introduced and analyzed the class of  quantum operations that can  be implemented with zero transfer of energy to the external environment.  Within this class,  we addressed the search for the optimal operations  implementing a desired state transition.   We considered operations that can generally be probabilistic, showing that the limitations arising from the preservation of the energy   can be lifted to a surprising extent at the price of a reduced probability of success.  
  Our investigation started from the problem of transforming a  given input  state into a desired output state.      To solve this problem, we developed two general techniques, dubbed   the  eigenstate alignment and the L\"uders reduction.   
   The eigenstate alignment provides the best deterministic way to transform pure states  at zero energy  cost.      
  The   L\"uders reduction applies more generally to the  optimization of energy-preserving quantum operations.  Essentially, it  allows one to break down every quantum operation into the product of a pure probabilistic part followed by a deterministic part.   
Employing these techniques,   we reduced  the search for the  best energy-preserving transformations of pure states  to a simple Lagrangian optimization.      

The characterization of the optimal energy-preserving transformations of pure states   allowed us to  construct  a multiple-round recursive protocol that  achieves maximum fidelity with maximum success probability in each round. 
 The probability of success of the protocol increases from one round to the next and, for a system with finite energy spectrum, the protocol terminates deterministically in a finite number of steps. 
  Our  protocol can be easily applied to every desired transformation of pure states, allowing one to find lower bounds to the optimal trade-off curve between fidelity and probability, whose exact expression is known only in a few cases.  
  As an illustration of the versatility of the protocol, we applied  it  to a number of concrete  tasks,  including quantum phase estimation, cloning of coherence, energy-preserving amplification, and ancilla-driven computation at zero energy cost.

To further improve the bounds on the fidelity-probability trade-off,  we applied the techniques of the L\"uders reduction and the eigenstate alignment  to the different histories resulting from subsequent rounds of the recursive protocol.    Specifically, we  introduced an operation of coherent coarse-graining, whereby a set of quantum operations are joined  into a pure quantum operation, with the same probability of occurrence of the original set and, typically, with a higher fidelity with the target.  
Remarkably, when  applied to the problem of  phase estimation with maximally coherent states,  coherent coarse-graining  yields points  that lie  exactly on  the optimal trade-off curve, provided that the number of  energy levels is sufficiently large.   Characterizing all the situations in which our method provides the optimal trade-off curve is an open problem.   In general,  we expect  optimality to be achieved asymptotically in  scenarios where the energy distribution of the state is sufficiently regular, including, e.g., the cases  of  
phase estimation    and quantum cloning in the asymptotic regime  \cite{ourNJP}.

In this paper we provided a comprehensive study of the optimal quantum information processing under the energy-preserving constraint.  
 A related avenue of future research is the study of optimal quantum information processing under general conservation laws. 
The techniques developed in this paper are already adapted to search for optimal quantum evolutions that preserve an \emph{algebra} of quantum observables, such as the algebra generated by the angular momentum operators. 
 Interactions that preserve the angular momentum have recently attracted attention in the implementation of quantum gates and quantum measurements \cite{ozawa,ozawaCNOT,ozawa2,spekkens-marvian-WAY, ahmadi}, although the characterization of the optimal operations is still an open problem.    In this context,  our result suggests a strategy to approach the optimization, by considering probabilistic modulation of the amplitudes of the wave-function in sectors with different angular momentum.  Also in this case, our results allow one to construct first a recursive  protocol and to increase its fidelity through the operation of coherent-coarse graining.     While such  generalizations are beyond the scope of the present paper, it is our hope that our work will pave the way to a systematic optimization of quantum operations under arbitrary conservation laws.   

\medskip

\acknowledgments{We thank  the four referees of this paper for a number of  comments that stimulated  presentation improvements and  the extension of the results to mixed states.   This work has been supported  by  the Hong Kong Research Grant Council through Grant No. 17326616, by the Foundational Questions Institute (FQXi-RFP3-1325),  by the National Natural Science Foundation of China (11675136, 11450110096, 11350110207),  by the 1000 Youth Fellowship Program of China, and by the Canadian Institute for Advanced Research (CIFAR). 
}

\bibliographystyle{apsrev}
\bibliography{recycle}

\begin{thebibliography}{106}
\expandafter\ifx\csname natexlab\endcsname\relax\def\natexlab#1{#1}\fi
\expandafter\ifx\csname bibnamefont\endcsname\relax
  \def\bibnamefont#1{#1}\fi
\expandafter\ifx\csname bibfnamefont\endcsname\relax
  \def\bibfnamefont#1{#1}\fi
\expandafter\ifx\csname citenamefont\endcsname\relax
  \def\citenamefont#1{#1}\fi
\expandafter\ifx\csname url\endcsname\relax
  \def\url#1{\texttt{#1}}\fi
\expandafter\ifx\csname urlprefix\endcsname\relax\def\urlprefix{URL }\fi
\providecommand{\bibinfo}[2]{#2}
\providecommand{\eprint}[2][]{\url{#2}}

\bibitem[{\citenamefont{Haroche}(2013)}]{nobel1}
\bibinfo{author}{\bibfnamefont{S.}~\bibnamefont{Haroche}},
  \bibinfo{journal}{Review of Modern Physics} \textbf{\bibinfo{volume}{85}},
  \bibinfo{pages}{1083} (\bibinfo{year}{2013}),
  \urlprefix\url{http://link.aps.org/doi/10.1103/RevModPhys.85.1083}.

\bibitem[{\citenamefont{Wineland}(2013)}]{nobel2}
\bibinfo{author}{\bibfnamefont{D.}~\bibnamefont{Wineland}},
  \bibinfo{journal}{Review of Modern Physics} \textbf{\bibinfo{volume}{85}},
  \bibinfo{pages}{1103} (\bibinfo{year}{2013}),
  \urlprefix\url{http://link.aps.org/doi/10.1103/RevModPhys.85.1103}.

\bibitem[{\citenamefont{Georgescu et~al.}(2014)\citenamefont{Georgescu, Ashhab,
  and Nori}}]{georgescu2014quantum}
\bibinfo{author}{\bibfnamefont{I.}~\bibnamefont{Georgescu}},
  \bibinfo{author}{\bibfnamefont{S.}~\bibnamefont{Ashhab}}, \bibnamefont{and}
  \bibinfo{author}{\bibfnamefont{F.}~\bibnamefont{Nori}},
  \bibinfo{journal}{Reviews of Modern Physics} \textbf{\bibinfo{volume}{86}},
  \bibinfo{pages}{153} (\bibinfo{year}{2014}).

\bibitem[{\citenamefont{Fuechsle et~al.}(2012)\citenamefont{Fuechsle, Miwa,
  Mahapatra, Ryu, Lee, Warschkow, Hollenberg, Klimeck, and
  Simmons}}]{fuechsle2012single}
\bibinfo{author}{\bibfnamefont{M.}~\bibnamefont{Fuechsle}},
  \bibinfo{author}{\bibfnamefont{J.}~\bibnamefont{Miwa}},
  \bibinfo{author}{\bibfnamefont{S.}~\bibnamefont{Mahapatra}},
  \bibinfo{author}{\bibfnamefont{H.}~\bibnamefont{Ryu}},
  \bibinfo{author}{\bibfnamefont{S.}~\bibnamefont{Lee}},
  \bibinfo{author}{\bibfnamefont{O.}~\bibnamefont{Warschkow}},
  \bibinfo{author}{\bibfnamefont{L.}~\bibnamefont{Hollenberg}},
  \bibinfo{author}{\bibfnamefont{G.}~\bibnamefont{Klimeck}}, \bibnamefont{and}
  \bibinfo{author}{\bibfnamefont{M.}~\bibnamefont{Simmons}},
  \bibinfo{journal}{Nature Nanotechnology} \textbf{\bibinfo{volume}{7}},
  \bibinfo{pages}{242} (\bibinfo{year}{2012}).

\bibitem[{\citenamefont{Barends et~al.}(2014)\citenamefont{Barends, Kelly,
  Megrant, Veitia, Sank, Jeffrey, White, Mutus, Fowler, Campbell
  et~al.}}]{barends2014superconducting}
\bibinfo{author}{\bibfnamefont{R.}~\bibnamefont{Barends}},
  \bibinfo{author}{\bibfnamefont{J.}~\bibnamefont{Kelly}},
  \bibinfo{author}{\bibfnamefont{A.}~\bibnamefont{Megrant}},
  \bibinfo{author}{\bibfnamefont{A.}~\bibnamefont{Veitia}},
  \bibinfo{author}{\bibfnamefont{D.}~\bibnamefont{Sank}},
  \bibinfo{author}{\bibfnamefont{E.}~\bibnamefont{Jeffrey}},
  \bibinfo{author}{\bibfnamefont{T.}~\bibnamefont{White}},
  \bibinfo{author}{\bibfnamefont{J.}~\bibnamefont{Mutus}},
  \bibinfo{author}{\bibfnamefont{A.}~\bibnamefont{Fowler}},
  \bibinfo{author}{\bibfnamefont{B.}~\bibnamefont{Campbell}},
  \bibnamefont{et~al.}, \bibinfo{journal}{Nature}
  \textbf{\bibinfo{volume}{508}}, \bibinfo{pages}{500} (\bibinfo{year}{2014}).

\bibitem[{\citenamefont{Hensen et~al.}(2015)\citenamefont{Hensen, Bernien,
  Dr{\'e}au, Reiserer, Kalb, Blok, Ruitenberg, Vermeulen, Schouten, Abell{\'a}n
  et~al.}}]{hensen2015loophole}
\bibinfo{author}{\bibfnamefont{B.}~\bibnamefont{Hensen}},
  \bibinfo{author}{\bibfnamefont{H.}~\bibnamefont{Bernien}},
  \bibinfo{author}{\bibfnamefont{A.}~\bibnamefont{Dr{\'e}au}},
  \bibinfo{author}{\bibfnamefont{A.}~\bibnamefont{Reiserer}},
  \bibinfo{author}{\bibfnamefont{N.}~\bibnamefont{Kalb}},
  \bibinfo{author}{\bibfnamefont{M.}~\bibnamefont{Blok}},
  \bibinfo{author}{\bibfnamefont{J.}~\bibnamefont{Ruitenberg}},
  \bibinfo{author}{\bibfnamefont{R.}~\bibnamefont{Vermeulen}},
  \bibinfo{author}{\bibfnamefont{R.}~\bibnamefont{Schouten}},
  \bibinfo{author}{\bibfnamefont{C.}~\bibnamefont{Abell{\'a}n}},
  \bibnamefont{et~al.}, \bibinfo{journal}{Nature}
  \textbf{\bibinfo{volume}{526}}, \bibinfo{pages}{682} (\bibinfo{year}{2015}).

\bibitem[{\citenamefont{Fettweis and Zimmermann}(2008)}]{fettweis2008ict}
\bibinfo{author}{\bibfnamefont{G.}~\bibnamefont{Fettweis}} \bibnamefont{and}
  \bibinfo{author}{\bibfnamefont{E.}~\bibnamefont{Zimmermann}}, in
  \emph{\bibinfo{booktitle}{Proceedings of the 11th International Symposium on
  Wireless Personal Multimedia Communications}}
  (\bibinfo{organization}{(Lapland}, \bibinfo{year}{2008}),
  vol.~\bibinfo{volume}{2}, p.~\bibinfo{pages}{6}.

\bibitem[{\citenamefont{Ionescu and Riel}(2011)}]{ionescu2011tunnel}
\bibinfo{author}{\bibfnamefont{A.~M.} \bibnamefont{Ionescu}} \bibnamefont{and}
  \bibinfo{author}{\bibfnamefont{H.}~\bibnamefont{Riel}},
  \bibinfo{journal}{Nature} \textbf{\bibinfo{volume}{479}},
  \bibinfo{pages}{329} (\bibinfo{year}{2011}).

\bibitem[{\citenamefont{Mukhanov}(2011)}]{mukhanov2011energy}
\bibinfo{author}{\bibfnamefont{O.~A.} \bibnamefont{Mukhanov}},
  \bibinfo{journal}{IEEE Transactions on Applied Superconductivity}
  \textbf{\bibinfo{volume}{21}}, \bibinfo{pages}{760} (\bibinfo{year}{2011}).

\bibitem[{\citenamefont{Berl et~al.}(2010)\citenamefont{Berl, Gelenbe,
  Di~Girolamo, Giuliani, De~Meer, Dang, and Pentikousis}}]{berl2010energy}
\bibinfo{author}{\bibfnamefont{A.}~\bibnamefont{Berl}},
  \bibinfo{author}{\bibfnamefont{E.}~\bibnamefont{Gelenbe}},
  \bibinfo{author}{\bibfnamefont{M.}~\bibnamefont{Di~Girolamo}},
  \bibinfo{author}{\bibfnamefont{G.}~\bibnamefont{Giuliani}},
  \bibinfo{author}{\bibfnamefont{H.}~\bibnamefont{De~Meer}},
  \bibinfo{author}{\bibfnamefont{M.~Q.} \bibnamefont{Dang}}, \bibnamefont{and}
  \bibinfo{author}{\bibfnamefont{K.}~\bibnamefont{Pentikousis}},
  \bibinfo{journal}{The computer journal} \textbf{\bibinfo{volume}{53}},
  \bibinfo{pages}{1045} (\bibinfo{year}{2010}).

\bibitem[{\citenamefont{Mittal}(2014)}]{mittal2014survey}
\bibinfo{author}{\bibfnamefont{S.}~\bibnamefont{Mittal}},
  \bibinfo{journal}{International Journal of Computer Aided Engineering and
  Technology} \textbf{\bibinfo{volume}{6}}, \bibinfo{pages}{440}
  (\bibinfo{year}{2014}).

\bibitem[{\citenamefont{Aspelmeyer et~al.}(2014)\citenamefont{Aspelmeyer,
  Kippenberg, and Marquardt}}]{aspelmeyer2014cavity}
\bibinfo{author}{\bibfnamefont{M.}~\bibnamefont{Aspelmeyer}},
  \bibinfo{author}{\bibfnamefont{T.}~\bibnamefont{Kippenberg}},
  \bibnamefont{and}
  \bibinfo{author}{\bibfnamefont{F.}~\bibnamefont{Marquardt}},
  \bibinfo{journal}{Reviews of Modern Physics} \textbf{\bibinfo{volume}{86}},
  \bibinfo{pages}{1391} (\bibinfo{year}{2014}).

\bibitem[{\citenamefont{Lodahl et~al.}(2015)\citenamefont{Lodahl, Mahmoodian,
  and Stobbe}}]{RMPsingle}
\bibinfo{author}{\bibfnamefont{P.}~\bibnamefont{Lodahl}},
  \bibinfo{author}{\bibfnamefont{S.}~\bibnamefont{Mahmoodian}},
  \bibnamefont{and} \bibinfo{author}{\bibfnamefont{S.}~\bibnamefont{Stobbe}},
  \bibinfo{journal}{Reviews of Modern Physics} \textbf{\bibinfo{volume}{87}},
  \bibinfo{pages}{347} (\bibinfo{year}{2015}).

\bibitem[{\citenamefont{Xiang et~al.}(2013)\citenamefont{Xiang, Ashhab, You,
  and Nori}}]{xiang2013hybrid}
\bibinfo{author}{\bibfnamefont{Z.}~\bibnamefont{Xiang}},
  \bibinfo{author}{\bibfnamefont{S.}~\bibnamefont{Ashhab}},
  \bibinfo{author}{\bibfnamefont{J.}~\bibnamefont{You}}, \bibnamefont{and}
  \bibinfo{author}{\bibfnamefont{F.}~\bibnamefont{Nori}},
  \bibinfo{journal}{Reviews of Modern Physics} \textbf{\bibinfo{volume}{85}},
  \bibinfo{pages}{623} (\bibinfo{year}{2013}).

\bibitem[{\citenamefont{Zhang et~al.}(2014)\citenamefont{Zhang, Aungskunsiri,
  Mart{\'\i}n-L{\'o}pez, Wabnig, Lobino, Nock, Munns, Bonneau, Jiang, Li
  et~al.}}]{chips}
\bibinfo{author}{\bibfnamefont{P.}~\bibnamefont{Zhang}},
  \bibinfo{author}{\bibfnamefont{K.}~\bibnamefont{Aungskunsiri}},
  \bibinfo{author}{\bibfnamefont{E.}~\bibnamefont{Mart{\'\i}n-L{\'o}pez}},
  \bibinfo{author}{\bibfnamefont{J.}~\bibnamefont{Wabnig}},
  \bibinfo{author}{\bibfnamefont{M.}~\bibnamefont{Lobino}},
  \bibinfo{author}{\bibfnamefont{R.}~\bibnamefont{Nock}},
  \bibinfo{author}{\bibfnamefont{J.}~\bibnamefont{Munns}},
  \bibinfo{author}{\bibfnamefont{D.}~\bibnamefont{Bonneau}},
  \bibinfo{author}{\bibfnamefont{P.}~\bibnamefont{Jiang}},
  \bibinfo{author}{\bibfnamefont{H.}~\bibnamefont{Li}}, \bibnamefont{et~al.},
  \bibinfo{journal}{Physical Review Letters} \textbf{\bibinfo{volume}{112}},
  \bibinfo{pages}{130501} (\bibinfo{year}{2014}).

\bibitem[{\citenamefont{Medford et~al.}(2013)\citenamefont{Medford, Beil,
  Taylor, Bartlett, Doherty, Rashba, DiVincenzo, Lu, Gossard, and
  Marcus}}]{sweet1}
\bibinfo{author}{\bibfnamefont{J.}~\bibnamefont{Medford}},
  \bibinfo{author}{\bibfnamefont{J.}~\bibnamefont{Beil}},
  \bibinfo{author}{\bibfnamefont{J.}~\bibnamefont{Taylor}},
  \bibinfo{author}{\bibfnamefont{S.}~\bibnamefont{Bartlett}},
  \bibinfo{author}{\bibfnamefont{A.}~\bibnamefont{Doherty}},
  \bibinfo{author}{\bibfnamefont{E.}~\bibnamefont{Rashba}},
  \bibinfo{author}{\bibfnamefont{D.}~\bibnamefont{DiVincenzo}},
  \bibinfo{author}{\bibfnamefont{H.}~\bibnamefont{Lu}},
  \bibinfo{author}{\bibfnamefont{A.}~\bibnamefont{Gossard}}, \bibnamefont{and}
  \bibinfo{author}{\bibfnamefont{C.~M.} \bibnamefont{Marcus}},
  \bibinfo{journal}{Nature nanotechnology} \textbf{\bibinfo{volume}{8}},
  \bibinfo{pages}{654} (\bibinfo{year}{2013}).

\bibitem[{\citenamefont{Fei et~al.}(2015)\citenamefont{Fei, Hung, Koh, Shim,
  Coppersmith, Hu, and Friesen}}]{sweet2}
\bibinfo{author}{\bibfnamefont{J.}~\bibnamefont{Fei}},
  \bibinfo{author}{\bibfnamefont{J.-T.} \bibnamefont{Hung}},
  \bibinfo{author}{\bibfnamefont{T.~S.} \bibnamefont{Koh}},
  \bibinfo{author}{\bibfnamefont{Y.-P.} \bibnamefont{Shim}},
  \bibinfo{author}{\bibfnamefont{S.}~\bibnamefont{Coppersmith}},
  \bibinfo{author}{\bibfnamefont{X.}~\bibnamefont{Hu}}, \bibnamefont{and}
  \bibinfo{author}{\bibfnamefont{M.}~\bibnamefont{Friesen}},
  \bibinfo{journal}{Physical Review B} \textbf{\bibinfo{volume}{91}},
  \bibinfo{pages}{205434} (\bibinfo{year}{2015}).

\bibitem[{\citenamefont{Rigetti et~al.}(2005)\citenamefont{Rigetti, Blais, and
  Devoret}}]{sweet3}
\bibinfo{author}{\bibfnamefont{C.}~\bibnamefont{Rigetti}},
  \bibinfo{author}{\bibfnamefont{A.}~\bibnamefont{Blais}}, \bibnamefont{and}
  \bibinfo{author}{\bibfnamefont{M.}~\bibnamefont{Devoret}},
  \bibinfo{journal}{Phys. Rev. Lett.} \textbf{\bibinfo{volume}{94}},
  \bibinfo{pages}{240502} (\bibinfo{year}{2005}),
  \urlprefix\url{http://link.aps.org/doi/10.1103/PhysRevLett.94.240502}.

\bibitem[{\citenamefont{Aharonov and Vaidman}(1990)}]{weakvalues}
\bibinfo{author}{\bibfnamefont{Y.}~\bibnamefont{Aharonov}} \bibnamefont{and}
  \bibinfo{author}{\bibfnamefont{L.}~\bibnamefont{Vaidman}},
  \bibinfo{journal}{Phys. Rev. A} \textbf{\bibinfo{volume}{41}},
  \bibinfo{pages}{11} (\bibinfo{year}{1990}),
  \urlprefix\url{http://link.aps.org/doi/10.1103/PhysRevA.41.11}.

\bibitem[{\citenamefont{Aaronson}(2005)}]{aaronson}
\bibinfo{author}{\bibfnamefont{S.}~\bibnamefont{Aaronson}},
  \bibinfo{journal}{Proceedings of the Royal Society A: Mathematical, Physical
  and Engineering Science} \textbf{\bibinfo{volume}{461}},
  \bibinfo{pages}{3473} (\bibinfo{year}{2005}).

\bibitem[{\citenamefont{Sun et~al.}(2001)\citenamefont{Sun, Hillery, and
  Bergou}}]{Bergou-Hillery1}
\bibinfo{author}{\bibfnamefont{Y.}~\bibnamefont{Sun}},
  \bibinfo{author}{\bibfnamefont{M.}~\bibnamefont{Hillery}}, \bibnamefont{and}
  \bibinfo{author}{\bibfnamefont{J.~A.} \bibnamefont{Bergou}},
  \bibinfo{journal}{Phys. Rev. A} \textbf{\bibinfo{volume}{64}},
  \bibinfo{pages}{022311} (\bibinfo{year}{2001}),
  \urlprefix\url{http://link.aps.org/doi/10.1103/PhysRevA.64.022311}.

\bibitem[{\citenamefont{Sun et~al.}(2002)\citenamefont{Sun, Bergou, and
  Hillery}}]{Bergou-Hillery2}
\bibinfo{author}{\bibfnamefont{Y.}~\bibnamefont{Sun}},
  \bibinfo{author}{\bibfnamefont{J.~A.} \bibnamefont{Bergou}},
  \bibnamefont{and} \bibinfo{author}{\bibfnamefont{M.}~\bibnamefont{Hillery}},
  \bibinfo{journal}{Phys. Rev. A} \textbf{\bibinfo{volume}{66}},
  \bibinfo{pages}{032315} (\bibinfo{year}{2002}),
  \urlprefix\url{http://link.aps.org/doi/10.1103/PhysRevA.66.032315}.

\bibitem[{\citenamefont{Sent\'is et~al.}(2013)\citenamefont{Sent\'is, Bagan,
  Calsamiglia, and Mu\~noz Tapia}}]{BaganDis}
\bibinfo{author}{\bibfnamefont{G.}~\bibnamefont{Sent\'is}},
  \bibinfo{author}{\bibfnamefont{E.}~\bibnamefont{Bagan}},
  \bibinfo{author}{\bibfnamefont{J.}~\bibnamefont{Calsamiglia}},
  \bibnamefont{and} \bibinfo{author}{\bibfnamefont{R.}~\bibnamefont{Mu\~noz
  Tapia}}, \bibinfo{journal}{Phys. Rev. A} \textbf{\bibinfo{volume}{88}},
  \bibinfo{pages}{052304} (\bibinfo{year}{2013}),
  \urlprefix\url{http://link.aps.org/doi/10.1103/PhysRevA.88.052304}.

\bibitem[{\citenamefont{Fiur\'a\ifmmode~\check{s}\else
  \v{s}\fi{}ek}(2006)}]{FiurasekEst}
\bibinfo{author}{\bibfnamefont{J.}~\bibnamefont{Fiur\'a\ifmmode~\check{s}\else
  \v{s}\fi{}ek}}, \bibinfo{journal}{New Journal of Physics}
  \textbf{\bibinfo{volume}{8}}, \bibinfo{pages}{192} (\bibinfo{year}{2006}).

\bibitem[{\citenamefont{Gendra et~al.}(2013{\natexlab{a}})\citenamefont{Gendra,
  Ronco-Bonvehi, Calsamiglia, Mu\~noz Tapia, and Bagan}}]{BaganPRA}
\bibinfo{author}{\bibfnamefont{B.}~\bibnamefont{Gendra}},
  \bibinfo{author}{\bibfnamefont{E.}~\bibnamefont{Ronco-Bonvehi}},
  \bibinfo{author}{\bibfnamefont{J.}~\bibnamefont{Calsamiglia}},
  \bibinfo{author}{\bibfnamefont{R.}~\bibnamefont{Mu\~noz Tapia}},
  \bibnamefont{and} \bibinfo{author}{\bibfnamefont{E.}~\bibnamefont{Bagan}},
  \bibinfo{journal}{Phys. Rev. A} \textbf{\bibinfo{volume}{88}},
  \bibinfo{pages}{012128} (\bibinfo{year}{2013}{\natexlab{a}}),
  \urlprefix\url{http://link.aps.org/doi/10.1103/PhysRevA.88.012128}.

\bibitem[{\citenamefont{Gendra et~al.}(2013{\natexlab{b}})\citenamefont{Gendra,
  Ronco-Bonvehi, Calsamiglia, Mu\~noz Tapia, and Bagan}}]{BaganPRL}
\bibinfo{author}{\bibfnamefont{B.}~\bibnamefont{Gendra}},
  \bibinfo{author}{\bibfnamefont{E.}~\bibnamefont{Ronco-Bonvehi}},
  \bibinfo{author}{\bibfnamefont{J.}~\bibnamefont{Calsamiglia}},
  \bibinfo{author}{\bibfnamefont{R.}~\bibnamefont{Mu\~noz Tapia}},
  \bibnamefont{and} \bibinfo{author}{\bibfnamefont{E.}~\bibnamefont{Bagan}},
  \bibinfo{journal}{Phys. Rev. Lett.} \textbf{\bibinfo{volume}{110}},
  \bibinfo{pages}{100501} (\bibinfo{year}{2013}{\natexlab{b}}),
  \urlprefix\url{http://link.aps.org/doi/10.1103/PhysRevLett.110.100501}.

\bibitem[{\citenamefont{Chiribella et~al.}(2013)\citenamefont{Chiribella, Yang,
  and Yao}}]{GiulioNat}
\bibinfo{author}{\bibfnamefont{G.}~\bibnamefont{Chiribella}},
  \bibinfo{author}{\bibfnamefont{Y.}~\bibnamefont{Yang}}, \bibnamefont{and}
  \bibinfo{author}{\bibfnamefont{A.~C.-C.} \bibnamefont{Yao}},
  \bibinfo{journal}{Nat. Commun.} \textbf{\bibinfo{volume}{4}}
  (\bibinfo{year}{2013}).

\bibitem[{\citenamefont{Chiribella and Yang}(2016)}]{chiribella2016quantum}
\bibinfo{author}{\bibfnamefont{G.}~\bibnamefont{Chiribella}} \bibnamefont{and}
  \bibinfo{author}{\bibfnamefont{Y.}~\bibnamefont{Yang}},
  \bibinfo{journal}{Frontiers of Physics} \textbf{\bibinfo{volume}{11}},
  \bibinfo{pages}{110304} (\bibinfo{year}{2016}).

\bibitem[{\citenamefont{Fiur\'a\ifmmode~\check{s}\else
  \v{s}\fi{}ek}(2004)}]{fiurasekclon}
\bibinfo{author}{\bibfnamefont{J.}~\bibnamefont{Fiur\'a\ifmmode~\check{s}\else
  \v{s}\fi{}ek}}, \bibinfo{journal}{Phys. Rev. A}
  \textbf{\bibinfo{volume}{70}}, \bibinfo{pages}{032308}
  (\bibinfo{year}{2004}),
  \urlprefix\url{http://link.aps.org/doi/10.1103/PhysRevA.70.032308}.

\bibitem[{\citenamefont{Ralph and Lund}(2009)}]{QCMC}
\bibinfo{author}{\bibfnamefont{T.}~\bibnamefont{Ralph}} \bibnamefont{and}
  \bibinfo{author}{\bibfnamefont{A.}~\bibnamefont{Lund}}, in
  \emph{\bibinfo{booktitle}{Quantum Communication Measurement and Computing
  Proceedings of 9th International Conference}}, edited by
  \bibinfo{editor}{\bibnamefont{A.Lvovsky}} (\bibinfo{organization}{AIP},
  \bibinfo{year}{2009}), pp. \bibinfo{pages}{155--160}.

\bibitem[{\citenamefont{Chiribella and Xie}(2013)}]{chiribella-xie}
\bibinfo{author}{\bibfnamefont{G.}~\bibnamefont{Chiribella}} \bibnamefont{and}
  \bibinfo{author}{\bibfnamefont{J.}~\bibnamefont{Xie}},
  \bibinfo{journal}{Phys. Rev. Lett.} \textbf{\bibinfo{volume}{110}},
  \bibinfo{pages}{213602} (\bibinfo{year}{2013}).

\bibitem[{\citenamefont{Zhao and Chiribella}(2017)}]{zhao2017quantum}
\bibinfo{author}{\bibfnamefont{X.}~\bibnamefont{Zhao}} \bibnamefont{and}
  \bibinfo{author}{\bibfnamefont{G.}~\bibnamefont{Chiribella}},
  \bibinfo{journal}{Phys. Rev. A} \textbf{\bibinfo{volume}{95}},
  \bibinfo{pages}{042303} (\bibinfo{year}{2017}),
  \urlprefix\url{https://link.aps.org/doi/10.1103/PhysRevA.95.042303}.

\bibitem[{\citenamefont{Levi and Mintert}(2014)}]{delocal}
\bibinfo{author}{\bibfnamefont{F.}~\bibnamefont{Levi}} \bibnamefont{and}
  \bibinfo{author}{\bibfnamefont{F.}~\bibnamefont{Mintert}},
  \bibinfo{journal}{New Journal of Physics} \textbf{\bibinfo{volume}{16}},
  \bibinfo{pages}{033007} (\bibinfo{year}{2014}),
  \urlprefix\url{http://stacks.iop.org/1367-2630/16/i=3/a=033007}.

\bibitem[{\citenamefont{Baumgratz et~al.}(2014)\citenamefont{Baumgratz, Cramer,
  and Plenio}}]{incoherence}
\bibinfo{author}{\bibfnamefont{T.}~\bibnamefont{Baumgratz}},
  \bibinfo{author}{\bibfnamefont{M.}~\bibnamefont{Cramer}}, \bibnamefont{and}
  \bibinfo{author}{\bibfnamefont{M.~B.} \bibnamefont{Plenio}},
  \bibinfo{journal}{Phys. Rev. Lett.} \textbf{\bibinfo{volume}{113}},
  \bibinfo{pages}{140401} (\bibinfo{year}{2014}),
  \urlprefix\url{http://link.aps.org/doi/10.1103/PhysRevLett.113.140401}.

\bibitem[{\citenamefont{Lostaglio et~al.}(2015)\citenamefont{Lostaglio,
  Korzekwa, Jennings, and Rudolph}}]{rudolphPRX}
\bibinfo{author}{\bibfnamefont{M.}~\bibnamefont{Lostaglio}},
  \bibinfo{author}{\bibfnamefont{K.}~\bibnamefont{Korzekwa}},
  \bibinfo{author}{\bibfnamefont{D.}~\bibnamefont{Jennings}}, \bibnamefont{and}
  \bibinfo{author}{\bibfnamefont{T.}~\bibnamefont{Rudolph}},
  \bibinfo{journal}{Phys. Rev. X} \textbf{\bibinfo{volume}{5}},
  \bibinfo{pages}{021001} (\bibinfo{year}{2015}),
  \urlprefix\url{http://link.aps.org/doi/10.1103/PhysRevX.5.021001}.

\bibitem[{\citenamefont{{\AA}berg}(2014)}]{aaberg-2014-prl}
\bibinfo{author}{\bibfnamefont{J.}~\bibnamefont{{\AA}berg}},
  \bibinfo{journal}{Physical Review Letters} \textbf{\bibinfo{volume}{113}},
  \bibinfo{pages}{150402} (\bibinfo{year}{2014}).

\bibitem[{\citenamefont{{\'C}wikli{\'n}ski
  et~al.}(2015)\citenamefont{{\'C}wikli{\'n}ski, Studzi{\'n}ski, Horodecki, and
  Oppenheim}}]{cwiklinski-2015-prl}
\bibinfo{author}{\bibfnamefont{P.}~\bibnamefont{{\'C}wikli{\'n}ski}},
  \bibinfo{author}{\bibfnamefont{M.}~\bibnamefont{Studzi{\'n}ski}},
  \bibinfo{author}{\bibfnamefont{M.}~\bibnamefont{Horodecki}},
  \bibnamefont{and}
  \bibinfo{author}{\bibfnamefont{J.}~\bibnamefont{Oppenheim}},
  \bibinfo{journal}{Physical Review Letters} \textbf{\bibinfo{volume}{115}},
  \bibinfo{pages}{210403} (\bibinfo{year}{2015}).

\bibitem[{\citenamefont{Marvian and Spekkens}(2016)}]{marvian2016quantify}
\bibinfo{author}{\bibfnamefont{I.}~\bibnamefont{Marvian}} \bibnamefont{and}
  \bibinfo{author}{\bibfnamefont{R.~W.} \bibnamefont{Spekkens}},
  \bibinfo{journal}{Physical Review A} \textbf{\bibinfo{volume}{94}},
  \bibinfo{pages}{052324} (\bibinfo{year}{2016}).

\bibitem[{\citenamefont{Cirac et~al.}(1999)\citenamefont{Cirac, Ekert, and
  Macchiavello}}]{cirac-ekert-macchiavello}
\bibinfo{author}{\bibfnamefont{J.~I.} \bibnamefont{Cirac}},
  \bibinfo{author}{\bibfnamefont{A.~K.} \bibnamefont{Ekert}}, \bibnamefont{and}
  \bibinfo{author}{\bibfnamefont{C.}~\bibnamefont{Macchiavello}},
  \bibinfo{journal}{Phys. Rev. Lett.} \textbf{\bibinfo{volume}{82}},
  \bibinfo{pages}{4344} (\bibinfo{year}{1999}),
  \urlprefix\url{http://link.aps.org/doi/10.1103/PhysRevLett.82.4344}.

\bibitem[{\citenamefont{Andersen et~al.}(2005)\citenamefont{Andersen, Filip,
  Fiur\'a\ifmmode~\check{s}\else \v{s}\fi{}ek, Josse, and Leuchs}}]{andersen}
\bibinfo{author}{\bibfnamefont{U.~L.} \bibnamefont{Andersen}},
  \bibinfo{author}{\bibfnamefont{R.}~\bibnamefont{Filip}},
  \bibinfo{author}{\bibfnamefont{J.}~\bibnamefont{Fiur\'a\ifmmode~\check{s}\else
  \v{s}\fi{}ek}}, \bibinfo{author}{\bibfnamefont{V.}~\bibnamefont{Josse}},
  \bibnamefont{and} \bibinfo{author}{\bibfnamefont{G.}~\bibnamefont{Leuchs}},
  \bibinfo{journal}{Phys. Rev. A} \textbf{\bibinfo{volume}{72}},
  \bibinfo{pages}{060301} (\bibinfo{year}{2005}),
  \urlprefix\url{http://link.aps.org/doi/10.1103/PhysRevA.72.060301}.

\bibitem[{\citenamefont{D'Ariano et~al.}(2005)\citenamefont{D'Ariano,
  Macchiavello, and Perinotti}}]{dariano-macchiavello-perinotti-prl}
\bibinfo{author}{\bibfnamefont{G.~M.} \bibnamefont{D'Ariano}},
  \bibinfo{author}{\bibfnamefont{C.}~\bibnamefont{Macchiavello}},
  \bibnamefont{and}
  \bibinfo{author}{\bibfnamefont{P.}~\bibnamefont{Perinotti}},
  \bibinfo{journal}{Phys. Rev. Lett.} \textbf{\bibinfo{volume}{95}},
  \bibinfo{pages}{060503} (\bibinfo{year}{2005}),
  \urlprefix\url{http://link.aps.org/doi/10.1103/PhysRevLett.95.060503}.

\bibitem[{\citenamefont{Cirac and Zoller}(1995)}]{ion1}
\bibinfo{author}{\bibfnamefont{J.~I.} \bibnamefont{Cirac}} \bibnamefont{and}
  \bibinfo{author}{\bibfnamefont{P.}~\bibnamefont{Zoller}},
  \bibinfo{journal}{Physical Review Letters} \textbf{\bibinfo{volume}{74}},
  \bibinfo{pages}{4091} (\bibinfo{year}{1995}).

\bibitem[{\citenamefont{H{\"a}ffner et~al.}(2008)\citenamefont{H{\"a}ffner,
  Roos, and Blatt}}]{ion2}
\bibinfo{author}{\bibfnamefont{H.}~\bibnamefont{H{\"a}ffner}},
  \bibinfo{author}{\bibfnamefont{C.~F.} \bibnamefont{Roos}}, \bibnamefont{and}
  \bibinfo{author}{\bibfnamefont{R.}~\bibnamefont{Blatt}},
  \bibinfo{journal}{Physics reports} \textbf{\bibinfo{volume}{469}},
  \bibinfo{pages}{155} (\bibinfo{year}{2008}).

\bibitem[{\citenamefont{Siegman}(2000{\natexlab{a}})}]{cavity1}
\bibinfo{author}{\bibfnamefont{A.~E.} \bibnamefont{Siegman}},
  \bibinfo{journal}{Selected Topics in Quantum Electronics, IEEE Journal of}
  \textbf{\bibinfo{volume}{6}}, \bibinfo{pages}{1380}
  (\bibinfo{year}{2000}{\natexlab{a}}), ISSN \bibinfo{issn}{1077-260X}.

\bibitem[{\citenamefont{Siegman}(2000{\natexlab{b}})}]{cavity2}
\bibinfo{author}{\bibfnamefont{A.~E.} \bibnamefont{Siegman}},
  \bibinfo{journal}{Selected Topics in Quantum Electronics, IEEE Journal of}
  \textbf{\bibinfo{volume}{6}}, \bibinfo{pages}{1389}
  (\bibinfo{year}{2000}{\natexlab{b}}), ISSN \bibinfo{issn}{1077-260X}.

\bibitem[{\citenamefont{Bloch}(2005)}]{lattice1}
\bibinfo{author}{\bibfnamefont{I.}~\bibnamefont{Bloch}}, \bibinfo{journal}{Nat
  Phys} \textbf{\bibinfo{volume}{1}}, \bibinfo{pages}{23}
  (\bibinfo{year}{2005}), \urlprefix\url{http://dx.doi.org/10.1038/nphys138}.

\bibitem[{\citenamefont{Derevianko and Katori}(2011)}]{lattice2}
\bibinfo{author}{\bibfnamefont{A.}~\bibnamefont{Derevianko}} \bibnamefont{and}
  \bibinfo{author}{\bibfnamefont{H.}~\bibnamefont{Katori}},
  \bibinfo{journal}{Rev. Mod. Phys.} \textbf{\bibinfo{volume}{83}},
  \bibinfo{pages}{331} (\bibinfo{year}{2011}),
  \urlprefix\url{http://link.aps.org/doi/10.1103/RevModPhys.83.331}.

\bibitem[{\citenamefont{Kippenberg and Vahala}(2008)}]{optomechanics1}
\bibinfo{author}{\bibfnamefont{T.~J.} \bibnamefont{Kippenberg}}
  \bibnamefont{and} \bibinfo{author}{\bibfnamefont{K.~J.}
  \bibnamefont{Vahala}}, \bibinfo{journal}{Science}
  \textbf{\bibinfo{volume}{321}}, \bibinfo{pages}{1172} (\bibinfo{year}{2008}).

\bibitem[{\citenamefont{Marquardt and Girvin}(2009)}]{optomechanics2}
\bibinfo{author}{\bibfnamefont{F.}~\bibnamefont{Marquardt}} \bibnamefont{and}
  \bibinfo{author}{\bibfnamefont{S.~M.} \bibnamefont{Girvin}},
  \bibinfo{journal}{Physics} \textbf{\bibinfo{volume}{2}}
  (\bibinfo{year}{2009}).

\bibitem[{\citenamefont{Bratteli and Robinson}(2003)}]{Bratteli_Robinson}
\bibinfo{author}{\bibfnamefont{O.}~\bibnamefont{Bratteli}} \bibnamefont{and}
  \bibinfo{author}{\bibfnamefont{D.~W.} \bibnamefont{Robinson}},
  \emph{\bibinfo{title}{Operator Algebras and Quantum Statistical Mechanics 1:
  $C*$- and $W*$-Algebras. Symmetry Groups. Decomposition of States}},
  Theoretical and Mathematical Physics (\bibinfo{publisher}{Springer},
  \bibinfo{year}{2003}), \bibinfo{edition}{2nd} ed.

\bibitem[{\citenamefont{Lindblad}(1999)}]{Lindblad}
\bibinfo{author}{\bibfnamefont{G.}~\bibnamefont{Lindblad}},
  \bibinfo{journal}{Letters in Mathemathical Physics}
  \textbf{\bibinfo{volume}{47}}, \bibinfo{pages}{189} (\bibinfo{year}{1999}).

\bibitem[{\citenamefont{Ozawa}(1984)}]{ozawa}
\bibinfo{author}{\bibfnamefont{M.}~\bibnamefont{Ozawa}},
  \bibinfo{journal}{Journal of Mathematical Physics}
  \textbf{\bibinfo{volume}{25}}, \bibinfo{pages}{79} (\bibinfo{year}{1984}).

\bibitem[{\citenamefont{Wigner}(1952)}]{W}
\bibinfo{author}{\bibfnamefont{E.~P.} \bibnamefont{Wigner}},
  \bibinfo{journal}{Zeitschrift f\"ur Physik} \textbf{\bibinfo{volume}{133}},
  \bibinfo{pages}{101} (\bibinfo{year}{1952}).

\bibitem[{\citenamefont{Araki and Yanase}(1960)}]{AY}
\bibinfo{author}{\bibfnamefont{H.}~\bibnamefont{Araki}} \bibnamefont{and}
  \bibinfo{author}{\bibfnamefont{M.~M.} \bibnamefont{Yanase}},
  \bibinfo{journal}{Phys. Rev.} \textbf{\bibinfo{volume}{120}},
  \bibinfo{pages}{622} (\bibinfo{year}{1960}),
  \urlprefix\url{http://link.aps.org/doi/10.1103/PhysRev.120.622}.

\bibitem[{\citenamefont{Ozawa}(2002{\natexlab{a}})}]{ozawa2}
\bibinfo{author}{\bibfnamefont{M.}~\bibnamefont{Ozawa}},
  \bibinfo{journal}{Phys. Rev. Lett.} \textbf{\bibinfo{volume}{88}},
  \bibinfo{pages}{050402} (\bibinfo{year}{2002}{\natexlab{a}}),
  \urlprefix\url{http://link.aps.org/doi/10.1103/PhysRevLett.88.050402}.

\bibitem[{\citenamefont{Ozawa}(2002{\natexlab{b}})}]{ozawaCNOT}
\bibinfo{author}{\bibfnamefont{M.}~\bibnamefont{Ozawa}},
  \bibinfo{journal}{Phys. Rev. Lett.} \textbf{\bibinfo{volume}{89}},
  \bibinfo{pages}{057902} (\bibinfo{year}{2002}{\natexlab{b}}),
  \urlprefix\url{http://link.aps.org/doi/10.1103/PhysRevLett.89.057902}.

\bibitem[{\citenamefont{Gea-Banacloche and Ozawa}(2006)}]{computing3}
\bibinfo{author}{\bibfnamefont{J.}~\bibnamefont{Gea-Banacloche}}
  \bibnamefont{and} \bibinfo{author}{\bibfnamefont{M.}~\bibnamefont{Ozawa}},
  \bibinfo{journal}{Phys. Rev. A} \textbf{\bibinfo{volume}{74}},
  \bibinfo{pages}{060301} (\bibinfo{year}{2006}),
  \urlprefix\url{http://link.aps.org/doi/10.1103/PhysRevA.74.060301}.

\bibitem[{\citenamefont{Marvian and Spekkens}(2012)}]{spekkens-marvian-WAY}
\bibinfo{author}{\bibfnamefont{I.}~\bibnamefont{Marvian}} \bibnamefont{and}
  \bibinfo{author}{\bibfnamefont{R.~W.} \bibnamefont{Spekkens}},
  \bibinfo{journal}{arXiv preprint arXiv:1212.3378}  (\bibinfo{year}{2012}).

\bibitem[{\citenamefont{Ahmadi et~al.}(2013)\citenamefont{Ahmadi, Jennings, and
  Rudolph}}]{ahmadi}
\bibinfo{author}{\bibfnamefont{M.}~\bibnamefont{Ahmadi}},
  \bibinfo{author}{\bibfnamefont{D.}~\bibnamefont{Jennings}}, \bibnamefont{and}
  \bibinfo{author}{\bibfnamefont{T.}~\bibnamefont{Rudolph}},
  \bibinfo{journal}{New Journal of Physics} \textbf{\bibinfo{volume}{15}},
  \bibinfo{pages}{013057} (\bibinfo{year}{2013}),
  \urlprefix\url{http://stacks.iop.org/1367-2630/15/i=1/a=013057}.

\bibitem[{\citenamefont{Navascu\'es and Popescu}(2014)}]{popescu}
\bibinfo{author}{\bibfnamefont{M.}~\bibnamefont{Navascu\'es}} \bibnamefont{and}
  \bibinfo{author}{\bibfnamefont{S.}~\bibnamefont{Popescu}},
  \bibinfo{journal}{Phys. Rev. Lett.} \textbf{\bibinfo{volume}{112}},
  \bibinfo{pages}{140502} (\bibinfo{year}{2014}),
  \urlprefix\url{http://link.aps.org/doi/10.1103/PhysRevLett.112.140502}.

\bibitem[{\citenamefont{Bush and Grabowski}(1997)}]{lathi_book}
\bibinfo{author}{\bibfnamefont{P.}~\bibnamefont{Bush}} \bibnamefont{and}
  \bibinfo{author}{\bibfnamefont{M.}~\bibnamefont{Grabowski}},
  \emph{\bibinfo{title}{Lahti, Operational Quantum Physics}}
  (\bibinfo{publisher}{Springer}, \bibinfo{year}{1997}).

\bibitem[{\citenamefont{Holevo}(2011)}]{holevo_stat_book}
\bibinfo{author}{\bibfnamefont{A.~S.} \bibnamefont{Holevo}},
  \emph{\bibinfo{title}{Probabilistic and statistical aspects of quantum
  theory}}, vol.~\bibinfo{volume}{1} (\bibinfo{publisher}{Springer},
  \bibinfo{year}{2011}).

\bibitem[{\citenamefont{Heinosaari and Ziman}(2012)}]{heinosaar_ziman}
\bibinfo{author}{\bibfnamefont{T.}~\bibnamefont{Heinosaari}} \bibnamefont{and}
  \bibinfo{author}{\bibfnamefont{M.}~\bibnamefont{Ziman}},
  \emph{\bibinfo{title}{The mathematical language of quantum theory: from
  uncertainty to entanglement}} (\bibinfo{publisher}{Cambridge University
  Press}, \bibinfo{year}{2012}).

\bibitem[{\citenamefont{Partanen et~al.}(2012)\citenamefont{Partanen,
  H\"ayrynen, Oksanen, and Tulkki}}]{preservingamp}
\bibinfo{author}{\bibfnamefont{M.}~\bibnamefont{Partanen}},
  \bibinfo{author}{\bibfnamefont{T.}~\bibnamefont{H\"ayrynen}},
  \bibinfo{author}{\bibfnamefont{J.}~\bibnamefont{Oksanen}}, \bibnamefont{and}
  \bibinfo{author}{\bibfnamefont{J.}~\bibnamefont{Tulkki}},
  \bibinfo{journal}{Phys. Rev. A} \textbf{\bibinfo{volume}{86}},
  \bibinfo{pages}{063804} (\bibinfo{year}{2012}),
  \urlprefix\url{http://link.aps.org/doi/10.1103/PhysRevA.86.063804}.

\bibitem[{\citenamefont{D'Ariano and Macchiavello}(2003)}]{qubit}
\bibinfo{author}{\bibfnamefont{G.~M.} \bibnamefont{D'Ariano}} \bibnamefont{and}
  \bibinfo{author}{\bibfnamefont{C.}~\bibnamefont{Macchiavello}},
  \bibinfo{journal}{Phys. Rev. A} \textbf{\bibinfo{volume}{67}},
  \bibinfo{pages}{042306} (\bibinfo{year}{2003}),
  \urlprefix\url{http://link.aps.org/doi/10.1103/PhysRevA.67.042306}.

\bibitem[{\citenamefont{L\"uders}(1950)}]{luders}
\bibinfo{author}{\bibfnamefont{G.}~\bibnamefont{L\"uders}},
  \bibinfo{journal}{Annalen der Physik} \textbf{\bibinfo{volume}{443}},
  \bibinfo{pages}{322} (\bibinfo{year}{1950}).

\bibitem[{\citenamefont{Busch et~al.}(1991)\citenamefont{Busch, Lahti, and
  Mittelstaedt}}]{busch}
\bibinfo{author}{\bibfnamefont{P.}~\bibnamefont{Busch}},
  \bibinfo{author}{\bibfnamefont{P.~J.} \bibnamefont{Lahti}}, \bibnamefont{and}
  \bibinfo{author}{\bibfnamefont{P.}~\bibnamefont{Mittelstaedt}},
  \emph{\bibinfo{title}{The quantum theory of measurement}}
  (\bibinfo{publisher}{Springer}, \bibinfo{year}{1991}).

\bibitem[{\citenamefont{Calsamiglia}(2014)}]{calsamiglia-2014-natphys}
\bibinfo{author}{\bibfnamefont{J.}~\bibnamefont{Calsamiglia}},
  \bibinfo{journal}{Nature Physics} \textbf{\bibinfo{volume}{10}},
  \bibinfo{pages}{91} (\bibinfo{year}{2014}).

\bibitem[{\citenamefont{Ying and Feng}(2010)}]{loop}
\bibinfo{author}{\bibfnamefont{M.}~\bibnamefont{Ying}} \bibnamefont{and}
  \bibinfo{author}{\bibfnamefont{Y.}~\bibnamefont{Feng}},
  \bibinfo{journal}{Acta Informatica} \textbf{\bibinfo{volume}{47}},
  \bibinfo{pages}{221} (\bibinfo{year}{2010}).

\bibitem[{\citenamefont{DiVincenzo}(1995)}]{divincenzo}
\bibinfo{author}{\bibfnamefont{D.~P.} \bibnamefont{DiVincenzo}},
  \bibinfo{journal}{Science} \textbf{\bibinfo{volume}{270}},
  \bibinfo{pages}{255} (\bibinfo{year}{1995}).

\bibitem[{\citenamefont{Paetznick and Svore}(2014)}]{RUS1}
\bibinfo{author}{\bibfnamefont{A.}~\bibnamefont{Paetznick}} \bibnamefont{and}
  \bibinfo{author}{\bibfnamefont{K.}~\bibnamefont{Svore}},
  \bibinfo{journal}{Quantum Info. Comput.} \textbf{\bibinfo{volume}{14}},
  \bibinfo{pages}{1277} (\bibinfo{year}{2014}).

\bibitem[{\citenamefont{Bocharov et~al.}(2015)\citenamefont{Bocharov,
  Roetteler, and Svore}}]{RUS2}
\bibinfo{author}{\bibfnamefont{A.}~\bibnamefont{Bocharov}},
  \bibinfo{author}{\bibfnamefont{M.}~\bibnamefont{Roetteler}},
  \bibnamefont{and} \bibinfo{author}{\bibfnamefont{K.~M.} \bibnamefont{Svore}},
  \bibinfo{journal}{Physical Review Letters} \textbf{\bibinfo{volume}{114}},
  \bibinfo{pages}{080502} (\bibinfo{year}{2015}).

\bibitem[{\citenamefont{Wiebe and Roetteler}(2014)}]{RUS3}
\bibinfo{author}{\bibfnamefont{N.}~\bibnamefont{Wiebe}} \bibnamefont{and}
  \bibinfo{author}{\bibfnamefont{M.}~\bibnamefont{Roetteler}},
  \bibinfo{journal}{arXiv preprint arXiv: 1406.2040}  (\bibinfo{year}{2014}).

\bibitem[{\citenamefont{Anders et~al.}(2010)\citenamefont{Anders, Oi, Kashefi,
  Browne, and Andersson}}]{ADQC}
\bibinfo{author}{\bibfnamefont{J.}~\bibnamefont{Anders}},
  \bibinfo{author}{\bibfnamefont{D.~K.~L.} \bibnamefont{Oi}},
  \bibinfo{author}{\bibfnamefont{E.}~\bibnamefont{Kashefi}},
  \bibinfo{author}{\bibfnamefont{D.~E.} \bibnamefont{Browne}},
  \bibnamefont{and}
  \bibinfo{author}{\bibfnamefont{E.}~\bibnamefont{Andersson}},
  \bibinfo{journal}{Phys. Rev. A} \textbf{\bibinfo{volume}{82}},
  \bibinfo{pages}{020301} (\bibinfo{year}{2010}),
  \urlprefix\url{http://link.aps.org/doi/10.1103/PhysRevA.82.020301}.

\bibitem[{\citenamefont{Ozols et~al.}(2013)\citenamefont{Ozols, Roetteler, and
  Roland}}]{rejectionsampling}
\bibinfo{author}{\bibfnamefont{M.}~\bibnamefont{Ozols}},
  \bibinfo{author}{\bibfnamefont{M.}~\bibnamefont{Roetteler}},
  \bibnamefont{and} \bibinfo{author}{\bibfnamefont{J.}~\bibnamefont{Roland}},
  \bibinfo{journal}{ACM Transactions on Computation Theory (TOCT)}
  \textbf{\bibinfo{volume}{5}}, \bibinfo{pages}{11} (\bibinfo{year}{2013}).

\bibitem[{\citenamefont{Nielsen and Chuang}(1997)}]{nielsen1997programmable}
\bibinfo{author}{\bibfnamefont{M.~A.} \bibnamefont{Nielsen}} \bibnamefont{and}
  \bibinfo{author}{\bibfnamefont{I.~L.} \bibnamefont{Chuang}},
  \bibinfo{journal}{Phys. Rev. Lett.} \textbf{\bibinfo{volume}{79}},
  \bibinfo{pages}{321} (\bibinfo{year}{1997}).

\bibitem[{\citenamefont{Sorkin}(1994)}]{sorkin1994quantum}
\bibinfo{author}{\bibfnamefont{R.~D.} \bibnamefont{Sorkin}},
  \bibinfo{journal}{Modern Physics Letters A} \textbf{\bibinfo{volume}{9}},
  \bibinfo{pages}{3119} (\bibinfo{year}{1994}).

\bibitem[{\citenamefont{Helstrom}(1976)}]{helstrom}
\bibinfo{author}{\bibfnamefont{C.~W.} \bibnamefont{Helstrom}},
  \emph{\bibinfo{title}{Quantum detection and estimation theory}}
  (\bibinfo{publisher}{Academic press}, \bibinfo{year}{1976}).

\bibitem[{\citenamefont{Bu\ifmmode~\check{z}\else \v{z}\fi{}ek
  et~al.}(1999)\citenamefont{Bu\ifmmode~\check{z}\else \v{z}\fi{}ek, Derka, and
  Massar}}]{sin}
\bibinfo{author}{\bibfnamefont{V.}~\bibnamefont{Bu\ifmmode~\check{z}\else
  \v{z}\fi{}ek}}, \bibinfo{author}{\bibfnamefont{R.}~\bibnamefont{Derka}},
  \bibnamefont{and} \bibinfo{author}{\bibfnamefont{S.}~\bibnamefont{Massar}},
  \bibinfo{journal}{Phys. Rev. Lett.} \textbf{\bibinfo{volume}{82}},
  \bibinfo{pages}{2207} (\bibinfo{year}{1999}),
  \urlprefix\url{http://link.aps.org/doi/10.1103/PhysRevLett.82.2207}.

\bibitem[{\citenamefont{Berry and Wiseman}(2000)}]{berry-wiseman-2000-prl}
\bibinfo{author}{\bibfnamefont{D.}~\bibnamefont{Berry}} \bibnamefont{and}
  \bibinfo{author}{\bibfnamefont{H.}~\bibnamefont{Wiseman}},
  \bibinfo{journal}{Physical Review Letters} \textbf{\bibinfo{volume}{85}},
  \bibinfo{pages}{5098} (\bibinfo{year}{2000}).

\bibitem[{\citenamefont{Scarani et~al.}(2005)\citenamefont{Scarani, Iblisdir,
  Gisin, and Ac\'in}}]{rmp}
\bibinfo{author}{\bibfnamefont{V.}~\bibnamefont{Scarani}},
  \bibinfo{author}{\bibfnamefont{S.}~\bibnamefont{Iblisdir}},
  \bibinfo{author}{\bibfnamefont{N.}~\bibnamefont{Gisin}}, \bibnamefont{and}
  \bibinfo{author}{\bibfnamefont{A.}~\bibnamefont{Ac\'in}},
  \bibinfo{journal}{Rev. Mod. Phys.} \textbf{\bibinfo{volume}{77}},
  \bibinfo{pages}{1225} (\bibinfo{year}{2005}),
  \urlprefix\url{http://link.aps.org/doi/10.1103/RevModPhys.77.1225}.

\bibitem[{\citenamefont{Cerf and Fiurasek}(2006)}]{cerfreview}
\bibinfo{author}{\bibfnamefont{N.~J.} \bibnamefont{Cerf}} \bibnamefont{and}
  \bibinfo{author}{\bibfnamefont{J.}~\bibnamefont{Fiurasek}},
  \bibinfo{journal}{Progress in Optics} \textbf{\bibinfo{volume}{49}},
  \bibinfo{pages}{455} (\bibinfo{year}{2006}).

\bibitem[{\citenamefont{Caves}(1982)}]{insensitive}
\bibinfo{author}{\bibfnamefont{C.~M.} \bibnamefont{Caves}},
  \bibinfo{journal}{Phys. Rev. D} \textbf{\bibinfo{volume}{26}},
  \bibinfo{pages}{1817} (\bibinfo{year}{1982}),
  \urlprefix\url{http://link.aps.org/doi/10.1103/PhysRevD.26.1817}.

\bibitem[{\citenamefont{Pandey et~al.}(2013)\citenamefont{Pandey, Jiang,
  Combes, and Caves}}]{caves}
\bibinfo{author}{\bibfnamefont{S.}~\bibnamefont{Pandey}},
  \bibinfo{author}{\bibfnamefont{Z.}~\bibnamefont{Jiang}},
  \bibinfo{author}{\bibfnamefont{J.}~\bibnamefont{Combes}}, \bibnamefont{and}
  \bibinfo{author}{\bibfnamefont{C.~M.} \bibnamefont{Caves}},
  \bibinfo{journal}{Phys. Rev. A} \textbf{\bibinfo{volume}{88}},
  \bibinfo{pages}{033852} (\bibinfo{year}{2013}),
  \urlprefix\url{http://link.aps.org/doi/10.1103/PhysRevA.88.033852}.

\bibitem[{\citenamefont{Namiki}(2011)}]{namikirapid}
\bibinfo{author}{\bibfnamefont{R.}~\bibnamefont{Namiki}},
  \bibinfo{journal}{Phys. Rev. A} \textbf{\bibinfo{volume}{83}},
  \bibinfo{pages}{042323} (\bibinfo{year}{2011}),
  \urlprefix\url{http://link.aps.org/doi/10.1103/PhysRevA.83.042323}.

\bibitem[{\citenamefont{Gregoratti and Werner}(2003)}]{LostFound}
\bibinfo{author}{\bibfnamefont{M.}~\bibnamefont{Gregoratti}} \bibnamefont{and}
  \bibinfo{author}{\bibfnamefont{R.~F.} \bibnamefont{Werner}},
  \bibinfo{journal}{Journal of Modern Optics} \textbf{\bibinfo{volume}{50}},
  \bibinfo{pages}{915} (\bibinfo{year}{2003}),
  \urlprefix\url{http://www.tandfonline.com/doi/abs/10.1080/09500340308234541}.

\bibitem[{\citenamefont{Oi}(2014)}]{unlearn}
\bibinfo{author}{\bibfnamefont{D.}~\bibnamefont{Oi}}, \bibinfo{journal}{The
  European Physical Journal D} \textbf{\bibinfo{volume}{68}},
  \bibinfo{eid}{259} (\bibinfo{year}{2014}), ISSN \bibinfo{issn}{1434-6060},
  \urlprefix\url{http://dx.doi.org/10.1140/epjd/e2014-50429-3}.

\bibitem[{\citenamefont{Owari et~al.}(2008)\citenamefont{Owari, Plenio, Polzik,
  Serafini, and Wolf}}]{polzik-NJP}
\bibinfo{author}{\bibfnamefont{M.}~\bibnamefont{Owari}},
  \bibinfo{author}{\bibfnamefont{M.~B.} \bibnamefont{Plenio}},
  \bibinfo{author}{\bibfnamefont{E.~S.} \bibnamefont{Polzik}},
  \bibinfo{author}{\bibfnamefont{A.}~\bibnamefont{Serafini}}, \bibnamefont{and}
  \bibinfo{author}{\bibfnamefont{M.~M.} \bibnamefont{Wolf}},
  \bibinfo{journal}{New Journal of Physics} \textbf{\bibinfo{volume}{10}},
  \bibinfo{pages}{113014} (\bibinfo{year}{2008}),
  \urlprefix\url{http://stacks.iop.org/1367-2630/10/i=11/a=113014}.

\bibitem[{\citenamefont{Bravy et~al.}(2008)\citenamefont{Bravy, DiVincenzo,
  Oliveira, and Terhal}}]{bravyi2008thecomplexity}
\bibinfo{author}{\bibfnamefont{S.}~\bibnamefont{Bravy}},
  \bibinfo{author}{\bibfnamefont{D.~P.} \bibnamefont{DiVincenzo}},
  \bibinfo{author}{\bibfnamefont{R.}~\bibnamefont{Oliveira}}, \bibnamefont{and}
  \bibinfo{author}{\bibfnamefont{B.~M.} \bibnamefont{Terhal}},
  \bibinfo{journal}{Quantum Information and Computation}
  \textbf{\bibinfo{volume}{8}}, \bibinfo{pages}{0361} (\bibinfo{year}{2008}).

\bibitem[{\citenamefont{Bravyi and Terhal}(2009)}]{bravyi2009complexity}
\bibinfo{author}{\bibfnamefont{S.}~\bibnamefont{Bravyi}} \bibnamefont{and}
  \bibinfo{author}{\bibfnamefont{B.}~\bibnamefont{Terhal}},
  \bibinfo{journal}{SIAM Journal on Computing} \textbf{\bibinfo{volume}{39}},
  \bibinfo{pages}{1462} (\bibinfo{year}{2009}).

\bibitem[{\citenamefont{Burkard et~al.}(2004)\citenamefont{Burkard, Koch, and
  DiVincenzo}}]{burkard2004multilevel}
\bibinfo{author}{\bibfnamefont{G.}~\bibnamefont{Burkard}},
  \bibinfo{author}{\bibfnamefont{R.~H.} \bibnamefont{Koch}}, \bibnamefont{and}
  \bibinfo{author}{\bibfnamefont{D.~P.} \bibnamefont{DiVincenzo}},
  \bibinfo{journal}{Physical Review B} \textbf{\bibinfo{volume}{69}},
  \bibinfo{pages}{064503} (\bibinfo{year}{2004}).

\bibitem[{\citenamefont{Ceperley}(1995)}]{ceperley1995path}
\bibinfo{author}{\bibfnamefont{D.~M.} \bibnamefont{Ceperley}},
  \bibinfo{journal}{Reviews of Modern Physics} \textbf{\bibinfo{volume}{67}},
  \bibinfo{pages}{279} (\bibinfo{year}{1995}).

\bibitem[{\citenamefont{D'Ariano et~al.}(2000)\citenamefont{D'Ariano,
  Macchiavello, Perinotti, and Sacchi}}]{d2000isotropic}
\bibinfo{author}{\bibfnamefont{G.~M.} \bibnamefont{D'Ariano}},
  \bibinfo{author}{\bibfnamefont{C.}~\bibnamefont{Macchiavello}},
  \bibinfo{author}{\bibfnamefont{P.}~\bibnamefont{Perinotti}},
  \bibnamefont{and} \bibinfo{author}{\bibfnamefont{M.~F.}
  \bibnamefont{Sacchi}}, \bibinfo{journal}{Physics Letters A}
  \textbf{\bibinfo{volume}{268}}, \bibinfo{pages}{241} (\bibinfo{year}{2000}).

\bibitem[{\citenamefont{Chiribella and Yang}(2014)}]{ourNJP}
\bibinfo{author}{\bibfnamefont{G.}~\bibnamefont{Chiribella}} \bibnamefont{and}
  \bibinfo{author}{\bibfnamefont{Y.}~\bibnamefont{Yang}}, \bibinfo{journal}{New
  Journal of Physics} \textbf{\bibinfo{volume}{16}}, \bibinfo{pages}{063005}
  (\bibinfo{year}{2014}),
  \urlprefix\url{http://stacks.iop.org/1367-2630/16/i=6/a=063005}.

\bibitem[{\citenamefont{Gour and Spekkens}(2008)}]{reference}
\bibinfo{author}{\bibfnamefont{G.}~\bibnamefont{Gour}} \bibnamefont{and}
  \bibinfo{author}{\bibfnamefont{R.~W.} \bibnamefont{Spekkens}},
  \bibinfo{journal}{New Journal of Physics} \textbf{\bibinfo{volume}{10}},
  \bibinfo{pages}{033023} (\bibinfo{year}{2008}),
  \urlprefix\url{http://stacks.iop.org/1367-2630/10/i=3/a=033023}.

\bibitem[{\citenamefont{Marvian and Spekkens}(2013)}]{marvian-spekkens1}
\bibinfo{author}{\bibfnamefont{I.}~\bibnamefont{Marvian}} \bibnamefont{and}
  \bibinfo{author}{\bibfnamefont{R.~W.} \bibnamefont{Spekkens}},
  \bibinfo{journal}{New Journal of Physics} \textbf{\bibinfo{volume}{15}},
  \bibinfo{pages}{033001} (\bibinfo{year}{2013}),
  \urlprefix\url{http://stacks.iop.org/1367-2630/15/i=3/a=033001}.

\bibitem[{\citenamefont{Marvian and Spekkens}(2014)}]{marvian-spekkens2}
\bibinfo{author}{\bibfnamefont{I.}~\bibnamefont{Marvian}} \bibnamefont{and}
  \bibinfo{author}{\bibfnamefont{R.~W.} \bibnamefont{Spekkens}},
  \bibinfo{journal}{Phys. Rev. A} \textbf{\bibinfo{volume}{90}},
  \bibinfo{pages}{062110} (\bibinfo{year}{2014}),
  \urlprefix\url{http://link.aps.org/doi/10.1103/PhysRevA.90.062110}.

\bibitem[{\citenamefont{Chiribella et~al.}(2009)\citenamefont{Chiribella,
  D'Ariano, and Perinotti}}]{chiribella-dariano-perinotti-jmp-2009}
\bibinfo{author}{\bibfnamefont{G.}~\bibnamefont{Chiribella}},
  \bibinfo{author}{\bibfnamefont{G.~M.} \bibnamefont{D'Ariano}},
  \bibnamefont{and}
  \bibinfo{author}{\bibfnamefont{P.}~\bibnamefont{Perinotti}},
  \bibinfo{journal}{Journal of Mathematical Physics}
  \textbf{\bibinfo{volume}{50}}, \bibinfo{eid}{042101} (\bibinfo{year}{2009}),
  \urlprefix\url{http://scitation.aip.org/content/aip/journal/jmp/50/4/10.1063/1.3105923}.

\bibitem[{\citenamefont{Narasimhachar and
  Gour}(2015)}]{narasimhachar-2015-natcomm}
\bibinfo{author}{\bibfnamefont{V.}~\bibnamefont{Narasimhachar}}
  \bibnamefont{and} \bibinfo{author}{\bibfnamefont{G.}~\bibnamefont{Gour}},
  \bibinfo{journal}{Nat Commun} \textbf{\bibinfo{volume}{6}}
  (\bibinfo{year}{2015}), \bibinfo{note}{article},
  \urlprefix\url{http://dx.doi.org/10.1038/ncomms8689}.

\bibitem[{\citenamefont{King et~al.}(2007)\citenamefont{King, Matsumoto,
  Nathanson, and Ruskai}}]{conjugate}
\bibinfo{author}{\bibfnamefont{C.}~\bibnamefont{King}},
  \bibinfo{author}{\bibfnamefont{K.}~\bibnamefont{Matsumoto}},
  \bibinfo{author}{\bibfnamefont{M.}~\bibnamefont{Nathanson}},
  \bibnamefont{and} \bibinfo{author}{\bibfnamefont{M.~B.}
  \bibnamefont{Ruskai}}, \bibinfo{journal}{Markov Process and Related Fields}
  \textbf{\bibinfo{volume}{13}}, \bibinfo{pages}{391} (\bibinfo{year}{2007}).

\bibitem[{\citenamefont{Ziman and Bu\ifmmode~\check{z}\else
  \v{z}\fi{}ek}(2005)}]{decoherence}
\bibinfo{author}{\bibfnamefont{M.}~\bibnamefont{Ziman}} \bibnamefont{and}
  \bibinfo{author}{\bibfnamefont{V.}~\bibnamefont{Bu\ifmmode~\check{z}\else
  \v{z}\fi{}ek}}, \bibinfo{journal}{Phys. Rev. A}
  \textbf{\bibinfo{volume}{72}}, \bibinfo{pages}{022110}
  (\bibinfo{year}{2005}),
  \urlprefix\url{http://link.aps.org/doi/10.1103/PhysRevA.72.022110}.

\bibitem[{\citenamefont{Buscemi et~al.}(2005)\citenamefont{Buscemi, Chiribella,
  and Mauro~D'Ariano}}]{buscemi-chiribella-prl}
\bibinfo{author}{\bibfnamefont{F.}~\bibnamefont{Buscemi}},
  \bibinfo{author}{\bibfnamefont{G.}~\bibnamefont{Chiribella}},
  \bibnamefont{and}
  \bibinfo{author}{\bibfnamefont{G.}~\bibnamefont{Mauro~D'Ariano}},
  \bibinfo{journal}{Phys. Rev. Lett.} \textbf{\bibinfo{volume}{95}},
  \bibinfo{pages}{090501} (\bibinfo{year}{2005}),
  \urlprefix\url{http://link.aps.org/doi/10.1103/PhysRevLett.95.090501}.

\bibitem[{\citenamefont{Buscemi et~al.}(2007)\citenamefont{Buscemi, Chiribella,
  and D'Ariano}}]{decoherence2}
\bibinfo{author}{\bibfnamefont{F.}~\bibnamefont{Buscemi}},
  \bibinfo{author}{\bibfnamefont{G.}~\bibnamefont{Chiribella}},
  \bibnamefont{and} \bibinfo{author}{\bibfnamefont{G.}~\bibnamefont{D'Ariano}},
  \bibinfo{journal}{Open Systems and Information Dynamics}
  \textbf{\bibinfo{volume}{14}}, \bibinfo{pages}{53} (\bibinfo{year}{2007}),
  ISSN \bibinfo{issn}{1230-1612},
  \urlprefix\url{http://dx.doi.org/10.1007/s11080-007-9028-4}.

\bibitem[{\citenamefont{Br\'adler et~al.}(2010)\citenamefont{Br\'adler, Hayden,
  Touchette, and Wilde}}]{hadamard}
\bibinfo{author}{\bibfnamefont{K.}~\bibnamefont{Br\'adler}},
  \bibinfo{author}{\bibfnamefont{P.}~\bibnamefont{Hayden}},
  \bibinfo{author}{\bibfnamefont{D.}~\bibnamefont{Touchette}},
  \bibnamefont{and} \bibinfo{author}{\bibfnamefont{M.~M.} \bibnamefont{Wilde}},
  \bibinfo{journal}{Phys. Rev. A} \textbf{\bibinfo{volume}{81}},
  \bibinfo{pages}{062312} (\bibinfo{year}{2010}),
  \urlprefix\url{http://link.aps.org/doi/10.1103/PhysRevA.81.062312}.

\bibitem[{\citenamefont{Wilde et~al.}(2014)\citenamefont{Wilde, Winter, and
  Yang}}]{hadamard2}
\bibinfo{author}{\bibfnamefont{M.~M.} \bibnamefont{Wilde}},
  \bibinfo{author}{\bibfnamefont{A.}~\bibnamefont{Winter}}, \bibnamefont{and}
  \bibinfo{author}{\bibfnamefont{D.}~\bibnamefont{Yang}},
  \bibinfo{journal}{Communications in Mathematical Physics}
  \textbf{\bibinfo{volume}{331}}, \bibinfo{pages}{593} (\bibinfo{year}{2014}).

\bibitem[{\citenamefont{Bartlett et~al.}(2006)\citenamefont{Bartlett, Rudolph,
  Spekkens, and Turner}}]{bartlett2006degradation}
\bibinfo{author}{\bibfnamefont{S.~D.} \bibnamefont{Bartlett}},
  \bibinfo{author}{\bibfnamefont{T.}~\bibnamefont{Rudolph}},
  \bibinfo{author}{\bibfnamefont{R.~W.} \bibnamefont{Spekkens}},
  \bibnamefont{and} \bibinfo{author}{\bibfnamefont{P.~S.}
  \bibnamefont{Turner}}, \bibinfo{journal}{New Journal of Physics}
  \textbf{\bibinfo{volume}{8}}, \bibinfo{pages}{58} (\bibinfo{year}{2006}).

\end{thebibliography}

\appendix

\section{Relation to covariant channels, Hadamard channels,  incoherent channels, and decoherence maps}\label{app:relations}

Here we highlight the relations between  energy-preserving channels and other important classes of channels considered in the literature. 
We start from the case of  channels that are covariant under time evolution  \cite{holevo_stat_book}, i.~e.,  channels that satisfy  the relation  
\begin{align*}
\map M(    U_t  \cdot    U_t^\dag  )  =   U_t  \,  \map M(\cdot)  \,  U_t^\dag \qquad \forall t\in  \R  \, .
\end{align*}    
 Covariant channels  are at the basis of the resource theory of asymmetry \cite{reference,marvian-spekkens1,marvian-spekkens2}. 
 Every covariant channel  can be realized via a  unitary evolution---as in Eq. (\ref{stine})---with the property that the initial state of the environment $|\phi_0\>$ is an eigenstate of the energy and the joint unitary evolution preserves the sum of the energies of the system and the environment:
\begin{align}\label{sum}
U^\dag \, (H_{\rm sys}  +  H_{\rm env})  \,  U  =      H_{\rm sys}  +  H_{\rm env}  
\end{align}
(see e.~g.~\cite{chiribella-dariano-perinotti-jmp-2009}).
Note that the interaction preserves the sum of the energies of the system and the environment, but may involve  an exchange of energy between them. This is the reason why not all covariant channels preserve the energy of the system individually.

A related subclass of covariant channels  is the class of  \emph{cooling maps} considered in Ref. \cite{narasimhachar-2015-natcomm} which arise from the exchange of energy between the system and an environment, originally initialized in the Gibbs state at near-zero temperature.   In this case, the exchange of energy can go only in the direction from the system to the environment, as the environment can acquire energy from the system but not vice-versa.


Energy-preserving channels are also closely related to Hadamard channels \cite{conjugate}, also known as decoherence maps \cite{decoherence,buscemi-chiribella-prl,decoherence2}, and to the class of incoherent channels \cite{delocal,incoherence}.    The relation can be seen in the case of   a Hamiltonian  $H_{\rm sys}$ with non-degenerate spectrum. In this case,  the energy-preserving channels have been characterized in Ref. \cite{buscemi-chiribella-prl}.      Denoting by $\{  \,  |E_n\>  \,\}_{n=1}^d$  the energy eigenbasis, it turns out that a channel is energy-preserving if and only if it is of the form
\begin{align}\label{hadamard}
\map M(\rho)   =   \sum_{m,n}    \,     C_{mn} \,      \<E_m|   \rho  |E_n\>  \,    |E_m\>\<  E_n|   \, ,   
\end{align}
where $C  =  [C_{mn}]$ is   a positive matrix with diagonal elements equal to 1.    Channels of this  are also known as \emph{Hadamard channels}   in the literature on quantum Shannon theory, where they represent one of the important classes of channels with  tractable capacity regions  \cite{hadamard,hadamard2}. Energy-preserving channels with non-degenerate Hamiltonian have a number of properties.      First, note that every energy eigenstate is a fixed point of the channel, and so is every  mixture of energy eigenstates.  Hence, in the case of non-degenerate Hamiltonian, the energy-preserving channels are a special case of \emph{incoherent channels}, i.~e., channels that transform incoherent mixtures into incoherent mixtures.   
Viewing coherence in the energy eigenbasis as a resource, it is clear that energy-preserving channels cannot be used to generate resourceful states from non-resourceful ones.  On the contrary,  typically they reduce quantum  coherence, by damping  down the off-diagonal elements of the density matrix \cite{buscemi-chiribella-prl}.
\begin{figure}
\centering
\includegraphics[width=0.9\linewidth]{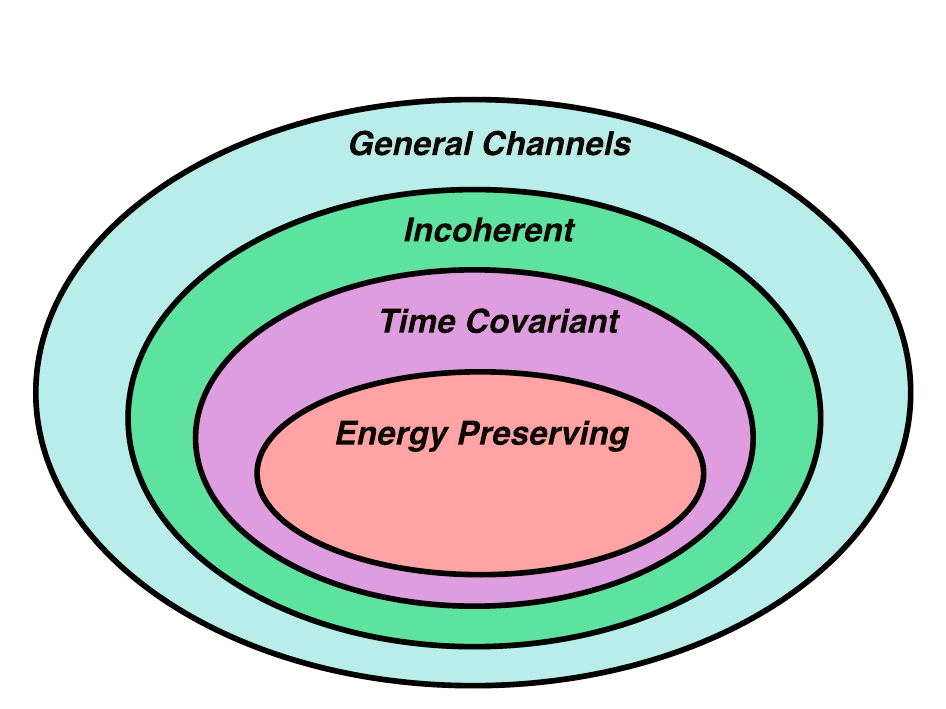}\\
\caption{{\bf Hierarchy of quantum channels.} Energy-preserving channels are contained in the class of time-covariant channels, corresponding to the special case of channels that arise from an interaction that leaves the environment inside a fixed eigenspace at every time.     For Hamiltonians with non-degenerate spectrum, covariant channels are a special case of incoherent channels,~i.~e.~channels that do not generate coherence across the eigenbasis of the energy.           }
\label{fig:hierarchy}
\end{figure}
The inclusion relations among energy-preserving, covariant and incoherent operations are illustrated in Fig. \ref{fig:hierarchy}.

\section{Proof of Theorems \ref{theo:enpres'}  and \ref{theo:enpresinst}}\label{app:proofV} 


{\bf Proof of Theorem \ref{theo:enpres'}.}   First, we show the implication $1  \Longrightarrow 2$:  every energy-preserving channel can be realized through a unitary interaction that separately preserves the energy of the system and the energy of the environment.  We start from Eq. (\ref{Vrealization}), namely $\map M(\rho)  = \Tr_{\rm env}  [ V \rho V^\dag]$ with the isometry $V$ given in  Eq. (\ref{block}). From  the definition it follows that  $V$ can be written as  
\begin{align}\label{VE}
V   =  \bigoplus_E  \,  V_E  \, \qquad V_E  :=  \sum_{k=1}^K  \,  P_E  M_k  P_E  \,  \otimes |\phi_k\> \, ,
\end{align}
having chosen the environment to be $K$-dimensional, with $\spc H_{\rm env}   =  \Span  \{  |\phi_k\>  ~|~  k=1,\dots, K\}$.  
Every operator $V_E$ in Eq. (\ref{VE})  is a unitary mapping from the eigenspace of $H_{\rm sys}$ with eigenvalue $E$, denoted by $\spc H_E$,  to a subspace of $\spc H_E\otimes \spc H_{\rm env}$, denoted by $\spc S_E$.  Since $\spc H_{\rm env}$ is $K$-dimensional,  $\spc H_E\otimes \spc H_{\rm env}$ can be decomposed as 
\begin{align}
\spc H_E\otimes \spc H_{\rm env}   =  \bigoplus_{k=1}^{K}    \spc S^{(k)}_E  \, ,
\end{align} 
where  $\left\{ \spc S^{(k)}_E\right\}_{k=1}^K$ are orthogonal subspaces isomorphic to  $\spc S_E$ and  $\spc S^{(1)}_E  \equiv \spc S_E$.  
 Let us denote by $U_{E,k}$ the unitary that maps the subspace $\spc S_E$   into  $\spc S_E^{(k)}$. 
 With this notation, we can   define the unitary operator $U_{E} $ as  
\begin{align}\label{definitionUE}
U_{E}  |\psi\>  |\phi_k\>   & :  =   
\left  \{  
\begin{array}{ll}   
V_E |\psi\>       \qquad   &  k=1  \\
 U_{E,k} \, V_E  |\psi\>   \qquad  & k  >1  \, .
 \end{array}  
\right.  
\end{align}
Note that, by construction, the operator $U_E$ maps  the subspace  $\spc H_E\otimes \spc H_{\rm env}$ into itself, and therefore satisfies the condition 
\begin{align}\label{projector}  
U_E^\dag \,  P_E  \,  U_E  =  P_E  \,.
\end{align} 
Now, consider the unitary $U  := \bigoplus_E \,  U_E$.  Clearly, $U$ preserves the energy of the system: indeed, we have 
\begin{align*}
U^\dag  \, H_{\rm sys}  \, U  &  =  U^\dag  \,  \left(  \bigoplus_E  \,  E  \,     P_E  \right)  \,  U  \\
  &  =  \bigoplus_E  \, E  \,     U_E^\dag   P_E U_E   \\
  &  =  \bigoplus_E  \, E  \,      P_E    \\
   &  =     H_{\rm sys} \, ,  
\end{align*}  
having used Eq. (\ref{projector}) and the definition of $U$.
On the other hand, choosing the Hamiltonian of the environment to be constant, e.~g.~$H_{\rm env}  =  0$, we trivially satisfy the condition $  U^\dag  \,  H_{\rm env} \,  U  =   H_{\rm env}$.   
Finally,  Eq. (\ref{definitionUE}) gives the relation   
\begin{align*}
U     |\psi \> |\phi_1  \>    &  =   \bigoplus_E  \,   U_E  \,   |\psi\>  |\phi_1  \>         \\
   &  =    \bigoplus_E  \,   V_E  \,   |\psi\>   \\
   &   =  V  |\psi\>    \qquad \forall |\psi  \> \in\spc H_{\rm sys}\, .
\end{align*}
In turn, this implies the condition 
\begin{align*}
\Tr_{\rm env}  [    U  \,  (  |\psi\>\<  \psi |  \otimes  |\phi_1\>\<\phi_1  |  )  \, U^\dag]   &  =   \Tr_{\rm env}  \,  [  V   |\psi\>\<\psi| V^\dag]  \\
&  =   \map M (  |\psi\>\<\psi|)\,  ,  
\end{align*}
meaning that the unitary $U$ implements the desired channel $\map M$ on every pure state, and, by linearity, on every mixed state.   

Let us now show the implication $2\Longrightarrow 3$: the unitary evolution $U$ can be realized through an interaction $H_{\rm int} (t)$ that commutes with    the Hamiltonians $H_{\rm sys}$ and $ H_{\rm env}$ at all times.  
 The proof is immediate from  the spectral decomposition 
\[U  =   \sum_{n}  \,  e^{-i\theta_n} \, |\phi_n\>\<\phi_n|  \, ,\qquad  \theta_n  \in [0,2\pi  ) \,  . \]
 Indeed, it is enough to define 
 \[H :  =  \frac{i  \hbar}{  t_1-t_0} \, \ln   U  \, ,  \qquad \ln U  =   \sum_{n}   -i\theta \,  |\varphi_n\>\<\varphi_n|  \, ,\]  and 
\[   H_{\rm int}   (t)  :   =   g(t)       \,   H     \, ,    \]
where $g(t)$ is an arbitrary function that   quantifies the strength of the interaction and satisfies the conditions  
\begin{align*}
g(t)&= 0   \qquad \forall t  \le  t_0 \, ,   \forall t\ge t_1  \, ,\\
 g(t) & \ge 0 \qquad  \forall t  \in  (t_0,t_1) \,, \\
\int_{t_0}^{t_1} \, \d t \,  g(t)  &=  1 \, .
\end{align*}
Finally, we note that the implication $3 \Longrightarrow 1$ has been already proven in the main text.    \qed 

\medskip   Note that the construction in the above proof allows one to engineer a \emph{minimal} realization of the desired channel, in which the Hamiltonian of the environment is fully degenerate. However,  nothing prevents us from regarding the  Hilbert space of the environment in the above construction as a degenerate eigenspace of a non-trivial Hamiltonian, acting on a larger space.  

\medskip 

{\bf Proof of Theorem \ref{theo:enpresinst}.}   For every $x$, let us pick a Kraus decomposition $\map M_x  (\rho)  =  \sum_{k  =1}^{m_x}  \,  M_{x,k}   \rho  M_{x,k}^\dag$.  By definition, the operators $\{  M_{x,k}~|~  x\in\set X,  k  = 1,\dots,  m_x\}$ are Kraus operators for the energy-preserving channel $\map M  =  \sum_x  \map M_x$.   Hence, the construction in the previous proof yields an environment with Hilbert space \[\spc H_{\rm env}  =   \Span \{   |\phi_{x,k}  \>  ~|~  x\in\set X,  k  = 1,\dots,  m_x\} \, ,\]
a constant Hamiltonian $H_{\rm env}$,  a unit vector $|\phi_{1,1}\>  \in\spc H_{\rm env}$,  and a unitary operator $U$ such that  
 \begin{align}\label{instrumentU}  U    \,  |\psi\>    |\phi_{1,1}  \>    =  \sum_{x,k}    \,   M_{x,k}   |\psi\>  |\phi_{x,k}  \> \qquad \forall |\psi\>\in  \spc H_{\rm sys} \end{align}
and $U$ satisfies the conditions $  U^\dag H_{\rm sys}  U  =  H_{\rm sys}$ and $    U^\dag H_{\rm env}  U  =  H_{\rm env}$.    Now, consider the projective measurement $\{Q_x\}_{x\in\set X}$  defined by $Q_x  =  \sum_{k=1}^{m_x}  \,  |\phi_{x,k}  \>\<  \phi_{x,k}|$.   By Eq. (\ref{instrumentU}), one has the desired condition  
\begin{align*}
&\Tr_{\rm env}  \left   [     (   I_{\rm sys}  \otimes Q_x)  \,    U  (  |\psi\>\< \psi|  \otimes  |\phi_{1,1}  \>\<  \phi_{1,1}| )  U^\dag \right ]  \\
&\qquad \qquad\qquad \qquad\qquad \qquad\qquad \qquad  \qquad   =    \map M_x  (  |\psi\>\<\psi|)  
\end{align*} 
valid for every $x$ and for every pure state  $|\psi\>$, and, by linearity, for every mixed state $\rho$.   The projective measurement   $\{Q_x\}_{x\in\set X}$ can be regarded as a measurement of an observable $O =  \sum_{x\in\set X}    f(x)  \,    Q_x$, where $f:  \set X  \to \R$ is a fixed (but otherwise arbitrary) injective function. Since the  Hamiltonian  of the environment is constant, the observable $O$ trivially satisfies  Yanase's condition (\ref{compatible}).  \qed

\section{Weak vs strong energy-preservation for probabilistic transformations}\label{appendix:weak}

Here we compare our notion of energy-preserving transformation---defined as transformations that can be part of an energy-preserving instrument---with a stronger notion.  For this reason, here we refer to our condition as  \emph{weak energy preservation}:  
\begin{defi}
A probabilistic transformation $\map M_0$ satisfies \emph{weak energy preservation} if there exists an energy-preserving instrument $\{\map M_x\}_{x\in\set X}$ and an outcome $x_0\in\set X$ such that   $\map M_0   =  \map M_{x_0}$.  
\end{defi}
The above  notion of weak energy preservation---on average over all the possible outcomes of a measurement---is fundamentally different from  an alternative, strong notion, of energy preservation, which  requires the output of the transformation to have the same energy distribution of the input:  
\begin{defi}
A probabilistic transformation  $\map M_0$ satisfies \emph{strong energy preservation} if, for every input state $\rho$, one has
\begin{align}\label{silly}
\Tr  [  \rho'  \,  H^n_{\rm sys}]    =  \Tr  [  \rho  \,  H_{\rm sys}^n]      \qquad \forall n\in\N \, ,
\end{align}
 where $\rho'$ is the conditional output state $\rho'  =  \map M_0  (\rho)/ \Tr  [\map M_0  (\rho)]$. 
\end{defi}

  While strong energy-preservation may seem more natural at first sight, a closer inspection reveals that the condition is quite restrictive.  For instance,  it even forbids  ideal energy measurements on the system, which  are the prototype of measurements that can be performed at no energy cost \cite{AY,ozawa2,popescu}.  Indeed, a  von Neumann measurement of the energy trivially collapses a mixture of different energy eigenstates into a single energy eigenstate, thus  violating Eq. (\ref{silly}).   Furthermore, it is not hard to see that, for non-degenerate Hamiltonians,  every quantum operation satisfying strong energy-preservation must be proportional to a \emph{deterministic}  energy-preserving transformation. In other words, the operation  can be realized by tossing a biased coin, and, if the coin turns out heads, by applying a deterministic transformation.  As a result,  the strong notion of energy-preservation would make the study of  probabilistic transformations  irrelevant.   
    In this work we focus on the more flexible notion of weak energy preservation,  which captures exactly the class of transformations that can be implemented at no energy cost. Since there is no ambiguity, we will omit the specification ``weak'' and refer to these transformations simply as energy-preserving.   

\section{Decomposition of stationary covariant instruments}\label{app:decompcov}  

{\bf Proof of Proposition \ref{prop:decompcov}.}    In Ref. \cite{GiulioNat} it was proven that every instrument $\{\map M_x\}_{x\in\set X}$ can decomposed as  $\map M_x   =  \map C_x \map P_x$, where   $\{\map P_x\}_{x\in\set X}$ is the  quantum instrument  defined by 
\begin{align}
\map P_x (\cdot)   =    \sqrt {P_x}  \cdot  \sqrt {P_x}  \, ,\qquad P_x   =  \map N_x^\dag (I) 
\end{align} 
and $\map C_x$ is a suitable quantum channel.  Now, suppose that $\map M_x$ is stationary and covariant, with 
$$   V_t  \map M_x  (\cdot)   V_t^\dag   =   \map M_x \left(   U_t  \cdot  U_t^\dag \right)  \qquad \forall  t \in  \R \, .$$ Then, one has 
\begin{align*}
   U^\dag_t   P_x      U_t  &   =     U_t^\dag   \map  M_x^\dag  (   I)    U_t  \\
   &  =  \map M^\dag_x   \left(    V_t^\dag V_t\right)\\
   &    =   \map M_x^\dag (I)  
  \\
  &    = P_x   \qquad \forall t\in \R \, .
\end{align*}
Taking derivatives with respect to $t$ on both sides and recalling the definition, one then obtains 
\begin{align}
[  P_x,   H_{\rm sys}]   =  0  
\end{align}
and consequently  $[  \sqrt{P_x},   H_{\rm sys}]   =  0 $.  Hence,  the quantum operation 
$$\map M_x  (\cdot)  =  \sqrt{  P_x }  \cdot   \sqrt {P_x}$$ is energy-preserving (Lemma \ref{lem:lindblad}).  
It remains to show that the channel $\map C_x$ is covariant.    This can be done easily when $P_x$ is invertible. In this case, one has $  \map C_x   =  \map M_x  \map P_x^{-1}$, where $\map P_x^{-1}$ is the   completely positive map  
$$  \map  P_x^{-1} (\cdot)  =   \sqrt{  P_x^{-1}}  \cdot \sqrt{P_x^{-1}}\, . $$ 
Using this fact, we obtain  
\begin{align*}  
\map C_x  ( U_t  \cdot   U_t^\dag  )  &  =   \map M_x    \map P_x^{-1}   ( U_t  \cdot   U_t^\dag  )  \\
&  =     \map M_x      \left(   \sqrt{P_x^{-1}}   U_t  \cdot   U^\dag _t   \sqrt{P_x^{-1}}  \right)\\
&  =     \map M_x      \left(U_t      \sqrt{P_x^{-1}}    \cdot      \sqrt{P_x^{-1}}  U^\dag _t  \right)\\
&  =   V_t \map M_x      \left(     \sqrt{P_x^{-1}}    \cdot      \sqrt{P_x^{-1}}  \right)  V_t^\dag\\
&  =  V_t  \map C_x  (\cdot)  V_t^\dag  \qquad \forall t\in \R \, ,
\end{align*}
that is, $\map C_x$ is covariant.    When $P_x$ is not invertible, the above reasoning shows that  the  restriction of $\map C_x$ to the support of $P_x$ is  covariant.  
On the orthogonal complement of the support of $P_x$, the action of the channel $\map C_x$ can be redefined arbitrarily in such a way that the covariance condition is guaranteed.  \qed

\section{Decomposition of non-stationary covariant instruments}\label{app:decompcovgen}

{\bf Proof of Proposition \ref{prop:decompcovgen}.}    The proof follows the same lines as the proof of Proposition \ref{prop:decompcov}:   Define the instrument $\{\map N_y\}$ as   
\[  \map N_y  :  =  \sum_{x\in\set X}  \,   \map M_x    \] 
(it is understood that, when the orbit is continuous, the sum has to be replaced by an integral over the orbit). Define the operators $  P_y:  = \map N_y^\dag (I)$ and the pure instrument $\{\map P_y\}$, with 
\[  \map P_y  (\cdot)    =   \sqrt{P_y}  \cdot \sqrt{P_y} \, .\]    
  Then, the same argument used in the proof of Proposition \ref{prop:decompcov} shows that the instrument $\{\map P_y\}$ is energy-preserving.   Then, for every $y\in\set Y$, define the instrument   $\{\map  M_x^{(y)} \}_{x\in\set O_y}$ as
  \[\map M_x^{(y)}  :  =    \map  M_x  \left(  P_y^{-1/2}  \cdot P_y^{-1/2}\right)    \, . \] 
  By definition, the quantum operation $\map M_x^{(y)}$ satisfies the condition 
  \begin{align}\label{justforyou}  \map M_x     =   \map M_x^{(y)}  \circ \map P_y \, .  \end{align}
  Moreover, the quantum operations  $\{\map  M_x^{(y)} \}_{x\in\set O_y}$ form a well-normalized  instrument transforming states with support of $P_y$ into states with support $\spc H_{\rm out}$.   When the support of $P_y$ is not the whole of $\spc H_{\rm in}$,  one can complete the quantum  operations $\{\map  M_x^{(y)} \}_{x\in\set O_y}$ by extending their action on the kernel of $P_y$.  As a result, one obtains a well-normalized instrument from $\spc H_{\rm in}$ to $\spc H_{\rm out} $, without affecting the validity of Eq. (\ref{justforyou}).  \qed

   \section{Optimality of eigenstate alignment}\label{app:EAoptimal}

{\bf Proof of Theorem \ref{theo:EAoptimal}. }
Let $\map M$ be an energy-preserving quantum operation such that takes place deterministically on $|\varphi\>$.  
 The, the fidelity of $\map M$ for the transition $|\varphi\> \to  |\psi\>$ can be computed as  
 \begin{align*}
 F   &  =  \< \psi|   \map M (|\varphi\>\<  \varphi|)  |\psi\>  \\
 &  =     \sum_{E, E'}  \sqrt{   p_E q_E   p_E'  q_E'}  \,   \<  \psi_E  |   \map M (   |\varphi_E  \>\<  \varphi_{E'}|)  |\psi_{E'}\>  \\  
 &  =     \sum_{E, E'}  \sqrt{   p_E q_E   p_E'  q_E'}  \,   \<  \psi_E  |   \map M (   |\varphi_E  \>\<  \varphi_{E'}|)  |\psi_{E'}\>     
 \end{align*}  

  Let $\map M$ be an energy-preserving quantum operation,  with Kraus decomposition $\map M (\rho)   =     \sum_k   M_k \, \rho\, M_k^\dag$.  
  be a Kraus decomposition for $\map M$.    Since $\map M$ is energy-preserving,  every  Kraus operator $M_k$ must be of the form
\begin{align}\label{blockM}
 M_k  =  \bigoplus_E     \,  P_E  M_k  P_E   \, ,
\end{align}
cf. Eq. (\ref{block}).  
The probability of success is then given by  
\begin{align}
\nonumber p_{\rm succ}   &  =  \Tr [  \map M  (|\varphi\>\<\varphi|)]  \\  
\nonumber    &   =   \sum_k   \< \varphi|    M_k^\dag  M_k  |\varphi\>   \\
\nonumber   &  = \sum_k   \sum_{E,E'}  \sqrt{p_E  p_E'}  \,   \<\varphi_E   |   M_k^\dag  M_k  |\varphi_{E'}\>   \\
\label{pEmix}  & =    \sum_k   \sum_{E}   p_E   \,   \<\varphi_E   |   M_k^\dag  M_k  |\varphi_{E}\>
  \, ,    
\end{align}
the third equality following from the decomposition of $|\varphi\>$   [Eq. (\ref{statedecomp})] and the fourth equality following from the block diagonal form of $M_k$.  Now, note that one has  
\begin{align*}  \sum_k   \<  \varphi_E  |  M_k^\dag M_k  |\varphi_E\>     =    \Tr  [  \map M (  |\varphi_E\>\<\varphi_E)]   \le 1 \, .  
\end{align*}  
Hence, the condition $p_{\rm succ}  =1$ can be satisfied by Eq. (\ref{pEmix}) only if the equality
\begin{align}\label{DetForAllE}  \sum_k   \<  \varphi_E  |  M_k^\dag M_k  |\varphi_E\>     = 1 
\end{align}   
holds for every $E$ such that $p_E \not  = 0$.  We will now optimize the fidelity subject to this constraint.

The fidelity for the transformation $|\varphi\> \to |\psi\>$ can be expressed as 
\begin{align*}
F  &=  \<\psi  |  \,\map C  (  |\varphi\>\<\varphi|) \, |\psi\>  \\
&  =  \sum_k   \,  |\< \psi  |  M_k |\varphi\> |^2 \\
&  =\sum_k    \,  \left|  \sum_E    \<\psi|  P_E   M_k   P_E  |\varphi\>    \right|^2 \\
&  =\sum_k    \,  \left|  \sum_E \sqrt{p_E  q_E}   \<\psi_E|   M_k    |\varphi_E\>    \right|^2 \, ,  
\end{align*}
having used the decompositions of Eq. (\ref{statedecomp}).     Then, one has  the bound
\begin{align*}
F  &\le   \sum_k    \,  \left(  \sum_E \sqrt{p_E  q_E}  \, | \<\psi_E|   M_k    |\varphi_E\>|\,    \right)^2   \\
&   \le   \sum_k    \,  \left(  \sum_E \sqrt{p_E  q_E}  \,   \|   M_k    |\varphi_E\> \|  \,    \right)^2 \\
&  \le \sum_k  \,\,   \left(  \sum_{E} \sqrt{p_{E}  q_{E}}  \,   \|   M_k    |\varphi_{E}\> \|^2  \,    \right)\,    \left(  \sum_{E'} \sqrt{p_{E'}  q_{E'}}   \,    \right)   \, ,
\end{align*}
the third bound coming from the Cauchy-Schwarz inequality.    Inserting the relation  $  \|  M_k  |\varphi_E\>  \|^2  =   \<  \varphi_E  |  \,  M_k^\dag M_k  \,  |\varphi_E\> $ one finally obtains  
\begin{align*}
F  &  \le \sum_k  \,\,   \left(  \sum_{E} \sqrt{p_{E}  q_{E}}  \,   \<  \varphi_E  |   M_k^\dag  M_k    |\varphi_{E}\>   \,    \right)\,    \left(  \sum_{E'} \sqrt{p_{E'}  q_{E'}}   \,    \right)  \\
  &=     \left(  \sum_{E} \sqrt{p_{E}  q_{E}}    \right)^2\, ,    
\end{align*}   
having used Eq. (\ref{DetForAllE}).       By direct inspection it is immediate to see that eigenstate alignment attains the bound. \qed 
\section{Decomposition of a quantum operation in terms of its L\"uders reduction.}\label{app:proofProp}

{\bf {Proof of Proposition \ref{prop:ludersdecomp}}.}    In the case when $P$ is a projector, the decomposition  $\map M  =  \map C  \circ \map P$ was proven  by  Bartlett \emph{et al}   in Ref. \cite{bartlett2006degradation}.   In general, the decomposition  follows from the decomposition of an instrument proven in Ref. \cite{GiulioNat}.   
 Now, suppose that $\map M$ is energy-preserving.  In this case, we have already seen that  $P$  commutes with $H_{\rm sys}$ and, therefore, the quantum operation $\map P(\cdot)   =  \sqrt{P}  \cdot  \sqrt P$ is energy-preserving (see  the proof of Proposition \ref{prop:decompcov} in Appendix \ref{app:decompcov}).  
 It only remains to prove that   $\map C$ is energy-preserving.  This is easily done when $P$ is invertible:  in this case, one has    
 \begin{align}\label{cfinale}\map C   =    \map M   \map P^{-1}  \, ,  \qquad\map P^{-1}  (\cdot)   =  \sqrt{ P^{-1}}  \cdot  \sqrt{  P^{-1}} \, ,
   \end{align} 
   meaning that every Kraus operator of $\map C$ is  the form $C_k  =  M_k \sqrt{  P^{-1}}$, where $M_k$ is a Kraus operator for $\map M$. 
 Now, since $\map M$ is energy-preserving, every operator $M_k$ commutes with $H_{\rm sys}$.
Hence,  also $C_k$  commutes with $H_{\rm sys}$. Since $C_k$ is a generic Kraus operator, Lemma \ref{lem:lindblad} implies that $\map C$ is energy-preserving.     When $P_x$ is not invertible, the above reasoning shows that  the  restriction of $\map C$ to the support of $P$ is  energy-preserving. 
On the orthogonal complement of the support of $P$, the action of the channel $\map C$ can be redefined arbitrarily in such a way that energy-preservation is guaranteed.  \qed

\section{The ultimate limits of probabilistic operations}\label{app:ultimate}

{\bf Proof of Proposition \ref{cor:maxfid}.}   By Proposition \ref{prop:better}, the maximum fidelity is achieved by a pure quantum operation $\map M  (\cdot)  =  M  \cdot M^\dag$, with success probability 
\begin{align}\label{psuccfinale}
p_{\rm succ}  =  \sum_E      \,p_E  \,  \<\varphi_E  |  P_{\rm succ}  |\varphi_E\> \, , \qquad  P_{\rm succ}  =   M^\dag M \, ,
\end{align} 
and fidelity 
\begin{align}\label{Ffinale}
F  =  \left(  \sum_E  \sqrt{ p_E'  q_E}  \right)^2\, , \qquad  p_E'    =   \frac{p_E   \, \<\varphi_E | \, P_{\rm succ}\, |\varphi_E\>}{p_{\rm succ}}  \, .
 \end{align} 
By the   Cauchy-Schwarz inequality, one immediately obtains the bound  $$F  \le  \sum_{E  \in\Sp(\varphi)  \cap \Sp(\psi)}   \,q_E    \, .$$
     The bound is achieved if and only if 
          \begin{align}\label{equazione}\< \varphi_E  |   P_{\rm succ}   |\varphi_E\>   =   c  \,\frac{  q_E}{p_E}  \, , \qquad \forall   E\in\Sp(\varphi)  \cap  \Sp  (\psi)  
     \end{align}
     for some constant $c\ge0$.  Note that, since the  $\< \varphi_E  |   P_{\rm succ}   |\varphi_E\>$ is the probability that $\map M$ takes place on the state $|\varphi_E\>$,   the constant $c$  must satisfy the relation   
     \begin{align}\label{cbound}
     c  \le \frac{p_E}{ q_E}  \qquad  \forall  E  \in  \Sp(\varphi)  \cap   \Sp(\psi) \,.
     \end{align}
  Now, recall that the quantum operation $\map M$ was constructed from L\"uders reduction and eigenstate alignment.   Hence,   its Kraus operator $M$ must satisfy  the condition
\begin{align}  M  |\varphi_E\>    \propto           |\psi_E\>   \qquad   \forall E  \in\Sp(\varphi) \cap \Sp(\psi)  \, . 
\end{align}  
Combining this fact with Eq. (\ref{equazione}) and recalling that $P_{\rm succ}  =  M^\dag M$, we obtain
\begin{align}\label{optM}  M  |\varphi_E\>    =           \sqrt{c\, \frac{ q_E }{p_E}}    |\psi_E\>   \qquad   \forall E  \in\Sp(\varphi) \cap \Sp(\psi)  \, . 
\end{align}  
 
 Now, inserting  Eq. (\ref{equazione}) into  Eq. (\ref{psuccfinale}),  
 the probability of success becomes   
\begin{align}\label{psuccfinalefinale}   p_{\rm succ}   =    c  \sum_{E \in  \Sp(\varphi)  \cap\Sp(\psi)}   q_E  \, . 
\end{align}  
  Given the constraint (\ref{cbound}),  the maximum success probability   is obtained by choosing 
  \begin{align}
  c    = \min_{E  \in  \Sp(\varphi)  \cap \Sp(\psi)}  \frac {p_E}{q_E} \, . 
  \end{align}  
Inserting this value into Eqs.  (\ref{psuccfinalefinale}) and (\ref{optM}),  we finally obtain the desired relations (\ref{psuccmax}) and (\ref{Moptimized}). \qed   
 

\section{Derivation of the optimal recursive protocol}\label{app:derivation}

The optimal filter in the $k$-th round is determined by induction from the requirements 1-3 in subsection \ref{subsec:RUS}. 
At the first round, the filter attempts at converting  $|\varphi\>$ into    $|\psi\>$. 
  According to  Proposition  \ref{cor:maxfid},      the maximum fidelity is given by 
 \begin{align}\label{F1}
F_{\max}^{(1)}   =   
      \sum_{E  \in\Sp(\varphi)  }     \,     q_E 
\end{align}   
and can be achieved with probability
\begin{align*}  
   p^{(1)}_{\rm succ}   = 
     \left(  \min_{E  \in  \Sp(\varphi)  \cap  \Sp(\psi)}  ~  \frac {p_E}{q_E}     \right)  \,    F^{(1)}_{\max}   \, .
   \end{align*}
The optimal  quantum operation  must be pure and  its Kraus operator   $   B_{\rm succ}^{(1)}$ must  satisfy the condition
 \begin{align*}
     B_{\rm succ}^{(1)}   \,   \left|\varphi_{E}\right\>    =         |\psi_E\>    \qquad \forall E  \in  \set R_ 1 \, ,
     \end{align*}
where $R_1$ is the set of energy values  in  $\Sp(\varphi)  \cap  \Sp(\psi)$ that minimize the ratio  $  p_{E}/q_E$     [cf. Eq. (\ref{optM})].    Writing the unsuccessful quantum operation in the Kraus form $\map B_{\rm fail}^{(1)}  (\cdot)  =      \sum_{t}    B_{{\rm fail},t}^{(1)}     \cdot B_{{\rm fail},t}^{(1)  \dag}$, we then obtain   
\begin{align}\label{killone}   B_{{\rm fail},t}^{(1)}     |  \varphi_E\>   =  0   \qquad    \forall E  \in  \set R_1   \end{align}
for every possible value of $t$.  
  At the second step, the filter attempts to produce the target state $|\psi\>$ from the  state 
$$  \rho^{(2)}    
=   \sum_t    p^{(2)}_t\,       \left|\varphi^{(2,t)}  \right  \>   \left \<    \varphi^{(2,t)}\right|     \, , $$
with   $p^{(2)}_t   : =   {   \|     B_{{\rm fail},t}^{(1)}     |  \varphi\>    \|^2}/{ \sum_{t'}      \|     B_{{\rm fail},t'}^{(1)}     |  \varphi\>    \|^2   } $   and     
\begin{align}\label{def2t}  \left |\varphi^{(2,t)}  \right\>  : =  \frac{B_{{\rm fail},t}^{(1)}     |  \varphi\> }{  \|  B_{{\rm fail},t}^{(1)}     |  \varphi\>  \|}   \, .
\end{align}
Clearly, the maximum fidelity achievable from the state $\rho^{(2)}$ cannot be larger than the maximum over $t$ of the fidelity achievable from $  \left |\varphi^{(2,t)}  \right\>$. 
Now, let us expand each state as
\begin{align*}  \left |\varphi^{(2,t)}  \right\>    =   \sum_E    \sqrt{  p_E^{(2,t)}}   \,   \left  |  \varphi_E^{(2,t)}   \right\>     \,  , 
\end{align*}  
for suitable probabilities     $\left\{  p_E^{(2,t)} \right\}$ and suitable  energy eigenstates $\left\{    \left  |  \varphi_E^{(2,t)}   \right\> \right\}$.   
Note that, due to  the condition in Eq. (\ref{killone}),   one has 
\begin{align}\label{specless}
\Sp  \left(\varphi^{(2,k)}\right)  \subseteq \Sp  (\varphi)  \setminus  \set R_1  \,  .
\end{align} 
Using this fact,  we can upper bound the fidelity achievable from the state    $|\varphi^{(2,t)}\>$---call it $F_{\max}^{(2,t)}$---as    \begin{align*}  F_{\max}^{(2,t)}   &\le \sum_{E  \in  \Sp(  \varphi^{(2,t)})  }   q_E     \\
&\le \sum_{E  \in  \Sp(  \varphi)  \setminus \set R_1 }   q_E
\end{align*}
the first inequality coming from  Proposition \ref{cor:maxfid}.  
In turn, this allows  us to upper bound the overall fidelity at the second step as 
 \begin{align}  
\nonumber F^{(2)}        & \le    \max_t   F^{(2,t)}_{\max}  \\
\nonumber  & \le   
\sum_{E  \in    \Sp(\varphi)  \setminus \set R_1}   q_E \\
 \label{F2}  &     =:    F^{(2)}_{\max}      \, .  
 \end{align}
    The bound  is attained  when the quantum operation
$\map B^{(1)}_{\rm fail}$  is pure and its  Kraus operator is given by
\begin{align}\label{failopt}
   B^{(1)}_{\rm fail}    :     =   \sqrt{     I  -      B_{\rm succ}^{(1)  \dag}     B_{\rm succ}^{(1)}  }  \, .
   \end{align}
Luckily, this choice maximizes not  only the fidelity at the second step, but also the \emph{probability} that  maximum  fidelity   is achieved:  indeed,  denoting   by $p_{\rm succ}^{(2)}$  the probability that the output has fidelity  $F^{(2)}_{\max}$ with the target and by $p_{\rm succ}^{(2,t)}$ the probability that the optimal transformation takes place on the state $ \left |\varphi^{(2,t)}  \right\> $,  we have the bound
\begin{align}
\nonumber p_{\rm succ}^{(2)}  &  \le      \sum_t       p^{(2)}_t  \,      p_{\rm succ}^{(2,t)}      \\
\nonumber   &   =   \sum_t  p^{(2)}_t  \,            \left[  \min_{E \in  \Sp(\varphi^{(2,t)})   \cap \Sp(\psi)    }    \frac {p^{(2,t)}_E}{q_E}     \right]  \,    F^{(2,t)}_{\max} \\
 \label{aaa} &  \le       \min_{E}              \left[      \sum_t   \,   \frac {   p^{(2)}_t  \, p^{(2,t)}_E   }{q_E}     \right]  \,    F^{(2)}_{\max}     \, , 
    \end{align}
the equality in the second line coming from Proposition \ref{cor:maxfid}.      It is easy to verify that the pure quantum operation of  Eq. (\ref{failopt}) reaches the bound:  indeed, its  output state     
   $$  \left|\varphi^{(2)}  \right\>    =   \frac{ B^{(1)}_{\rm fail}   |\varphi\>}{  \|    B^{(1)}_{\rm fail}   |\varphi\> \|       }     
   \, , $$
  can be converted optimally into the state $|\psi\>$ with   probability given by  Proposition \ref{cor:maxfid}, which now yields   
     \begin{align}\label{bbb}
   p_{\rm succ}^{(2)}    &  =    \min_{E  \in  \Sp(\varphi^{(2)})  \cap \Sp(\psi)  }              \left[       \,   \frac {     p^{(2)}_E   }{q_E}     \right]  \,    F^{(2)}_{\max}     
 \end{align} 
with  
$p_E^{(2)}   :=     \left \|    P_E  \left|\varphi^{(2)}   \right \>    \right \|^2  
   \equiv      \sum_t     p^{(2)}_t  \, p^{(2,t)}_E $.    Inserting this equality in Eq. (\ref{bbb}) we then obtain that the bound of Eq. (\ref{aaa}) is attained.  

Summarizing, we have proven that the ``best way to fail" is via a pure quantum operation.   Iterating the same argument, we obtain that the optimal strategy  at each step is described by a binary filter consisting of two pure quantum operations, with   Kraus operators  $B^{(k)}_{\rm succ}$ and $B^{(k)}_{\rm fail}$, respectively.     Expanding the state at the $k$-th step as   
$$\left|\varphi^{(k)}\right\>   =   \sum_{E}   \sqrt{  p^{(k)}_{E}}   \,  \left |\varphi^{(k)}_{E} \right \> \, , $$
the successful Kraus operator is determined in an essentially unique way by Proposition \ref{cor:maxfid}, which yields the condition
\begin{align}\label{bksucc1}
  B^{(k)}_{\rm succ}   \left| \varphi^{(k)}_E  \right  \>       =           \left[     \min_{E' \in  \Sp(\varphi^{(k)})   \cap  \Sp(\psi)  }   \sqrt{     \,   \frac {     p^{(k)}_{E'}   }{q_{E'}}     }\right]   \,              \sqrt{\frac{q_E}{     p^{(k)}_E   } }   ~ |\psi_E\>     
\end{align}  
for every energy $E$ in  $\Sp(\varphi^{(k)})$.    The unsuccessful Kraus operator is then given by
\begin{align}\label{bkfail1} 
B^{(k)}_{\rm fail}    :=   \sqrt{  I-    B^{(k)  \dag}_{\rm succ} B^{(k)}_{\rm succ} } \, ,
\end{align} 
and its definition is essentially unique, up to the  application of an energy-preserving unitary on the output and to a possible re-definition of $B^{(k)}_{\rm fail}$ outside the relevant subspace.  

Applying iteratively  Eqs. (\ref{bksucc1}) and (\ref{bkfail1}) it is easy to obtain that the eigenstates $|\varphi_E^{(k)}\>$ are independent of $k$,~i.~e.~one has 
$$  |\varphi^{(k)}_E\>  \equiv   |\varphi_E\>  \qquad    \forall  k   =  1,\dots, K \, ,  \forall  E   \in\Sp(\varphi^{(k)}) \, . $$
This condition implies that Eq. (\ref{bksucc1}) can be rewritten as 
\begin{align*}
  B^{(k)}_{\rm succ}   \left| \varphi_E  \right  \>       =           \left[     \min_{E' \in  \Sp(\varphi^{(k)})   \cap  \Sp(\psi)  }   \sqrt{     \,   \frac {     p^{(k)}_{E'}   }{q_{E'}}     }\right]   \,              \sqrt{\frac{q_E}{     p^{(k)}_E   } }   ~ |\psi_E\>     
\end{align*}  
for every energy $E$ in  $\Sp(\varphi^{(k)})$.

\section{Kraus operators of the recursive protocol}\label{app:Mk}

Here we characterize the form of the successful Kraus operators $M_k$, with $k\le K  \le L$.  
Physically, the operator $M_k$  corresponds to the event that one succeeds at the $k$-th round, after having  failed in the first $k-1$ rounds, namely
\begin{align}\label{defiMk}  M_k   =     B^{(k)}_{\rm succ}   B^{(k-1)}_{\rm fail}   \cdots B^{(1)}_{\rm fail}\,  .
\end{align} 

To characterize $M_k$ we first analyze the operators   $B^{(i)}_{\rm fail}$, with $i   =1,\dots,  k-1$.  Combining Eqs.  (\ref{bksucc1}) and  (\ref{bkfail1}), we obtain that      $B^{(i)}_{\rm fail}$ satisfies the condition  
\begin{align}\label{Bk0}
B^{(i)}_{\rm fail}     |\varphi_E\>    =    0    \qquad \forall  E  \in  \set R_1^{(i)} \, ,
\end{align}
where $\set R_1^{(i)}$ is the set  defined  by 
\begin{align}\label{R1}
\set R^{(i)}_1     :=   \left \{   E  \in  \Sp(\varphi^{(i)})  \cap \Sp (\psi)  ~  \left |~  \frac {p^{(i)}_E}{q_E}   =   r^{(i)}_1 \right. \right\} \, , 
\end{align}  
$r_1^{(i)}$ being the minimum non-zero value of the ratio  $r_E^{(i)}   =   {p^{(i)}_E}/{q_E}$.   
Now, the key observation is provided by the following 

\begin{lemma}
The set $\set R_1^{(i)}$ coincides with the set $\set R_i$  defined in Eq. (\ref{R}). 
\end{lemma}  
\Proof     The proof is by recursion over $i$, based on  the relation   
\begin{align}
\nonumber \frac{ p_E^{(i+1)}}{q_E}   & =     \frac{  \left\|    P_E   B_{\rm fail}^{(i)}  \,  |\varphi^{(i)}  \> \right\|^2   }{   q_E \left\|      B_{\rm fail}^{(i)}  \,  |\varphi^{(i)}  \> \right\|^2  }  \\
\nonumber  &=     \frac{  \left \<  \varphi^{(i)}  \right|  \,    P_E    \,  \left(  I  -    B_{\rm succ}^{(i)\dag}  B_{\rm succ}^{(i)}    \right)  \,   P_E    \,   \left |\varphi^{(i)}  \right\>    }{   q_E \left\|      B_{\rm fail}^{(i)}  \,  |\varphi^{(i)}  \> \right\|^2  }  \\
\nonumber  &=     \frac{    p_E^{(i)}  \,    \left\{  1  -      \left[  \min_{E'  \in  \Sp (\varphi^{(i)})  \cap \Sp(\psi) }  \frac{p_{E'}^{(i)}}{q_{E'}}    \right]  \frac  {q_E}     { p_E^{(i)}}\right\}  
}{   q_E \left\|      B_{\rm fail}^{(i)}  \,  |\varphi^{(i)}  \> \right\|^2  }  \\
\label{ccc}  &=   \left[      \frac{p_E^{(i)}}{q_E}   -  \min_{E'  \in  \Sp(\varphi^{(i)})  \cap  \Sp(\psi)  }  \frac{p_{E'}^{(i)}}{q_{E'}}      \right]  \,   \left\|      B_{\rm fail}^{(i)}  \,  |\varphi^{(i)}  \> \right\|^{-2} \, ,
\end{align}
where the first equality follows from the definition $|\varphi^{(i+1)}\>  :=  B^{(i)}_{\rm fail}   |  \varphi^{(i)}\> /  \|  B^{(i)}_{\rm fail}   |  \varphi^{(i)}\>  \|$, the second from the relation  $  B^{(i)}_{\rm fail}    :=   \sqrt{  I-    B^{(i)  \dag}_{\rm succ} B^{(i)}_{\rm succ} }$, and the third from Eq. (\ref{bksucc1}).  Eq. (\ref{ccc})   shows that the set of energies for which the ratio $p^{(i+1)}_E/q_E$ has the smallest value coincides with the set of energies for which the ratio $p^{(i)}_E/q_E$ has the second smallest value, which in turn coincides with the set of energies for which the ratio $p_E/q_E$ has the $(i+1)$-th smallest value.  By definition of the sets $R_1^{(i)}$ and $\set R_i$, this proves the thesis.  \qed

Using the above lemma, Eq. (\ref{Bk0}) becomes
\begin{align}\label{Bk00}
B^{(i)}_{\rm fail}     |\varphi_E\>    =    0    \qquad \forall  E  \in  \set R_i \, . 
\end{align}
We now use this relation to determine the form of the Kraus operator $M_k$.    Setting $m_E^{(k)}:=\|M_k|\varphi_E\>\|^2$, the definition of $M_k$  [Eq. (\ref{defiMk})] gives the bound 
$$   m_E^{(k)}    \le     \left\|    B_{\rm fail}^{(i)}    \,  |\varphi_E\> \right\|^2   \qquad \forall i  =  1,\dots,   k-1 \,.  $$
Combining this bound with Eq. (\ref{Bk00})  we obtain   the condition  
\begin{align}\label{zero}
m_{E}^{(k)} =    0    \qquad \forall E\in\bigcup_{i=1}^{k-1}\set{R}_i   \equiv \set U_{k-1} \, ,
\end{align}
which shows that the operator $M_k$ annihilates the subspace spanned by the energy eigenstates with eigenvalues in $\set U_{k-1}$.   

The action of $M_k$ on the  remaining eigenstates   is determined by the fact that, when $M_k$ is decomposed as in Eq. (\ref{defiMk}),   the last operator acting in the sequence is $B^{(k)}_{\rm succ}$.  Hence, we know that 
 the initial amplitude $\sqrt{p_E}$  should be modulated to $\sqrt{q_E}$ for all the energy eigenvalues that survived the first $k-1$ rounds, that is,   
 \begin{align}\label{proportional}
m_{E}^{(k)}=
c_k   {\frac{q_E}{p_E}} \qquad \forall E\in  \Sp(\varphi)  \setminus  \set U_{k-1}  
\end{align}
where $c_k>0$ is a suitable proportionality constant. 
To determine $c_k$, note that the trace-preserving condition for the instrument   $\{\map M_k\}_{k=1}^{K+1}$ is equivalent to  
$$\sum_{k=1}^{K+1}   m_E^{(k)}  =1  \qquad \forall E \, .$$
 Combining this fact with Eqs. (\ref{zero}) and (\ref{proportional}) we then obtain the recursion relations 
\begin{align*}
\begin{array}{rcll}
c_1     {\frac{q_E}{p_E}}     & =  &    1    \quad  &\forall E\in\set R_1  \\  \\
(c_1+ c_2)   {\frac{q_E}{p_E}}  &   = & 1       & \forall E\in\set R_2 \\
  &\vdots& &     \\
\left (\sum_{k=1}^K   c_k   \right)\,    {\frac{q_E}{p_E}}  &= & 1     & \forall E\in\set R_K  \, .
\end{array}
\end{align*}
Finally, using the definition of the sets $\set R_k$  [Eq. (\ref{R})] we can solve the system of equations, obtaining   $c_k  =   r_k-r_{k-1}$ for every $k\in\{1,\dots, K\}$, having set $r_0  :=  0$.  In conclusion, the action of the successful Kraus operators is 
given  by
\begin{align}\label{solution1}
M_k  |\varphi_E\>    = 
\left\{  
\begin{array}{ll}  
0    \qquad &  E\in  \set U_{k-1}   \\ \\
\sqrt{   (r_k -r_{k-1})  \,  \frac{q_E}{p_E}   }\,   |\psi_E\>   &  E  \in\Sp(\varphi)  \setminus \set U_{k-1}  \, .
\end{array}
\right.   
\end{align}

The expressions of the fidelity and of the probability of success, anticipated in Eqs. (\ref{fidelity1}) and (\ref{prob1}), can be easily derived from the above equation.  Indeed, we have 
  \begin{align*}
  F_{\max}^{(k)}   &  =  \frac{  |  \<\psi  |  M_k   |\varphi\>|^2}{   \|  M_k  |\varphi\>  \|^2}\\
  &  =    \sum_{E  \in  \Sp  (\varphi)   \setminus  \set U_{k-1}}  \,  q_E 
  \end{align*}   
and
\begin{align*}
p_{\rm succ}^{(k)}    &=  \|  M_k  |\varphi\>  \|^2  \\
  &  =    (r_k  -  r_{k-1}) \,   \left(   \sum_{E  \in  \Sp  (\varphi)   \setminus  \set U_{k-1}}  \,  q_E  \right)   \\
  &  =      (r_k  -  r_{k-1}) \,   F_{\max}^{(k)} \, .   
  \end{align*} 
 
\section{The ultimate limit for probabilistic transitions from   mixed  to pure states}\label{app:optprob_mixed}

{\bf Proof of Theorem \ref{theo:EAoptimal_mixed}.}  

Let $\map M$ be an energy-preserving quantum channel that acts on the input state $\rho$. Its fidelity for the transition $\rho\to|\psi\>\<\psi|$ is given by
\begin{align*}
F&=\<\psi|\map M(\rho)|\psi\>\\
&=\sum_{E,E'}\sqrt{q_E q_{E'}}\<\psi_E|\map{M}(P_E\rho P_{E'})|\psi_{E'}\>\\
&\le\sum_{E,E'}\sqrt{q_E q_{E'}}|\<\psi_E|\map{M}(\rho_{E,E'})|\psi_{E'}\>|\\
&\le\sum_{E,E'}\sqrt{q_E q_{E'}}\|\map{M}(\rho_{E,E'})\|_1,
\end{align*}
the last inequality following from the relation $|\Tr[AB]|\le\|A\|_{\infty}\|B\|_1$ applied to the operators $A=|\psi_{E'}\>\<\psi_E|$ and $B=\map{M}(\rho_{E,E'})$. Since $\map{M}$ is a trace-preserving quantum operation, we then obtain the bound
\begin{align*}
F\le\sum_{E,E'}\sqrt{q_E q_{E'}}\|\rho_{E,E'}\|_1.
\end{align*}
The bound can be achieved when the state $\rho$ is block positive, namely when there exists an orthonormal basis $  \{|\varphi_{E,k}\>  ~|~  k =  1,\dots,  d_E\}$ for each eigenspace with energy $E$ such that  all matrices  
\[   [  \rho_{E,E'}]   =    [\<  \varphi_{E,k}  |    \rho  |  \varphi_{E',  l   } \> ]  \qquad k,l  \le \min \{d_E,d_{E'}\}\] 
are (square and) positive semidefinite.  Then, one can define the Kraus operators  
\[   A_k     =  \sum_{E  :   k \le d_E}    |\psi_E\>   \<  \varphi_{E,  k}| \, ,   \] 
and the quantum channel  
$\map A  (\cdot )   =  \sum_k  \,  A_k  \cdot  A_k^\dag $.  
  With this definition, one has  
  \begin{align*}
  F  &  =  \<   \psi|  \map A  (\rho)   |  \psi\>  \\
  &  = \sum_k \sum_{E:  k\le  d_E}  \sum_{E' :  k\le d_{E'} }\, \sqrt{q_E   \,  q_{E'}}   \,  \<  \varphi_{E,k}|   \,  \rho  \,  |  \varphi_{E',  k}  \>    \\
  &  =  \sum_{E, E'}    \sum_{k  \le  \min  \{ d_E,d_{E'}\}}  \,   \sqrt{q_E   \,  q_{E'}}   \,  \<  \varphi_{E,k}|   \,  \rho  \,  |  \varphi_{E',  k}  \> \\
   &  =  \sum_{E, E'}  \,   \sqrt{q_E   \,  q_{E'}}   \,  \Tr  [  \rho_{E,E'}   ]   \\
     &  =  \sum_{E, E'}  \,   \sqrt{q_E   \,  q_{E'}}   \,  \|  \rho_{E,E'}   \|_1 \, ,   
  \end{align*}
  having used the fact that $\rho_{E,E'}$ is positive semidefinite, and therefore $\Tr  [  \rho_{E,E'}]   =   \|  \rho_{E,E'}  \|_1$.  
\qed 

 {\bf Proof of Theorem \ref{prop:optprob_mixed}.} Let $\map M$ be an energy-preserving quantum channel and let $M$ be its Choi operator, defined as \[M   =   (\map M  \otimes \map I)   (  |   I\>\!\>\<\!\<  I|  )  \, \]  
with  $|I\>\!\>  =  \sum_{n}  \,  |n\>| n\>$ being the unnormalized maximally entangled state  on $\spc H_1\otimes \spc H_2$,  $\spc H_1  \simeq  \spc H_2 \simeq \spc H$.  
 In the Choi representation, the energy-preserving condition    is equivalent to the requirement 
  \[  \Pi_0   \, M\,  \Pi_0   =   M  \,  ,   \qquad   \Pi_0  =  \bigoplus_E  \,  \left(   P_E\otimes P_E\right) \, . \]
 Now, the fidelity can be expressed as 
 \begin{align*}
 F  &  =   \frac{ \<\psi|   \map M  (\rho)  |\psi\>}{\Tr [\map M(\rho)]}\\
 &  =  \frac{      \Tr \left [  M   \, \left  (   |\psi\>\<\psi|   \otimes \rho^T\right)  \right ]}{       \Tr  \left[  M   \,  \left(  I   \otimes \rho^T\right) \right] } \\
 &  =  \frac{      \Tr \left [  M   \, \left  (   |\psi\>\<\psi|   \otimes \rho^T\right)  \right ]}{       \Tr  \left[  \Pi_0 M  \Pi_0   \,  \left(  I   \otimes \rho^T\right) \right] } \\
 &  =    \Tr  [  \sigma     \,  R^{-1/2}     \left( |\psi\>\<\psi|   \otimes \rho^T \right)    R^{-1/2}]   \, ,
   \end{align*}
  with   
  \begin{align}
  \nonumber R   & =   \Pi_0     \,(I   \otimes \rho^T )\,   \Pi_0      \\
  &   =   \bigoplus_E   \,  \left(   P_E  \otimes \rho_{E,E}^T \right)\,  \label{R}
  \end{align}
and   
\[\sigma  =  \frac{   \sqrt R    M \sqrt R}{\Tr  [  M R]} \, .  \]

Since $\sigma$ is a density matrix,  we have the achievable upper bound  
\begin{align*}   F    &\le   \sup_{  \sigma  :  \sigma  \ge 0  \,  , \Tr [\sigma]  =1}  \,  \Tr [  \sigma   R^{-1/2}     \left( |\psi\>\<\psi|   \otimes \rho^T \right)    R^{-1/2}   \,  ]     \\
&  =   \|   R^{-1/2}     \left( |\psi\>\<\psi|   \otimes \rho^T \right)    R^{-1/2}  \|_\infty \\
&  = \left\|   \sum_{E,E'}     \, \sqrt{q_E q_{E'} }\,  |\psi_E\>\<\psi_{E'}|  \otimes    \left(  \rho_{E,E}^{-\frac 12} \right)^T  \rho^T_{E,E'}  \left(  \rho_{E',E'}^{-\frac 12} \right)^T  \right\|_\infty  \\
& \equiv  \|  A\|_\infty  \, ,  
\end{align*}
achieved if and only if  the support of $\sigma$ is contained in the eigenspace of   $A$  with  maximum eigenvalue.    Hence, we must have  
\[  A\,   \sigma    =     F_{\max}  \,  \sigma\]
and 
\begin{align}\label{mgamma} M  =    \gamma  \,      R^{-1/2}  \, \sigma  \,     R^{-1/2}   \, ,   
\end{align}
for a suitable constant $\gamma>  0$.  Note that $\gamma$ is equal to the success probability:  indeed, we have
\begin{align*}
p_{\rm succ}  &   =  \Tr  [   M   \,   (I\otimes \rho^T)]  \\
&  =  \gamma  \,    \Tr[    R^{-1/2} \, \sigma  \, R^{-1/2}   \,      (I\otimes \rho^T)   ]  \\
&  =   \gamma\,  \sum_{E,E'} \Tr  [    (  P_E  \otimes  \rho_{E,E}^{T})^{-1/2}         \,\sigma      \\
&  \qquad \qquad    \times  (  P_{E'}  \otimes  \rho_{E',E'}^{T})^{-1/2}      \,      (I\otimes \rho^T)      ]\\
&  =   \gamma\,  \sum_{E} \Tr  \left\{ \sigma    \right  .           \\
& \qquad \qquad \times  \,   \left.  \left[P_E\otimes    (\rho_{E,E}^{T})^{-1/2}     \rho^T      (\rho_{E,E}^{T})^{-1/2}      \right]\right\} \\
&  =   \gamma\,  \sum_{E} \Tr  \left [ \sigma \,    (P_E\otimes       Q_E    )  \right] \, ,
\end{align*}  
where  $Q_E$ is the projector onto the support of $\rho_{E,E}$.  Now, note that the support of $A$ is contained in the support of the projector $P   =   \sum_E  P_E\otimes Q_E$.  Since the support of $\sigma$ is contained in the support of  $A$,  we conclude 
\begin{align*}  p_{\rm succ}     &=  \gamma  \, \Tr  \left [           \sigma         \,    \left( \sum_{E}  P_E\otimes       Q_E    \right)  \right]    \\
&  =  \gamma \, .  
\end{align*}
 Finally, the maximum value of $\gamma$ can be derived from the trace non-increasing property of the quantum operation $\map M$.  In terms of the Choi operator, the non-increasing property reads   $\Tr_{\rm 1} [  M]  \le I$, where $\Tr_{\rm out}$ denotes the trace on the output Hilbert space.  Using Eqs. (\ref{mgamma})  and (\ref{R}) we  obtain the relation  
 \begin{align*}
 \Tr_{1} [ M ]&   =  \gamma  \,     \sum_{E,E'} \Tr_{1} \left[    (  P_{E}  \otimes  \rho_{E,E}^{T})^{-1/2}     \,  \sigma   
 (  P_{E'}  \otimes  \rho_{E',E'}^{T})^{-1/2}  \right] \\
   &  =  \gamma  \,  \sum_E    \left(    \rho_{E,E}^T \right)^{-1/2}     \,  \sigma_E  \,       \left(    \rho_{E,E}^T \right)^{-1/2} \, ,
\end{align*} 
where $\sigma_E   :  =  \Tr_{\rm 1}   [   (P_E\otimes  I)       \,  \sigma  ]$ are the  unnormalized states on the input system, conditional to the outcomes of an energy measurement on the output system.     Hence, the trace non-increasing condition $\Tr_{ \rm 1}  [  M]  \le I$ is guaranteed if and only if   
\[  \gamma   \le      \min_E   \frac 1  {    \,  \left\|      \left(    \rho_{E,E}^T \right)^{-1/2}     \,  \sigma_E  \,       \left(    \rho_{E,E}^T \right)^{-1/2}    \right\|_\infty  }\]
In conclusion, the maximum probability of success is given by  
\[   p_{\rm succ}    =       \max_\sigma  \min_E  \,      \frac 1  {    \,  \left\|      \left(    \rho_{E,E}^T \right)^{-1/2}     \,  \sigma_E  \,       \left(    \rho_{E,E}^T \right)^{-1/2}    \right\|_\infty  }  \, .\]
\qed

\end{document}